\documentclass[twocolumn,aps,prl,showpacs,superscriptaddress,tightenlines]{revtex4-1}
\usepackage{array}
\usepackage{amsmath}
\usepackage{graphicx}
\usepackage{color}
\usepackage{amsfonts}
\usepackage{color}
\usepackage{txfonts}
\usepackage[colorlinks,citecolor=blue]{hyperref}
\usepackage{multirow}
\newcommand{\tabincell}[2]{\begin{tabular}{@{}#1@{}}#2\end{tabular}}
\def\narrowtext{\par\global\columnwidth20.5pc
\global\hsize\columnwidth\global\linewidth\columnwidth
\global\displaywidth\columnwidth}

\begin{document}
\title{Dark-Mode Theorems for Quantum Networks}
\author{Jian Huang}
\affiliation{Key Laboratory of Low-Dimensional Quantum Structures and Quantum Control of Ministry of Education, Key Laboratory for Matter Microstructure and Function of Hunan Province, Department of Physics and Synergetic Innovation Center for Quantum Effects and Applications, Hunan Normal University, Changsha 410081, China}

\author{Cheng Liu}
\affiliation{Key Laboratory of Low-Dimensional Quantum Structures and Quantum Control of Ministry of Education, Key Laboratory for Matter Microstructure and Function of Hunan Province, Department of Physics and Synergetic Innovation Center for Quantum Effects and Applications, Hunan Normal University, Changsha 410081, China}

\author{Xun-Wei Xu}
\affiliation{Key Laboratory of Low-Dimensional Quantum Structures and Quantum Control of Ministry of Education, Key Laboratory for Matter Microstructure and Function of Hunan Province, Department of Physics and Synergetic Innovation Center for Quantum Effects and Applications, Hunan Normal University, Changsha 410081, China}

\author{Jie-Qiao Liao}
\email{Corresponding author: jqliao@hunnu.edu.cn}
\affiliation{Key Laboratory of Low-Dimensional Quantum Structures and Quantum Control of Ministry of Education, Key Laboratory for Matter Microstructure and Function of Hunan Province, Department of Physics and Synergetic Innovation Center for Quantum Effects and Applications, Hunan Normal University, Changsha 410081, China}
\affiliation{Institute of Interdisciplinary Studies, Hunan Normal University, Changsha, 410081, China}

\begin{abstract}
We propose and prove two theorems for determining the number of dark modes in linear two-component quantum networks composed of two types of bosonic modes. This is achieved by diagonalizing the two sub-networks of the same type of modes, mapping the networks to either a standard or a thick arrowhead matrix, and analyzing the linear dependence and independence between the column vectors associated with degenerate normal modes in  the coupling matrix. We confirm the two theorems by checking the simultaneous ground-state cooling of the mechanical modes in linearized optomechanical networks. These results also work for linear fermionic networks and other networks described by quadratic coupled-mode Hamiltonian.  The present method can be extended to study the dark-state effect in driven atom systems and to construct large decoherence-free subspaces for processing quantum information. This work will initiate the studies on dynamical, transport, and statistical properties of linear networks with decoupled subspaces.
\end{abstract}
\maketitle
\narrowtext

\emph{Introduction.}---Quantum networks~\cite{Kimble2008,Wehner2018} are typically composed by quantum nodes and quantum connections. The nodes are realized by quantum systems such as atoms and ions~\cite{Cirac1997,Raimond2001,Hammerer2010,Duan2010,Reiserer2015,Reiserer2015,Covey2023}, bosonic modes~\cite{Carusotto2020,Pino2022,Chen2023,Saxena2023}, and even synthetic dimensions~\cite{Regensburger2012,Fang2012,Celi2014,Schmidt2015,Lustig2019,Ozawa2019,Lustig2021}. The connections could be either practical quantum channels such as optical fiber and quantum data bus or the direct couplings existing between different nodes~\cite{Cirac1997,Regensburger2011,Patel2018,Flamini2019}. Bosonic-mode network has become an important physical research object, because it can describe various physical setups such as coupled cavities and waveguides~\cite{Hartmann2006,Greentree2006,Houck2012,Habraken2012,Vermersch2017,Kollar2019}, cold atoms in optical lattices~\cite{Jaksch1998,Greiner2002,Gross2017}, coupled-oscillator networks~\cite{Pietras2019,Ren2010,Martens2013,Okamoto2013,Kuzyk2018,Csaba2020}, optomechanical networks~\cite{Aspelmeyer2014,Heinrich2011,Stannigel2011,Safavi-Naeini2011,Xuereb2012,Armstrong2012,Tomadin2012,Bochmann2013,Mari2013,Ludwig2013,Xuereb2014,Zhang2015,Dong2015,Moore2016,Cai2017,Peterson2017,Wengerowsky2018,Han2020,Arnold2020}, synthetic photonic lattices~\cite{Regensburger2012,Fang2012,Celi2014,Schmidt2015,Lustig2019,Ozawa2019,Lustig2021}, various bosonic lattices in condensed matter physics and statistical physics~\cite{Morsch2006,Lewenstein2007,Bhattacharya2008,Cazalilla2011}.

Quantum interference plays an inherent role in quantum networks because there exist many possible transition paths in networks. One of the prestigious physical effects caused by quantum interference is the dark-mode effect~\cite{Dong2012,Wang2012a,Tian2012,Genes2008a,Massel2012,Shkarin2014,Kuzyk2017,Ockeloen-Korppi2019,Sommer2019,Sommer2020,Naseem2021,Lai2020,Lai2022,Huang2022,Liu2022}.
The dark modes decouple from other subsystems of the networks and hence will affect the transport of excitation energy, quantum state, and quantum information in the networks~\cite{Genes2008a,Lai2020,Lai2022,Huang2022,Liu2022}. In linearized optomechanical networks, for example,  the optical dark mode has been positively used to realize adiabatic quantum state transfer between two optical modes coupled a common mechanical mode~\cite{Dong2012,Wang2012a,Tian2012}. Moreover, the mechanical dark modes can also affect negatively the simultaneous ground-state cooling of multiple mechanical modes, further suppress the generation of various quantum resources~\cite{Genes2008a,Lai2020,Lai2022,Huang2022,Liu2022}. Optomechanical networks provide a fruitful platform for studying novel quantum effects~\cite{Bose1997,Marshall2003,Liao2016,Riedinger2018,Ockeloen-Korppi2018,Kotler2021,Lepinay2021}, modern quantum applications~\cite{Massel2011,Huang2013,Peano2015b,Barzanjeh2022}, many-body physics~\cite{Heinrich2011,Ludwig2013,Xuereb2014}, topological physics~\cite{Xu2016}, and other important topics in condensed matter physics and statistical physics~\cite{Fradkin2013}. Therefore, as a prerequisite,  how to judge the existence of dark modes in optomechanical networks becomes a significant task in this field. Currently, most studies on the dark-mode effect focus on few-mode optomechanical systems~\cite{Dong2012,Wang2012a,Tian2012,Genes2008a,Massel2012,Shkarin2014,Kuzyk2017,Ockeloen-Korppi2019,Sommer2019,Sommer2020,Naseem2021}. A general method for determining the  number of dark modes in an optomechanical network and even a  general linear quantum network keeps unclear.



In this Letter, we propose the arrowhead-matrix method to explore the parameter conditions for determining the number of dark modes in a linear two-component quantum network composed of two types of bosonic modes. By diagonalizing the two sub-networks of the same type of modes, the network can be  transformed to a bipartite-graph network and hence the system can be described by either an arrowhead matrix or a thick arrowhead matrix. Based on the linear dependence and independence of the coupling vectors in the degenerate coupling sub-matrices, we propose and prove two theorems for determining the number of dark modes. The theorems are confirmed by checking the  simultaneous ground-state cooling of these mechanical modes in optomechanical networks including either single optical mode or multiple optical modes coupled to these mechanical modes.

\emph{Linear quantum networks.}---Consider a linear quantum network consisting of  two types of bosonic modes ($M$ type-$a$ modes denoted by $a_{k=1\text{-}M}$ and $N$ type-$b$ modes denoted by $b_{j=1\text{-}N}$). These type-$a$ modes and type-$b$ modes form two sub-networks, and  there exist inter and intra excitation-hopping interactions between these modes in the two sub-networks~\cite{SMaterial}.  The Hamiltonian of the network reads~\cite{SMaterial}
\begin{eqnarray}
\label{LinearMN}
H=(\boldsymbol{a}^{\dag},\boldsymbol{b}^{\dag})\mathbf{H}_{ab}(\boldsymbol{a},\boldsymbol{b})^{T},
\end{eqnarray}
where  $\boldsymbol{a}=(a_{1},a_{2},\cdots,a_{M})^{T}$ $[\boldsymbol{a}^{\dag}=(a_{1}^{\dag},a_{2}^{\dag},\cdots,a_{M}^{\dag})]$ and $\boldsymbol{b}=(b_{1},b_{2},\cdots,b_{N})^{T}$ $[\boldsymbol{b}^{\dag}=(b_{1}^{\dag},b_{2}^{\dag},\cdots,b_{N}^{\dag})]$ are the annihilation (creation)-operator  column (row)  vectors of the type-$a$ and type-$b$ modes, respectively. The operators satisfy the nonzero commutation relations: $[a_{k},a^{\dag}_{k^{\prime}}]=\delta_{kk^{\prime}}$ and $[b_{j},b^{\dag}_{j^{\prime}}]=\delta_{jj^{\prime}}$. The coefficient matrix in the bare-mode ($a_{k}$, $b_{j}$) representation is given by
\begin{equation}
\label{matrixmMN}
\mathbf{H}_{ab}=\left(
\begin{array}{cc}
\textbf{H}_{a} &\textbf{C}_{ab}  \\
\textbf{C}_{ab}^{\dag} & \textbf{H}_{b}
\end{array}\right),
\end{equation}
where $\textbf{H}_{a}$  and $\textbf{H}_{b}$ are, respectively, the matrices related to the type-$a$ and type-$b$ sub-networks, defined by $(\textbf{H}_{a})_{kk^{\prime}}=(\textbf{H}_{a})^{\ast}_{k^{\prime}k}=\xi_{kk^{\prime}}$ for $k<k^{\prime}$ and $(\textbf{H}_{a})_{kk}=\delta_{k}$ ($k,k^{\prime}=1,2,\cdots,M$),  $(\textbf{H}_{b})_{jj^{\prime}}=(\textbf{H}_{b})^{\ast}_{j^{\prime}j}=\eta_{jj^{\prime}}$ for $j<j^{\prime}$ and $(\textbf{H}_{b})_{jj}=\omega_{j}$ ($j,j^{\prime}=1,2,\cdots,N$). Here, $\xi_{kk^{\prime}}$ ($\eta_{jj^{\prime}}$) are the type-$a$ ($b$) excitation-hopping coupling strengths, $\delta_{k}$ and $\omega_{j}$ are the resonance frequencies of modes $a_{k}$ and $b_{j}$, respectively. The coupling matrix $\textbf{C}_{ab}$ describes the coupling between the two types of modes, defined by the elements $(\textbf{C}_{ab})_{kj}=g_{kj}$, which is the coupling strength between modes $a_{k}$ and $b_{j}$~\cite{SMaterial}.

To analyze the dark-mode existence condition in this linear network, we first diagonalize the type-$a$ mode sub-network with the unitary matrix $\mathbf{U}_{a}$ as $\mathbf{U}_{a}\mathbf{H}_{a}\mathbf{U}_{a}^{\dag}=\mathbf{H}_{A}=\text{diag}\{\Delta_{1},\Delta_{2},\ldots ,\Delta_{M}\}$, where $\Delta_{k}$ is the effective resonance frequency of the $k$th type-$a$ normal mode $A_{k}=\sum_{k^{\prime}}(\textbf{U}_{a})_{kk^{\prime}}a_{k^{\prime}}$, with $[A_{k},A^{\dag}_{k^{\prime}}]=\delta_{kk^{\prime}}$. We also introduce a unitary matrix $\textbf{U}_{b}$ to diagonalize the type-$b$ sub-network  $\textbf{H}_{b}$, i.e., $\textbf{U}_{b}\textbf{H}_{b}\textbf{U}_{b}^{\dag}=\textbf{H}_{B}=\text{diag}\{\Omega_{1},\Omega_{2},\ldots,\Omega_{N}\}$,  where $\Omega_{j}$ is the effective frequency of the $j$th type-$b$ normal mode $B_{j}=\sum_{j^{\prime}}(\textbf{U}_{b})_{jj^{\prime}}b_{j^{\prime}}$, with $[B_{j},B^{\dag}_{j^{\prime}}]=\delta_{jj^{\prime}}$.  In this case, the  network is transformed to a bipartite-graph-coupling configuration, i.e., there are no direct interactions between any two type-$a$ ($b$) normal modes~\cite{SMaterial}.  The  Hamiltonian of this  bipartite-graph network can be expressed as $H=(\mathbf{A}^{\dag},\mathbf{B}^{\dag})\mathbf{H}_{AB}(\mathbf{A},\mathbf{B})^{T}$, where the matrix in the ($A_{k}$, $B_{j}$) representation reads $\textbf{H}_{AB}=\{\{\textbf{H}_{A},\textbf{C}_{AB}\},\{\textbf{C}_{AB}^{\dag},\textbf{H}_{B}\}\}$,
with $(\textbf{C}_{AB})_{kj}\equiv G_{kj}=\sum_{k^{\prime}j^{\prime}}(\textbf{U}_{a})_{kk^{\prime}}g_{k^{\prime}j^{\prime}}(\textbf{U}^{\dag}_{b})_{jj^{\prime}}$ being the effective coupling strength between modes $A_{k}$ and $B_{j}$. Below we analyze the existence condition for dark modes in this $(M+N)$-normal-mode network.

\emph{Dark-mode theorems.}---We first present the definition of dark modes: In the bipartite-graph network formed by type-$a$ and type-$b$ normal modes, if a certain superposition of type-$b$  modes is decoupled from all the type-$a$ modes, then this superposed type-$b$ mode is regarded as a dark mode with respect to all these type-$a$ modes. Conversely, we can also define a type-$a$ dark mode with respect to those type-$b$ modes.

(1) \emph{The case of  $M=1$ and $N\geq2$ }. To expound the dark-mode existence condition more clearly, we first consider the case of one type-$a$ mode $a_{1}$ and $N$ type-$b$ modes $b_{j=1\text{-}N}$. Follow the above steps, this $(1+N)$-mode network can be transformed to a star-coupling network, i.e.,  the mode $A_{1}$ is coupled to modes $B_{j=1\text{-}N}$. In the case $M=1$, we have $A_{1}=a_{1}$, $\Delta_{1}=\delta_{1}$, and $\textbf{U}_{a}=I_{a}$. The dark-mode effect can be determined by the following Theorem~$1$.

\textbf{Theorem 1.}---\emph{Consider a $(1+N)$-mode linear quantum network described by the Hamiltonian $H=(\mathbf{A}^{\dag},\mathbf{B}^{\dag})\boldsymbol{H}_{AB}(\mathbf{A},\mathbf{B})^{T}$, the coefficient matrix is given by an arrowhead matrix $\textbf{H}_{AB}=\{\{\Delta_{1},\textbf{C}_{AB}\},\{\textbf{C}_{AB}^{\dag},\textbf{H}_{B}\}\}$,
where the coupling matrix is a row vector $\textbf{C}_{AB}=(G_{11},G_{12},\cdots,G_{1N})$ and $\textbf{H}_{B}=\text{diag}\{\Omega_{1},\Omega_{2},\ldots,\Omega_{N}\}$ with an ascending order.}

(i) \emph{If the element $G_{1j}=0$, then the corresponding mode $B_{j}$ is a dark mode.}

(ii) \emph{If all the elements of $\textbf{C}_{AB}$ are nonzero and $l$ type-$b$ modes ($B_{j=1\text{-}l}$) are degenerate, i.e., $\Omega_{j=1\text{-}l}=\Omega$, then in the mode space spanned by these degenerate modes, there are one bright mode $B_{l+}=(\vert G_{1l}\vert ^{2}+\vert G_{1( l-1) +}\vert ^{2})^{-1/2}(G_{1l}B_{l}+G_{1(l-1) +}B_{( l-1) +})=(\sum_{j=1}^{l}\vert G_{1j}\vert ^{2})^{-1/2}\sum_{j=1}^{l}G_{1j}B_{j}$ and $l-1$ dark modes: $B_{2-}$, $B_{3-}$, ..., and $B_{l-}$, where $B_{j-} =(\vert G_{1( j-1) +}\vert
^{2}+\vert G_{1j}\vert ^{2})^{-1/2}( G_{1j}^{\ast }B_{(j-1) +}-G_{1( j-1) +}B_{j})$ with $B_{(j-1) +}=(\vert G_{1( j-2) +}\vert^{2}+\vert G_{1(j-1)}\vert ^{2})^{-1/2}(G_{1( j-2)+}B_{( j-2) +}+G_{1(j-1)}B_{j-1})$ and $G_{1j+}=(\vert G_{1j}\vert ^{2}+\vert G_{1( j-1)+}\vert ^{2})^{1/2}$. Note that the forms of these dark modes are not unique, depending on the grouping order of these degenerate type-$b$ modes.}

(iii) \emph{If all the coupling strengths are nonzero, and all the type-$b$ modes are nondegenerate, i.e., $G_{1j}\neq0$ for $j=1\text{-}N$ and $\Omega_{j}\neq\Omega_{j^{\prime}}$ for all $j\neq j^{\prime}$, then there is no type-$b$ dark mode in the network.}

\emph{Proof of T1}(i): When $G_{1j}=0$, the mode $B_{j}$ decouples from $a_{1}$ and other modes $B_{j^{\prime}}$ (for $j^{\prime}\neq j$), so $B_{j}$ becomes a dark mode.

\emph{Proof of T1}(ii): Firstly, when $\Omega_{1}=\Omega_{2}=\Omega$, we choose the two modes $B_{1}$ and $B_{2}$ to  be hybridized as $B_{2+}=(G^{2}_{11}+G^{2}_{12})^{-1/2}(G_{11}B_{1}+ G_{12}B_{2})$ and  $B_{2-}=(G^{2}_{11}+G^{2}_{12})^{-1/2}(G_{12}B_{1}- G_{11}B_{2})$. Then we have $\Omega(B_{2+}^{\dag}B_{2+}+B_{2-}^{\dag}B_{2-})=\Omega(B^{\dag}_{2}B_{1}+B_{1}^{\dag}B_{2})$ and $A_{1}^{\dag}(G_{11}B_{1}+G_{12}B_{2})+\text{H.c.}=(G^{2}_{11}+G^{2}_{12})^{1/2}A^{\dag}_{1}B_{2+}+\text{H.c.}$. Therefore, the mode $B_{2-}$ decouples from modes $A_{1}$, $B_{2+}$, and $B_{l\neq1,2}$, then we obtain the bright mode $B_{2+}$ and dark mode $B_{2-}$. Secondly, we group the modes $B_{2+}$ with the next degenerate mode $B_{3}$ to form a bright mode $B_{3+}$ and a dark mode $B_{3-}$. Repeat  this process many times, we can group modes $B_{j+}$ and $B_{j+1}$ to form a bright mode $B_{(j+1)+}$ and a dark mode $B_{(j+1)-}$. Finally, by grouping the modes $B_{(l-1)+}$ and $B_{l}$, we obtain the bright mode $B_{l+}$ and dark mode $B_{l-}$. Therefore, there are one bright mode $B_{l+}$ and $l-1$ dark modes $B_{j-}$ for $j=2$, $3$, $...$, $l$. The detailed proof is presented in~\cite{SMaterial}.

\emph{Proof of T1}(iii): When $G_{1j}\neq0$ and $\Omega_{j}\neq\Omega_{j^{\prime}}$, the eigenvalues $\lambda_{i}$ of $\mathbf{H}_{AB}$ satisfy the interlacing property~\cite{leary1990,Gu1995,Vandebril2008,Stor2015}, i.e., $\lambda_{1}<\Omega_{1}<\lambda_{2}<\Omega_{2}<\ldots<\lambda_{N}<\Omega_{N}<\lambda_{N+1}$. Moreover, $\lambda_{i}$ can be obtained by solving the secular equation $f(\lambda) =\lambda-\Delta_{1}+\sum_{i=1}^{N}\vert G_{1i}\vert^{2}/(\Omega_{i}-\lambda)=0$. The eigenvector of $\textbf{H}_{AB}$ corresponding to $\lambda_{i}$ is given by $\textbf{v}_{i}=[ -1,G^{\ast}_{11}/(\Omega _{1}-\lambda _{i}),G^{\ast}_{12}/(\Omega_{2}-\lambda_{i}),\ldots,G^{\ast}_{1N}/(\Omega_{N}-\lambda_{i})]^{T}/[1+\sum_{j=1}^{N}\vert G_{1j}\vert^{2}/(\Omega _{j}-\lambda_{i})^{2}]^{-1/2}$. The eigenvectors show that the coefficient related to the mode $A_{1}$ is $-1$, i.e., the $N$ type-$b$ normal modes are always coupled to the mode $A_{1}$, and hence type-$b$ dark mode does not exist  in this case.

(2) \emph{The case of $M\geq2$ and $N\geq2$}. We can generalize the arrowhead-matrix method to a general network consisting of $M\geq2$ type-$a$ modes and $N\geq2$ type-$b$ modes.  The $(M+N)$-mode network can be transformed to the bipartite-graph-coupling configuration~\cite{SMaterial},  which can be described by a thick (the coupling matrix $\textbf{C}_{AB}$ has a dimension $M\times N$ for $M\geq2$ ) arrowhead matrix $\textbf{H}_{AB}$. The dark-mode effect in this $(M+N)$-mode network can be confirmed by the following Theorem~$2$.

\textbf{Theorem~2.}---\emph{Consider a $(M+N)$-mode linear quantum network described by the Hamiltonian $H=(\mathbf{A}^{\dag},\mathbf{B}^{\dag})\boldsymbol{H}_{AB}(\mathbf{A},\mathbf{B})^{T}$, the coefficient matrix  is given by a thick arrowhead matrix $\textbf{H}_{AB}=\{\{\textbf{H}_{A},\textbf{C}_{AB}\},\{\textbf{C}_{AB}^{\dag},\textbf{H}_{B}\}\}$, where $\textbf{C}_{AB}=[\textbf{G}_{1},\textbf{G}_{2},\cdots,\textbf{G}_{N}]$ with column vectors $\textbf{G}_{j}=[G_{1j},G_{2j},\cdots,G_{Mj}]^{T}$}, $\mathbf{H}_{A}=\text{diag}\{\Delta_{1},\Delta_{2},\ldots ,\Delta_{M}\}$, and $\textbf{H} _{B}=\text{diag}\{\Omega_{1},\Omega_{2},\ldots,\Omega_{N}\}$.

(i) \emph{If the $j$th column vector $\textbf{G}_{j}=\textbf{0}$, then the corresponding mode $B_{j}$ is a type-$b$ dark mode.}

(ii) \emph{If $l$ (for $l=2-N$) type-$b$ modes are degenerate (for example $\Omega_{j=1\text{-}l}=\Omega$), and all these corresponding column vectors in $\textbf{C}_{AB}$ are linearly dependent, then in the degenerate-mode subspace, there are one type-$b$ bright mode and ($l-1$) type-$b$ dark modes.}

(iii) \emph{In the case of $M$ type-$a$ normal modes and $N$ degenerate type-$b$ normal modes ($\Omega_{j=1\text{-}N}=\Omega$),  there are at least $N-M$ type-$b$ dark modes when $N>M$~\cite{SMaterial}.}

(iv) \emph{When all the $N$ type-$b$ normal modes can be divided into different groups with the same resonance frequency (each group corresponds to a coupling sub-matrix formed by the corresponding coupling column vectors $\textbf{G}_{j}$, the dimension of the sub-matrix could be $M\times1$ for a single non-degenerate mode). The total number of type-$b$ bright modes in the network is $R=\sum_{s} R_{s}$, where $R_{s}$ is the rank of the $s$th submatrix in the coupling matrix $\textbf{C}_{AB}$, and  the number of the type-$b$ dark modes is $N-R$.}

(v) \emph{The change of the coupling strengths $\xi_{kk^{\prime}}$  between these type-$a$ modes $a_{k}$ and $a_{k^{\prime}}$ will not change the number of the type-$b$ dark modes, but will change the forms of these type-$b$ dark modes.}

(vi) \emph{If $\textbf{G}_{j}\neq\textbf{0}$ for $j=1\text{-}N$ and $\Omega_{j}\neq\Omega_{j^{\prime}}$ for all  $j\neq j^{\prime}$, then there is no type-$b$ dark mode in the system.}

\emph{Proof of T2}(i): When the $j$th column vector $\textbf{G}_{j}=\textbf{0}$, the type-$b$ normal mode $B_{j}$  is decoupled from all these type-$a$ normal modes, so it becomes a  dark mode. The proofs of Theorems 2(ii)-2(v)  are presented in~\cite{SMaterial}.

\emph{Proof of T2}(vi): When $\textbf{G}_{j}\neq\textbf{0}$, the mode $B_{j}$ will not decouple from all the type-$a$ modes, then $B_{j}$ is not a dark mode. When $\textbf{G}_{j}\neq\textbf{0}$ is satisfied for all $j=1\text{-}N$, then all these normal modes $B_{j=1\text{-}N}$ are not dark modes. In addition, when $\Omega_{j}\neq\Omega_{j^{\prime}}$ for all $j\neq j^{\prime}$, then the superposition of $B_{j}$ will not be dark modes.

\begin{figure}[tbp]
\center
\includegraphics[width=0.47 \textwidth]{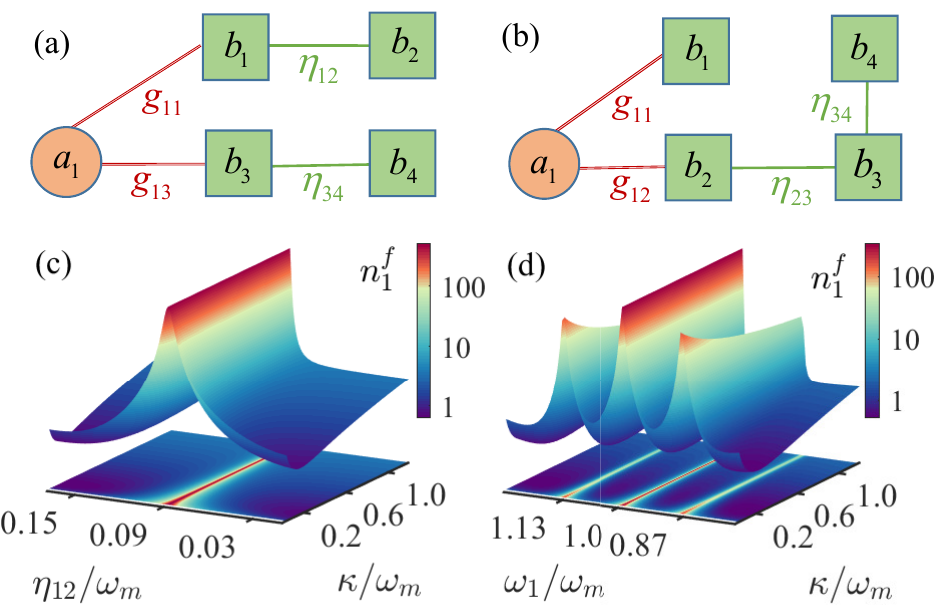}
\caption{Two different $[1,4]$-mode  optomechanical systems consisting of  an optical mode and four mechanical modes. (a) [(b)] There are two (one) mechanical modes and two (three) mechanical modes on the left and right of the optical mode, respectively.  (c) Final mean phone number $n_{1}^{f}$ of the mechanical mode $b_{1}$ in Fig.~\ref{Fig2}(a) versus $\eta_{12}/\omega_{m}$ and $\kappa/\omega_{m}$ when $\omega_{1}/\omega_{m}=1$,  $g_{11}/\omega_{m}=g_{13}/\omega_{m}=0.1$, and $\eta_{34}/\omega_{m}=0.09$. (d) $n_{1}^{f}$ in the Fig.~\ref{Fig2}(b) versus $\omega_{1}/\omega_{m}$ and $\kappa/\omega_{m}$ when $g_{11}/\omega_{m}=g_{12}/\omega_{m}=0.1$ and $\eta_{23}/\omega_{m}=\eta_{34}/\omega_{m}=0.09$. Other parameters used are $\delta_{1}/\omega_{m}=1$,  $\omega_{l=2\text{-}4}/\omega_{m}=1$, $\kappa/\omega_{m}=0.1$, $\gamma_{l=1\text{-}4}/\omega_{m}=10^{-5}$, and $\bar{n}_{l=1\text{-}4}=10^{3}$.}
\label{Fig2}
\end{figure}

\emph{Example: Dark-mode effect in linearized optomechanical networks.}---To check the two dark-mode theorems, we analyze the mechanical dark-mode effect in  linearized optomechanical networks consisting of either one or many optical modes and multiple mechanical modes. These optical (mechanical) modes could couple with each other through photon (phonon)-hopping interactions, and the optical modes are coupled to mechanical modes via the linearized optomechanical interactions. In optomechanical networks, the optical modes are the cooling channels of these mechanical modes. Therefore, the dark mode cannot be cooled because it decouples from the optical modes, and hence the existence of dark modes can be checked by examining the simultaneous ground-state cooling of these mechanical modes. Below, as example, we first study the cooling of mechanical modes in two different $[1,4]$-mode  optomechanical networks [Figs.~\ref{Fig2}(a) and~\ref{Fig2}(b)]. The linearized Hamiltonians of these two configurations are, respectively, described by $H_{(a)}^{[1,4]}=\delta_{1}a_{1}^{\dagger}a_{1}+\sum_{j=1}^{4}\omega_{j}b_{j}^{\dagger}b_{j}+\sum_{j=1,3}(g_{1j}a_{1}^{\dagger}b_{j}+\mathrm{H.c.})+(\eta_{12}b_{1}^{\dagger}b_{2}+\eta_{34}b_{3}^{\dagger}b_{4}+\mathrm{H.c.})$ and  $H_{(b)}^{[1,4]}=\delta_{1}a_{1}^{\dagger}a_{1}+\sum_{j=1}^{4}\omega_{j}b_{j}^{\dagger}b_{j}+\sum_{j=1,2}(g_{1j}a_{1}^{\dagger}b_{j}+\mathrm{H.c.})+(\eta_{23}b_{2}^{\dagger}b_{3}+\eta_{34}b_{3}^{\dagger}b_{4}+\mathrm{H.c.})$,  where the operators and parameters have been defined in Eq.~(\ref{LinearMN}).

By diagonalizing the mechanical-mode sub-network,  we obtain the total coupling matrix, and find that the coupling strengths $G_{1j}$ in these two networks are nonzero. Based on Theorem 1(ii), the existence conditions of the dark modes depend on the resonance frequencies $\Omega_{j}$ of the mechanical normal modes. For the system depicted in Fig.~\ref{Fig2}(a), we assume $\omega_{j=1\text{-}4}=\omega_{m}$ and obtain $\textbf{H}_{B}^{(a)}=\text{diag}\{\omega_{m}-\eta_{12},\omega_{m}+\eta_{12},\omega_{m}-\eta_{34},\omega_{m}+\eta_{34}\}$.  Therefore, the dark modes appear only when $\eta_{12}=\eta_{34}$. For the other system depicted in Fig.~\ref{Fig2}(b), we  assume $\omega_{j=2\text{-}4}=\omega_{m}$ and obtain $\textbf{H}_{B}^{(b)}=\text{diag}\{\omega_{1},\omega_{m},\omega_{m}-\sqrt{2}\eta,\omega_{m}+\sqrt{2}\eta\}$ when $\eta_{23}=\eta_{34}=\eta$.  In this case, we can sweep $\omega_{1}$ to obtain two degenerate normal modes in three cases, i.e., when $\omega_{1}=\omega_{m}$ or $\omega_{m}\pm\sqrt{2}\eta$, there is a mechanical dark mode in this system.

In Figs.~\ref{Fig2}(c) and~\ref{Fig2}(d) we plot $n_{1}^{f}$ as functions of ($\eta_{12}/\omega_{m}$, $\kappa/\omega_{m}$) and ($\omega_{1}/\omega_{m}$, $\kappa/\omega_{m})$ corresponding to the configuration in Figs.~\ref{Fig2}(a) and~\ref{Fig2}(b), respectively. Without loss of generality, here we only plot $n_{1}^{f}$ for the mechanical mode $b_{1}$. Figure~\ref{Fig2}(c) shows that the cooling of $b_{1}$ is greatly suppressed in a finite-detuning window when  $\eta_{12}/\omega_{m}=\eta_{34}/\omega_{m}=0.09$. This phenomenon is caused by the dark-mode effect. When $\eta_{12}=\eta_{34}$, the interference between modes $B_{1}$ and $B_{3}$ leads to a dark mode, while that for modes $B_{2}$ and $B_{4}$ leads to another dark mode. In this case, phonon excitations cannot be extracted from the dark modes through the optomechanical-cooling channel, and hence the simultaneous ground-state cooling for the near-degenerate and degenerate normal modes is unfeasible. In Fig.~\ref{Fig2}(d) we see that the cooling of mode $b_{1}$ is suppressed around $\omega_{1}/\omega_{m}=1.13$, $\omega_{1}/\omega_{m}=1$, and $\omega_{1}/\omega_{m}=0.87$, which means that the system has a dark mode at these three points.  By combining the effective frequencies of normal modes in $\textbf{H}_{B}^{(b)}$, we find that the normal mode $B_{1}$ resonates with $B_{2}$, $B_{3}$, and $B_{4}$ at $\omega_{m}$, $\omega_{m}-\sqrt{2}\eta$, and $\omega_{m}+\sqrt{2}\eta$, respectively, i.e., $\omega_{1}/\omega_{m}=1$, $0.87$, and $1.13$ in Fig.~\ref{Fig2}(d). These results confirm the validity of Theorem 1. Note that we also checked Theorem 1 in the $(1+N)$-mode optomechanical networks by considering all the configurations when $N=2$, $3$, and $4$~\cite{SMaterial}.

\begin{figure}[tbp]
\center
\includegraphics[width=0.48 \textwidth]{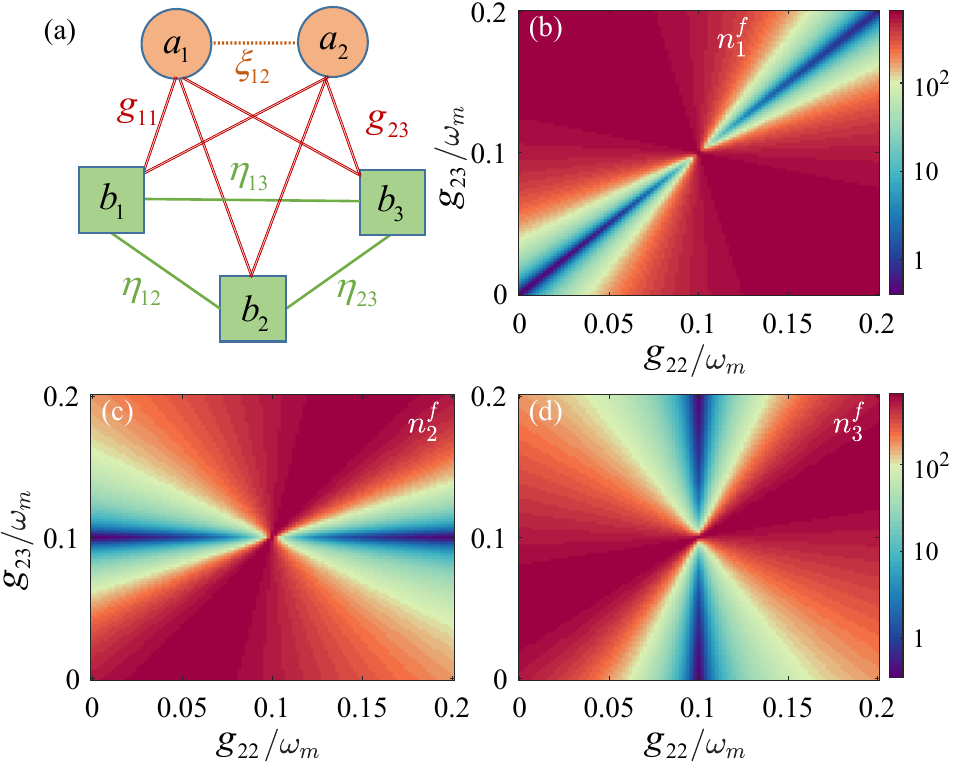}
\caption{(a) $[2,3]$-mode optomechanical network consisting of two optical modes coupled to three mechanical modes, and any two optical (mechanical) modes are coupled with each other. (b) $n_{1}^{f}$,  (c) $n_{2}^{f}$, and (d) $n_{3}^{f}$ versus the coupling strengths $g_{22}/\omega_{m}$ and  $g_{23}/\omega_{m}$.  Other parameters are $\delta_{k=1\text{-}2}/\omega_{m}=1$,  $\omega_{j=1\text{-}3}/\omega_{m}=1$, $\kappa_{k=1\text{-}2}/\omega_{m}=0.1$, $\gamma_{j=1\text{-}3}/\omega_{m}=10^{-5}$, $g_{11}/\omega_{m}=g_{12}/\omega_{m}=g_{13}/\omega_{m}=g_{21}/\omega_{m}=0.1$, $\eta_{12}/\omega_{m}=\eta_{13}/\omega_{m}=\eta_{23}/\omega_{m}=0.09$, $\xi_{12}/\omega_{m}=0.08$, and  $\bar{n}_{j=1\text{-}3}=10^{3}$.}
\label{Fig3}
\end{figure}

We also check the cooling performance in the $[2,3]$-mode  optomechanical network [Fig.~\ref{Fig3}(a)]. The linearized Hamiltonian  reads $H^{[2,3]}=\sum_{k=1}^{2}\delta_{k}a_{k}^{\dagger}a_{k}+\sum_{j=1}^{3}\omega_{j}b^{\dagger}_{j}b_{j}+(\xi_{12}a_{1}^{\dagger}a_{2}+\mathrm{H.c.})+\sum_{j,j^{\prime}=1,j<j^{\prime}}^{3}(\eta_{jj^{\prime}}b_{j}^{\dagger}b_{j^{\prime}}+\mathrm{H.c.})+\sum_{k=1}^{2}\sum_{j=1}^{3}(g_{kj}a_{k}^{\dagger}b_{j}++\mathrm{H.c.})$. Follow the above steps, we consider $\delta_{1}=\delta_{2}=\delta$, $\omega_{j=1\text{-}3}=\omega_{m}$, $\eta_{12}=\eta_{13}=\eta_{23}=\eta$, and $g_{11}=g_{12}=g_{13}=g_{21}=0.1\omega_{m}$,   then the thick arrowhead matrix  can be written as $\textbf{H}_{AB}=\{\{\textbf{H}_{A},\textbf{C}_{AB}\},\{\textbf{C}_{AB}^{\dag},\textbf{H}_{B}\}\}$, with $\textbf{H}_{A}=\text{diag}\{\delta-\xi_{12},\delta+\xi_{12}\}$, $\textbf{H}_{B}=\text{diag}\{\omega_{m}-\eta,\omega_{m}-\eta,\omega_{m}+2\eta\}$, and
\footnotesize
\begin{eqnarray}
\label{matrixm23G}
&&\textbf{C}_{AB}=\left(
\begin{array}{ccc}
g_{23}-0.1\omega_{m} & \frac{\sqrt{3}}{6}\left(2g_{22}-g_{23}-0.1\omega_{m}\right) & \frac{\sqrt{6}}{6}\left(g_{22}+g_{23}-0.2\omega_{m}\right)  \\
g_{23}-0.1\omega_{m} & \frac{\sqrt{3}}{6}\left(2g_{22}-g_{23}-0.1\omega_{m}\right) & \frac{\sqrt{6}}{6}\left(g_{22}+g_{23}+0.4\omega_{m}\right)
\end{array}
\right).\nonumber\\
\end{eqnarray}
\normalsize
Here the mechanical normal modes are expressed as $B_{1}=(-b_{1}+b_{3})/\sqrt{2}$, $B_{2}=(-b_{1}+2b_{2}-b_{3})/\sqrt{6}$, and $B_{3}=(b_{1}+b_{2}+b_{3})/\sqrt{3}$.

In Figs.~\ref{Fig3}(b), ~\ref{Fig3}(c), and ~\ref{Fig3}(d) we display the final mean phonon numbers $n_{1}^{f}$, $n_{2}^{f}$, and   $n_{3}^{f}$ versus $g_{22}/\omega_{m}$ and $g_{23}/\omega_{m}$.  We observe from Fig.~\ref{Fig3}(b) that the ground-state cooling of mode $b_{1}$ is accessible when $g_{22}/\omega_{m}=g_{23}/\omega_{m}\neq0.1$, but it is suppressed in other areas. Figure~\ref{Fig3}(c)~[\ref{Fig3}(d)]  shows that the mode $b_{2}$ ($b_{3}$) can be cooled into ground state when $g_{23}/\omega_{m}=0.1$ and $g_{22}/\omega_{m}\neq0.1$ ($g_{22}/\omega_{m}=0.1$ and $g_{23}/\omega_{m}\neq0.1$). This phenomenon can be explained based on the coefficient matrix~(\ref{matrixm23G}) and the used parameters.

 In Fig.~\ref{Fig3}(b), we find that the ground-state cooling of the mode $b_{1}$ is strongly suppressed in most areas, but can be realized when $g_{22}/\omega_{m}=g_{23}/\omega_{m}\neq0.1$.  Since the elements in the first and second and columns are linearly dependent, we  know from Theorem 2 that one of the normal mode $B_{2-}=[3( g_{23}-0.1\omega_{m}) ^{2}+(2g_{22}-g_{23}-0.1\omega_{m}) ^{2}]^{-1/2}[ (2g_{22}-g_{23}-0.1\omega_{m}) B_{1}-\sqrt{3}(g_{23}-0.1\omega_{m}) B_{2}] $ becomes a dark mode, thus the cooling of mode $b_{1}$ is strongly suppressed in most areas. However, when $g_{22}/\omega_{m}=g_{23}/\omega_{m}\neq0.1$, the dark mode becomes $B_{2-}=(-b_{2}+b_{3})/\sqrt{2}$, which means that the mode $b_{1}$  is connected to optical modes, so it can be cooled to the ground state.  In Fig.~\ref{Fig3}(c), the results show that the mode $b_{2}$ can be cooled into ground state when $g_{23}/\omega_{m}=0.1$ and $g_{22}/\omega_{m}\neq0.1$. In this case, all the elements in first column become zero and the normal mode $B_{1}=(-b_{1}+b_{3})/\sqrt{2}$ becomes a dark mode, but mode $b_{2}$  is connected to optical modes, and hence can be cooled to the ground state. In Fig.~\ref{Fig3}(d), when $g_{22}/\omega_{m}=0.1$ and $g_{23}/\omega_{m}\neq0.1$,  the dark mode becomes $B_{2-}=(-b_{1}+b_{2})/\sqrt{2}$, which means that the mode $b_{3}$ is connected to optical modes, so the ground-state cooling of mode $b_{3}$ can be realized. However,  the cooling of $b_{3}$ is strongly suppressed in other areas due to the dark-mode effect.

\emph{Discussions and conclusion.}---The linear quantum networks under consideration is general, and can be implemented with various physical platforms. In particular, the coupled optomechanical cavities are good candidates to realize the present networks. In optomechanical networks, the linearized optomechanical coupling strengths can be tuned via changing the driven amplitude of the cavity field. Therefore, it is feasible to exhibit the dark-mode effect via sweeping the coupling strength. Moreover, we can choose the red-sideband resonance such that the rotating-wave approximation made in the optomechanical coupling is valid. These analyses indicate that the present theorems can be demonstrated with current experimental techniques.

In conclusion, we have proposed the arrowhead-matrix method to study the dark-mode effect in linear quantum networks consisting of two types of bosonic modes. In the normal-mode representation of the two sub-networks for the same-type modes, the network has been transformed  into a bipartite-graph network. We have proposed and proved two theorems for determining the dark modes in linear networks. We have also confirmed the validity of the theorems by examining  the mechanical cooling in linearized optomechanical networks. The theorems have wide potential applications in the study of quantum effects, energy and heat transport, and quantum state and information transfer on networks. They can also be extended to judge the dark-state (coherent trapping) effect in driven atomic systems because the coefficient matrix also works for them~\cite{Alzetta1976,Arimondo1976,Gray1978,Arimondo1996,Zhao2023}. The method can be used to study block-decoupling in bosonic networks, dynamical and statistical properties of linear systems possessing decoupled subspace, and construction of large decoherence-free subspace~\cite{Lidar1998,Lidar2003} for processing quantum information. It can also be extended to study more networks with one or multiple types of nodes~\cite{Newman2010}.

\begin{acknowledgments}
J.H. thanks Ye-Xiong Zeng for discussions. J.-Q.L. was supported in part by National Natural Science Foundation of China (Grants No.~12175061, No.~12247105, and No.~11935006), the Science and Technology Innovation Program of Hunan Province (Grant No.~2021RC4029), and Hunan Provincial Major Science and Technology Program (Grant No.~2023ZJ1010). X.-W.X. was supported by the National Natural Science Foundation of China (NSFC) (Grants No.~12064010 and No.~12247105), Natural Science Foundation of Hunan Province of China (Grant No.~2021JJ20036), and the science and technology innovation Program of Hunan Province (Grant No.~2022RC1203).
\end{acknowledgments}


\onecolumngrid
\newpage
\setcounter{equation}{0} \setcounter{figure}{0}
\setcounter{table}{0}
\setcounter{page}{1}\setcounter{secnumdepth}{3} \makeatletter
\renewcommand{\theequation}{S\arabic{equation}}
\renewcommand{\thefigure}{S\arabic{figure}}
\renewcommand{\bibnumfmt}[1]{[S#1]}
\renewcommand{\citenumfont}[1]{S#1}
\renewcommand\thesection{S\arabic{section}}

\begin{center}
{\large \bf Supplementary Material for ``Dark-Mode Theorems for Quantum Networks"}
\end{center}

This supplementary material is composed of seven Sections. In Sec.~\ref{transfor}, we introduce the two-component linear quantum networks expressed in both the bare- and normal-mode representations. In Sec.~\ref{twodarkmode}, we propose and prove two dark-mode theorems for linear quantum networks. In Sec.~\ref{modelmn}, we show the implementation of the type-($a, b$) ($M+N$)-mode linear quantum networks with linearized optomechanical networks. In Sec.~\ref{CooloneN}, we show the cooling of multiple mechanical modes in the ($1+N$)-mode optomechanical networks including one optical mode and $N$ mechanical modes.
In Sec.~\ref{Nmodel}, we show the cooling of multiple mechanical modes in the ($M+N$)-mode optomechanical networks including $M$ optical modes and $N$ mechanical modes. In Sec.~\ref{thedarkstate}, we show that our method can be extended to analyze the dark-state effect in driven atom systems. In Sec.~\ref{DFS}, we present the analyses concerning of the decoherence-free subspace of two two-level atoms coupled to a common bath based on the dark-mode theorems.

\section{Two-component linear quantum network expressed in both the bare- and normal-mode representations \label{transfor}}

\begin{figure}[tbp]
	\center
	\includegraphics[width=0.8 \textwidth]{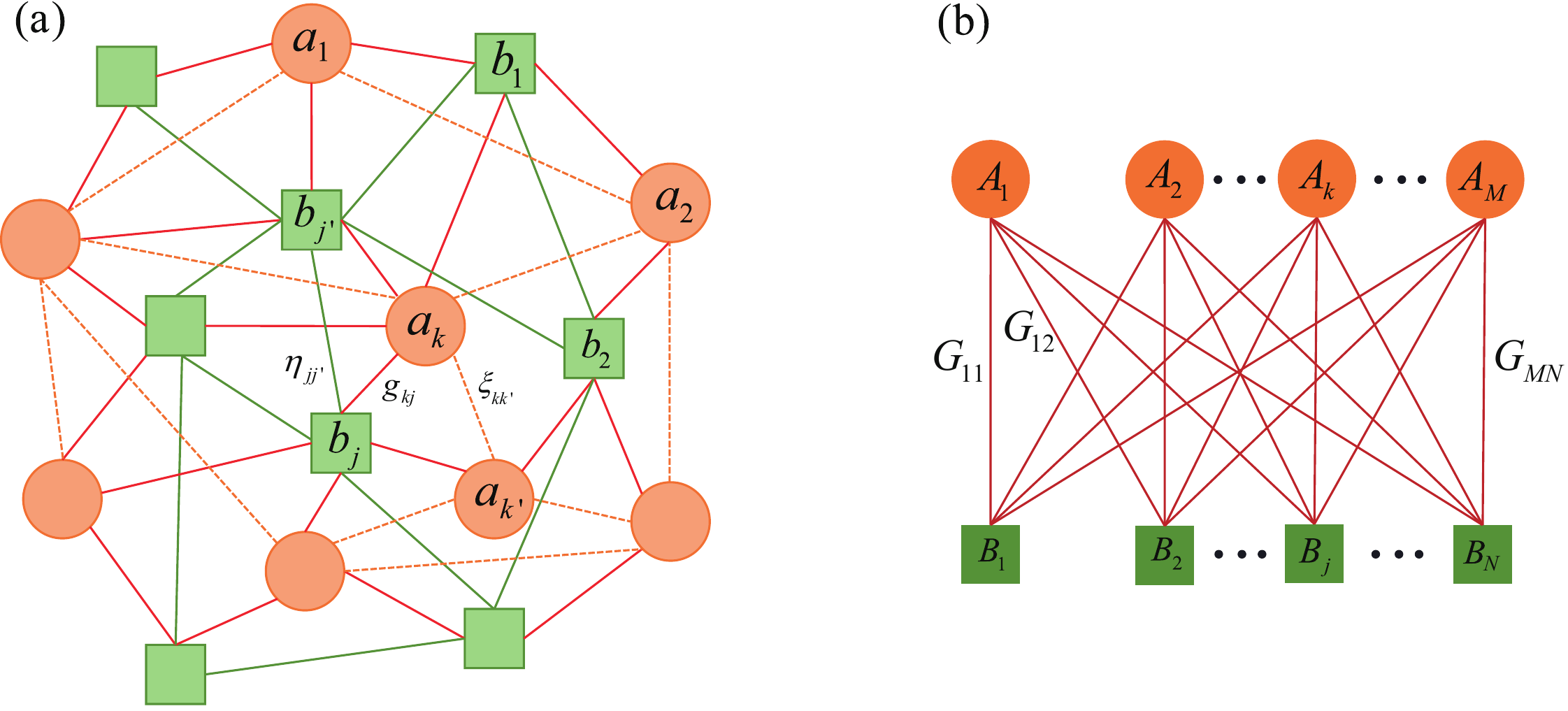}
	\caption{(a) Schematic of the two-component (modes $a_k$ and $b_j$) ($M+N$)-mode linear quantum network consisting of $M$ type-$a$ modes ($a_1$, $a_2$, $\dotsb$, $a_{M}$) and $N$ type-$b$ modes ($b_1$, $b_2$, $\dotsb$, $b_N$).  The modes in the network are coupled together through excitation-hopping interactions, $\xi_{kk^{\prime}}$ ($\eta_{jj^{\prime}}$) is the coupling strength between the modes $a_{k}$ and $a_{k^{\prime}}$ ($b_j$ and $b_{j^{\prime}}$), and $g_{kj}$ is the coupling strength between modes $a_k$ and $b_j$. (b) Schematic of the bipartite-graph network consisting of $M$ type-$a$ normal modes ($A_1$, $A_2$, $\dotsb$, $A_M$) coupled to $N$ type-$b$ normal modes ($B_1$, $B_2$, $\dotsb$, $B_N$). $G_{kj}$ is the coupling strength between the modes $A_k$ and $B_j$.}
	\label{FigS1}
\end{figure}

In this section, we introduce the two-component ($M+N$)-mode linear quantum network and present its Hamiltonian. We also present the detailed calculations concerning the transformation between the bare- and normal-mode representations of the $(M+N)$-mode linear network. As shown in Fig. \ref{FigS1}(a), the network consists of $M$ type-$a$ modes ($a_{1},a_{2},\dotsb,a_{M}$) and $N$ type-$b$ modes ($b_{1},b_{2},\dotsb,b_{N}$). The Hamiltonian of this ($M+N$)-mode linear quantum network can be written as ($\hbar=1$)
\begin{equation}
H=\sum_{k=1}^{M}\delta _{k}a_{k}^{\dagger}a_{k}+\sum_{j=1}^{N}\omega _{j} b_{j}^{\dagger} b_{j}+\sum_{\textcolor{black}{k,k^{\prime}=1,k<k^{\prime}}}^{M}(\xi_{kk^{\prime}}a_{k}^{\dagger}a_{k^{\prime}}+\xi_{kk^{\prime}}^{\ast}a_{k^{\prime}}^{\dagger}a_{k})+\sum_{\textcolor{black}{j,j^{\prime}=1,j<j^{\prime}}}^{N}(\eta_{jj^{\prime}}b_{j}^{\dagger}b_{j^{\prime}}+\eta_{jj^{\prime}}^{\ast}b_{j^{\prime}}^{\dagger}b_{j})+\sum_{k=1}^{M}\sum_{j=1}^{N}(g_{kj} a_{k}^{\dagger}b_{j}+g_{kj}^{\ast}b_{j}^{\dagger} a_{k} ),~\label{HMNMN11}
\end{equation}
where $a_{k}$ ($a_{k}^{\dagger}$) and $b_{j}$ ($b_{j}^{\dagger}$) are, respectively, the annihilation (creation) operators of the $k$th type-$a$ mode and the $j$th type-$b$ mode, with the corresponding resonance frequencies  $\delta_{k}$ and $\omega_{j}$. The parameter $\xi_{kk^{\prime}}$ ($\eta_{jj^{\prime}}$) is the coupling strength between the $k$th and $k^{\prime}$th ($j$th and $j^{\prime}$th) type-$a$ ($b$) modes, and $g_{kj}$ is the coupling strength between the $k$th type-$a$ mode $a_{k}$ and the $j$th type-$b$ mode $b_j$.

To keep conciseness, we introduce the vectors of annihilation and creation operators for type-$a$ and -$b$ modes,
\begin{eqnarray}
\label{vectorab}
\boldsymbol{a}&=&(a_{1},a_{2},a_{3},\cdots,a_{M})^{T},\nonumber\\
\boldsymbol{a}^{\dag}&=&(a^{\dag}_{1},a^{\dag}_{2},a^{\dag}_{3},\cdots,a^{\dag}_{M}),\nonumber\\
\boldsymbol{b}&=&(b_{1},b_{2},b_{3},\dotsb,b_{N})^{T},\nonumber\\
\boldsymbol{b}^{\dag}&=&(b^{\dag}_{1},b^{\dag}_{2},b^{\dag}_{3},\cdots,b^{\dag}_{N}),
\end{eqnarray}
where the superscript “$T$” denotes matrix transpose. Then Hamiltonian (\ref{HMNMN11}) can be expressed as
\begin{eqnarray}
	H=(\boldsymbol{a}^{\dag},\boldsymbol{b}^{\dag})\mathbf{H}_{ab}\left(\begin{array}{c}
		\boldsymbol{a} \\
		\boldsymbol{b}
	\end{array}
	\right), \label{HMNRWA}
\end{eqnarray}
where we introduce the coefficient matrix in the bare-mode ($a_k,b_j$) representation as
\begin{eqnarray}
\mathbf{H}_{ab}=\left(\begin{array}{cc}
	\mathbf{H}_{a} & \mathbf{C}_{ab}\\
	\mathbf{C}_{ab}^{\dagger} & \mathbf{H}_{b}
\end{array}\right)=\left(\begin{array}{cccc|cccc}
	\delta_{1} & \xi_{12} & \cdots & \xi_{1M} & g_{11} & g_{12} & \cdots & g_{1N}\\
	\xi_{12}^{\ast} & \delta_{2} & \cdots & \xi_{2M} & g_{21} & g_{22} & \cdots & g_{2N}\\
	\vdots & \vdots & \ddots & \vdots & \vdots & \vdots & \ddots & \vdots\\
	\xi_{1M}^{\ast} & \xi_{2M}^{\ast} & \cdots & \delta_{M} & g_{M1} & g_{M2} & \cdots & g_{MN}\\
	\hline g_{11}^{\ast} & g_{21}^{\ast} & \cdots & g_{M1}^{\ast} & \omega_{1} & \eta_{12} & \cdots & \eta_{1N}\\
	g_{12}^{\ast} & g_{22}^{\ast} & \cdots & g_{M2}^{\ast} & \eta_{12}^{\ast} & \omega_{2} & \cdots & \eta_{2N}\\
	\vdots & \vdots & \ddots & \vdots & \vdots & \vdots & \ddots & \vdots\\
	g_{1N}^{\ast} & g_{2N}^{\ast} & \cdots & g_{MN}^{\ast} & \eta_{1N}^{\ast} & \eta_{2N}^{\ast} & \cdots & \omega_{N}
\end{array}\right).~\label{HabMNMN}
\end{eqnarray}
Here, $\textbf{H}_a$ and $\textbf{H}_b$ characterize the on-site and interaction parts for the type-$a$ and type-$b$ sub-networks, respectively, and $\textbf{C}_{ab}$ describes the couplings between the two sub-networks. Note that the coefficient matrix $\textbf{H}_{ab}$ has the dimension $(M+N)\times(M+N)$.

To clearly analyze the parameter conditions for the appearance of dark modes, we introduce the unitary operators $\textbf{U}_a$ and $\textbf{U}_{a}^{\dagger}$ to diagonalize the sub-matrix $\textbf{H}_a$ associated with the type-$a$ mode sub-network,  $\mathbf{U}_{a}\mathbf{H}_{a}\mathbf{U}_{a}^{\dagger}=\mathbf{H}_{A}=\textrm{diag}\{\Delta_{1},\Delta_{2},\Delta_{3},\cdots,\Delta_{M}\}$. We also introduce the unitary operators $\mathbf{U}_{b}$ and $\mathbf{U}_{b}^{\dag}$ to diagonalize the sub-matrix $\textbf{H}_b$ associated with the type-$b$ mode sub-network,
$\mathbf{U}_{b}\mathbf{H}_{b}\mathbf{U}_{b}^{\dag}=\mathbf{H}_{B}=\textrm{diag}\{\Omega_{1},\Omega_{2},\Omega_{3},\cdots,\Omega_{N}\}$. Hereafter, we use the subscripts ``$A$ ($B$)" and ``$a$ ($b$)" to mark the normal- and bare-mode representations of type-$a$ ($b$) modes, respectively. After diagonalization of the two sub-networks, the general ($M+N$)-mode linear network can be transformed to a bipartite-graph network [as shown in Fig.~\ref{FigS1}(b)].
The Hamiltonian of the bipartite-graph network can be expressed in the normal-mode ($A_k$,$B_j$) representation as
\begin{equation}
\label{HamiltABMNS}
H=(\mathbf{A}^{\dag},\mathbf{B}^{\dag})\mathbf{H}_{AB}
\left(\begin{array}{c}
\mathbf{A} \\
\mathbf{B}
\end{array}
\right),
\end{equation}
where the vectors of annihilation and creation operators for type-$a$ and -$b$ normal modes are introduced as
\begin{eqnarray}
\mathbf{A}&=&(A_1,A_2,\dotsb,A_M)^{T}=\textbf{U}_{a}\boldsymbol{a},\nonumber\\ \mathbf{A}^{\dagger}&=&(A_1^{\dagger},A_2^{\dagger},\dotsb,A_M^{\dagger})=\boldsymbol{a}^{\dagger}\textbf{U}_{a}^{\dagger},\nonumber\\ \mathbf{B}&=&(B_1,B_2,\dotsb,B_N)^{T}=\mathbf{U}_{b}\boldsymbol{b},\nonumber\\ \mathbf{B}^{\dag}&=&(B_1^{\dagger},B_2^{\dagger},\dotsb,B_N^{\dagger})=\boldsymbol{b}^{\dag}\mathbf{U}_{b}^{\dag}.
\end{eqnarray}
In Eq. (\ref{HamiltABMNS}), the coefficient matrix can be obtained via the following transformation
\begin{equation}
	\mathbf{H}_{AB}=\left(
	\begin{array}{c c}
		\textbf{U}_{a}\mathbf{H}_{a}\textbf{U}^{\dagger}_{a} & \textbf{U}_{a}\mathbf{C}_{ab}\textbf{U}^{\dagger}_{b} \\
		\textbf{U}_{b}\mathbf{C}_{ab}^{\dagger}\textbf{U}^{\dagger}_{a} & \textbf{U}_{b}\mathbf{H}_{b}\textbf{U}^{\dagger}_{b}
	\end{array}\right)=\left(
	\begin{array}{c c}
		\mathbf{H}_{A} & \mathbf{C}_{AB} \\
		\mathbf{C}_{AB}^{\dagger} & \mathbf{H}_{B}
	\end{array}\right)=(\textbf{U}_a\oplus\textbf{U}_b)\textbf{H}_{ab}(\textbf{U}_{a}^{\dagger}\oplus\textbf{U}_b^{\dagger})=\left(\begin{array}{cc}
		\textbf{U}_{a}&0 \\
		0&\textbf{U}_{b}
	\end{array}
	\right)\left(
	\begin{array}{c c}
		\mathbf{H}_{a} & \mathbf{C}_{ab} \\
		\mathbf{C}_{ab}^{\dagger} & \mathbf{H}_{b}  \\
	\end{array}\right)
	\left(\begin{array}{c c}
		\textbf{U}_{a}^{\dag}&0 \\
		0&\textbf{U}_{b}^{\dag}
	\end{array}
	\right).
\end{equation}
Here, the coupling matrix in the normal-mode ($A_k$,$B_j$) representation is given by
\begin{equation}
	\textbf{C}_{AB}=\textbf{U}_{a}\textbf{C}_{ab}\textbf{U}_{b}^{\dagger}.
\end{equation}
The coefficient matrix in Eq. (\ref{HamiltABMNS}) can be expressed as either an arrowhead matrix (for the case $M=1$) or a thick (here the word ``thick" means that the width of the arrowhead has $M$ rows with $M>1$) arrowhead matrix
\begin{eqnarray}
\label{matrixABMNS}
\mathbf{H}_{AB}=\left(
\begin{array}{c c}
\mathbf{H}_{A} & \mathbf{C}_{AB} \\
\mathbf{C}_{AB}^{\dagger} & \mathbf{H}_{B}
\end{array}\right)
=\left(
\begin{array}{cccc|cccc}
\Delta _{1} & 0 & \cdots  & 0 &G_{11}& G_{12} & \cdots  & G_{1N} \\
0 & \Delta _{2} & \cdots  & 0 & G_{21}& G_{22} & \cdots  & G_{2N} \\
\vdots  & \vdots  & \ddots  & \vdots  & \vdots  & \vdots  & \ddots  & \vdots\\
0 & 0 & \cdots  & \Delta _{M} & G_{M1} &G_{M2} & \cdots  & G_{MN} \\  \hline
G^{\ast}_{11} & G^{\ast}_{21} & \cdots  & G^{\ast}_{M1} & \Omega _{1} & 0 & \cdots  & 0 \\
G^{\ast}_{12} & G^{\ast}_{22} & \cdots  & G^{\ast}_{M2} & 0 & \Omega _{2} & \cdots  & 0 \\
\vdots  & \vdots  & \ddots  & \vdots  & \vdots  & \vdots  & \ddots  & \vdots \\
G^{\ast}_{1N} & G^{\ast}_{2N} & \cdots  & G^{\ast}_{MN} & 0 & 0 & \cdots & \Omega _{N}
\end{array}\right).~\label{HABMNMN}
\end{eqnarray}
Here, $G_{kj}$ is the effective coupling strength between the type-$a$ normal mode $A_{k}$ and the type-$b$ normal mode $B_{j}$, which can be expressed as
\begin{equation} G_{kj}=\sum_{k^{\prime},j^{\prime}}(\textbf{U}_a)_{kk^{\prime}}g_{k^{\prime}j^{\prime}}(\textbf{U}_b^{\dagger})_{j^{\prime}j}.~\label{GJKprime}
\end{equation}
Equation (\ref{GJKprime}) shows the relation between the coupling strengths $G_{kj}$ and $g_{kj}$.
Below we study the parameter conditions for the existence of dark modes in this $(M + N)$-mode linear quantum network by analyzing properties of the thick arrowhead matrix $\mathbf{H}_{AB}$.

\section{Two dark-mode theorems for linear quantum networks and their proofs~\label{twodarkmode}}

In this section, we present two theorems for determining the number of dark modes in both the ($1+N$)- and ($M+N$)-mode linear quantum networks. We also exhibit the detailed proofs of the two theorems.

\subsection{Dark-mode theorems for the ($1+N$)-mode linear quantum networks}

We consider a $(1+N)$-mode linear quantum network consisting of one type-$a$ mode and $N$ type-$b$ modes, which is a special case (i.e., $M=1$) of the ($M+N$)-mode network introduced in Sec.~\ref{transfor}. The Hamiltonian of the ($1+N$)-mode linear quantum network reads
\begin{equation}
\label{linRWAHamit1NS}
H=\delta_{1}a_{1}^{\dagger} a_{1}+\sum_{j=1}^{N}\omega_{j} b_{j}^{\dagger} b_{j}
+\sum_{j,j^{\prime}=1,j<j^{\prime}}^{N}(\eta_{jj^{\prime}}b_{j}^{\dagger}b_{j^{\prime}}+\eta_{jj^{\prime}}^{\ast}b_{j^{\prime}}^{\dagger} b_{j})+\sum_{j=1}^{N}( g_{1j}a_{1}^{\dagger} b_{j}+g_{1j}^{\ast}a_{1}b^{\dagger}_{j}),
\end{equation}
which can be further expressed as
\begin{eqnarray}
\label{HamiltabMNS}
H=(a_{1}^{\dag},\boldsymbol{b}^{\dag})\mathbf{H}_{ab}\left(\begin{array}{c}
a_{1} \\
\boldsymbol{b}
\end{array}
\right),
\end{eqnarray}
where the operator vectors $\boldsymbol{b}$ and $\boldsymbol{b}^{\dagger}$ have been defined in Eq.~(\ref{vectorab}). The coefficient matrix for this network in the bare-mode $(a_{1},b_{j})$ representation is given by
\begin{eqnarray}
\mathbf{H}_{ab}=\left(
\begin{array}{c c}
\delta_{1} & \mathbf{C}_{ab} \\
\mathbf{C}_{ab}^{\dagger} & \mathbf{H}_{b}
\end{array}\right)
= \left(
\begin{array}{c|cccc}
\delta_{1} & g_{11} & g_{12} &\cdots & g_{1N}\\  \hline
g_{11}^{\ast} & \omega_{1} & \eta_{12} &\cdots & \eta_{1N}\\
g_{12}^{\ast} & \eta^{\ast}_{12} & \omega_{2} &\cdots & \eta_{2N}\\
\vdots & \vdots & \vdots &\ddots & \vdots\\
g_{1N}^{\ast} & \eta^{\ast}_{1N} & \eta^{\ast}_{2N} &\cdots & \omega_{N}\\
\end{array}
\right), \nonumber
\end{eqnarray}
which has the dimension $(1+N)\times(1+N)$. We introduce the unitary operators $\textbf{U}_{b}$ and $\textbf{U}_{b}^{\dag}$ to diagonalize the type-$b$ mode sub-network, and then the Hamiltonian of this ($1+N$)-mode network can be expressed as
\begin{equation}
H=(A_{1}^{\dag},\mathbf{B}^{\dag})\mathbf{H}_{AB}
\left(\begin{array}{c}
A_{1} \\
\mathbf{B}
\end{array}
\right),
\end{equation}
where $A_1=a_1$, $\Delta_1=\delta_1$, and $\textbf{B}=(B_1,B_2,\dotsb,B_N)^{T}$ with $B_j=\sum_{j^{\prime}}(\textbf{U}_b)_{jj^{\prime}}b_{j^{\prime}}$.
The coefficient matrix in the normal-mode ($A_1,B_j$) representation is  given by
\begin{eqnarray}
\mathbf{H}_{AB}=\left(
\begin{array}{c c}
\Delta_{1} & \mathbf{C}_{AB} \\
\mathbf{C}_{AB}^{\dagger} & \mathbf{H}_{B}
\end{array}\right)
=\left(
\begin{array}{c|cccc}
\Delta_{1} & G_{11} & G_{12} &\cdots & G_{1N}\\  \hline
G^{\ast}_{11} & \Omega_{1} &0 &\cdots & 0\\
G^{\ast}_{12} & 0 & \Omega_{2} &\cdots &0\\
\vdots & \vdots & \vdots &\ddots & \vdots\\
G^{\ast}_{1N} & 0 & 0 &\cdots & \Omega_{N}\\
\end{array}
\right),~\label{H1NAB}
\end{eqnarray}
where $\mathbf{H}_{B}=\textbf{U}_b\textbf{H}_b\textbf{U}_b^{\dagger}=\text{diag}\{\Omega_{1},\Omega_{2},\ldots,\Omega_{N}\}$ is a diagonal matrix with $\Omega _{j}$ being the resonance frequency of the $j$th type-$b$ normal mode $B_j$. $\mathbf{C}_{AB}=(G_{11},G_{12},\cdots,G_{1N})$ is a row vector with $G_{1j}$ being the effective coupling strength between the type-$a$ normal mode $A_{1}$ and the $j$th type-$b$ normal mode $B_{j}$. Then the theorem for determining the number of type-$b$ dark modes in this ($1+N$)-mode network can be proposed by analyzing the properties of the arrowhead matrix $\mathbf{H}_{AB}$.

\textbf{Theorem 1.}---\emph{Consider a $(1+N)$-mode linear quantum network described by the Hamiltonian $H=(\mathbf{A}^{\dag},\mathbf{B}^{\dag})\boldsymbol{H}_{AB}(\mathbf{A},\mathbf{B})^{T}$, the coefficient matrix is given by an arrowhead matrix $\textbf{H}_{AB}=\{\{\Delta_{1},\textbf{C}_{AB}\},\{\textbf{C}_{AB}^{\dag},\textbf{H}_{B}\}\}$,
	where the coupling matrix is a row vector $\textbf{C}_{AB}=(G_{11},G_{12},\cdots,G_{1N})$ and $\textbf{H}_{B}=\text{diag}\{\Omega_{1},\Omega_{2},\ldots,\Omega_{N}\}$ with an ascending order.}

(i) \emph{If the element $G_{1j}=0$, then the corresponding mode $B_{j}$ is a dark mode.}

(ii) \emph{If all the elements of $\textbf{C}_{AB}$ are nonzero and $l$ type-$b$ modes ($B_{j=1\text{-}l}$) are degenerate, i.e., $\Omega_{j=1\text{-}l}=\Omega$, then in the mode space spanned by these degenerate modes, there are one bright mode $B_{l+}=(\vert G_{1l}\vert ^{2}+\vert G_{1( l-1) +}\vert ^{2})^{-1/2}(G_{1l}B_{l}+G_{1(l-1) +}B_{( l-1) +})=(\sum_{j=1}^{l}\vert G_{1j}\vert ^{2})^{-1/2}\sum_{j=1}^{l}G_{1j}B_{j}$ and $l-1$ dark modes: $B_{2-}$, $B_{3-}$, ..., and $B_{l-}$, where $B_{j-} =(\vert G_{1( j-1) +}\vert
	^{2}+\vert G_{1j}\vert ^{2})^{-1/2}( G_{1j}^{\ast }B_{(j-1) +}-G_{1( j-1) +}B_{j})$ with $B_{(j-1) +}=(\vert G_{1( j-2) +}\vert^{2}+\vert G_{1(j-1)}\vert ^{2})^{-1/2}(G_{1( j-2)+}B_{( j-2) +}+G_{1(j-1)}B_{j-1})$ and $G_{1j+}=(\vert G_{1j}\vert ^{2}+\vert G_{1( j-1)+}\vert ^{2})^{1/2}$. Note that the forms of these dark modes are not unique, depending on the grouping order of these degenerate type-$b$ modes.}

(iii) \emph{If all the coupling strengths are nonzero, and all the type-$b$ modes are nondegenerate, i.e., $G_{1j}\neq0$ for $j=1\text{-}N$ and $\Omega_{j}\neq\Omega_{j^{\prime}}$ for all $j\neq j^{\prime}$, then there is no type-$b$ dark mode in the network.}

The proofs of  Theorems 1(i) and 1(iii) have been given in the main text. Below we present the detailed proofs of Theorem 1(ii).
\subsubsection{Proof of Theorem 1(ii)}
In this section, we adopt the mathematical induction method to prove  Theorem 1(ii). In particular, we analyze the number of dark modes in a given degenerate-mode subspace because the modes in different degenerate-mode subspaces are orthogonal to each other.

(i) Step 1: When $l=1$, the Hamiltonian is given by
\begin{equation}
H^{[1,1]}=\Delta_1 A_1^{\dagger }A_1+\Omega B_{1}^{\dagger}B_{1}+( G_{11}A_1^{\dagger }B_{1}+G_{11}^{\ast }B_{1}^{\dagger }A_1).~\label{H11ab}
\end{equation}
For avoiding confusion, hereafter we add the superscript ``[$M,N$]'' to denote the involved mode numbers $M$ and $N$ for type-$a$ and type-$b$ modes, respectively. The superscript ``[1,1]'' in Eq.~(\ref{H11ab}) indicates that there are one type-$a$ mode and one type-$b$ mode.
Obviously, there is one type-$b$ bright mode $B_{1}$ and no type-$b$ dark mode in this ($1+1$)-mode network.

(ii) Step 2: When $l=2$, the Hamiltonian reads
\begin{eqnarray}
H^{[1,2]} &=&H^{[1,1]}+\Omega B_{2}^{
\dagger }B_{2}+G_{12}A_1^{\dagger }B_{2}+G_{12}^{\ast
}B_{2}^{\dagger }A_1 \nonumber\\
&=&\Delta_1 A_1^{\dagger }A_1+\Omega ( B_{1}^{\dagger
}B_{1}+B_{2}^{\dagger }B_{2}) +A_1^{\dagger }\left(
G_{11}B_{1}+G_{12}B_{2}\right) +( G_{11}^{\ast }B_{1}^{\dagger
}+G_{12}^{\ast }B_{2}^{\dagger }) A_1.
\end{eqnarray}
To find out the bright mode and dark mode, we introduce the new bosonic annihilation and creation operators as
\begin{eqnarray}
B_{2+} &=&\frac{1}{G_{12+}}
( G_{11}B_{1}+G_{12}B_{2}) ,\hspace{1cm}B_{2+}^{\dagger } =\frac{1}{G_{12+}}( G_{11}^{\ast
}B_{1}^{\dagger }+G_{12}^{\ast }B_{2}^{\dagger }),\nonumber\\
B_{2-} &= &
\frac{1}{G_{12+}}( G_{12}^{\ast }B_{1}-G_{11}^{\ast }B_{2}),
 \hspace{1cm}B_{2-}^{\dagger } =\frac{1}{G_{12+}}( G_{12}B_{1}^{\dagger
}-G_{11}B_{2}^{\dagger }),\label{b222}
\end{eqnarray}
with $G_{12+}=\sqrt{\vert G_{11}\vert ^{2}+\vert G_{12}\vert^{2}}>0$. We can show the relation
\begin{equation}
B_{1}^{\dagger }B_{1}+B_{2}^{\dagger }B_{2}=B_{2+}^{\dagger }B_{2+}+B_{2-}^{\dagger }B_{2-}.
\end{equation}
Then the Hamiltonian can be expressed as
\begin{eqnarray}
H^{[1,2]} =\Delta_1 A_1^{\dagger }A_1+\Omega B_{2-}^{\dagger
}B_{2-}+\Omega B_{2+}^{\dagger }B_{2+}+G_{12+}( A_1^{\dagger
}B_{2+}+B_{2+}^{\dagger }A_1) .
\end{eqnarray}
Here we can see that there are one type-$b$ bright mode $B_{2+}$ and one type-$b$ dark mode $B_{2-}$.

(iii) Step 3: We assume that Theorem 1(ii) is valid for $l=j$, namely the Hamiltonian corresponding to $l=j$ can be expressed as
\begin{eqnarray}
H^{[1, j] }&=&H^{[1, (j-1)] }+\Omega B_{j}^{\dagger}B_{j}+G_{1j}A_{1}^{\dagger}B_{j}+G_{1j}^{\ast}B_{j}^{\dagger}A_{1} \nonumber\\
&=&\Delta_1 A_1^{\dagger }A_1+\Omega \sum_{s=2}^{j}B_{s-}^{\dagger}B_{s-}+\Omega B_{j+}^{\dagger }B_{j+}+G_{1j+}(A_1^{\dagger
}B_{j+}+B_{j+}^{\dagger }A_1),~\label{H1jj1}
\end{eqnarray}
where the introduced operators are defined by
\begin{eqnarray}
	B_{j+} &=&\frac{1}{G_{1j+}}( G_{1\left( j-1\right)
		+}B_{\left(j-1\right) +}+G_{1j}B_{j}),\hspace{1cm}	B^{\dagger}_{j+} =\frac{1}{G_{1j+}}( G_{1\left( j-1\right)
		+}B^{\dagger}_{\left( j-1\right) +}+G^{\ast}_{1j}B^{\dagger}_{j}),\nonumber\\
		B_{j-} &=&\frac{1}{G_{1j+}}( G^{\ast}_{1j
		}B_{\left( j-1\right) +}-G_{1(j-1)+}B_{j})
,\hspace{1cm}	B^{\dagger}_{j-} =\frac{1}{G_{1j+}}( G_{1j
	}B^{\dagger}_{\left( j-1\right) +}-G_{1(j-1)+}B^{\dagger}_{j}),~\label{BJBJ}
\end{eqnarray}
with $G_{1j+}=\sqrt{G_{1(j-1)+}^{2}+\vert G_{1j}\vert ^{2}}=\sqrt{\vert G_{11}\vert ^{2}+\left\vert G_{12}\right\vert ^{2}+\dotsb+\vert G_{1j}\vert ^{2}}>0$.
Equation (\ref{H1jj1}) indicates that there are one type-$b$ bright mode $B_{j+}$ and $j-1$ type-$b$ dark modes: $B_{2-},B_{3-},B_{4-},\dotsb$, and $B_{j-}$.

(iv) Step 4: We show that Theorem 1(ii) is valid for $l=j+1$. When $l=j+1$, the Hamiltonian can be expressed as
\begin{eqnarray}
H^{[1,(j+1)]} & = & H^{[1,j]}+\Omega B_{j+1}^{\dagger}B_{j+1}+G_{1(j+1)}A_{1}^{\dagger}B_{j+1}+G_{1(j+1)}^{\ast}B_{j+1}^{\dagger}A_{1}\nonumber\\
& = & \Delta A_{1}^{\dagger}A_{1}+\Omega B_{2-}^{\dagger}B_{2-}+\Omega B_{3-}^{\dagger}B_{3-}+...+\Omega B_{j-}^{\dagger}B_{j-}+\Omega B_{j+}^{\dagger}B_{j+}+\Omega B_{j+1}^{\dagger}B_{j+1}\nonumber\\
&  & +A_{1}^{\dagger}(G_{1j+}B_{j+}+G_{1(j+1)}B_{j+1})+(G_{1j+}B_{j+}^{\dagger}+G_{1(j+1)}^{\ast}B_{j+1}^{\dagger})A_{1}.~\label{H1j1j1}
\end{eqnarray}
 To analyze the dark-mode effect in this [$1+(j+1)$]-mode network, we introduce the new bosonic annihilation and creation operators as
\begin{eqnarray}
B_{(j+1)+} & = & \frac{1}{G_{1(j+1)+}}(G_{1j+}B_{j+}+G_{1(j+1)}B_{j+1}),\hspace{1cm}B_{(j+1)+}^{\dagger}  =  \frac{1}{G_{1(j+1)+}}(G_{1j+}B_{j+}^{\dagger}+G_{1(j+1)}^{\ast}B_{j+1}^{\dagger}),\nonumber\\
B_{(j+1)-}  &=& \frac{1}{G_{1(j+1)+}}(G_{1(j+1)}^{\ast}B_{j+}-G_{1j+}B_{j+1}),
\hspace{1cm}B_{(j+1)-}^{\dagger} = \frac{1}{G_{1(j+1)+}}(G_{1(j+1)}B_{j+}^{\dagger}-G_{1j+}B_{j+1}^{\dagger}),
\end{eqnarray}
with $G_{1( j+1) +} =\sqrt{\vert G_{1j+}\vert
^{2}+\vert G_{1(j+1)}\vert ^{2}}>0$.
It can be shown that
\begin{equation}
B_{j+}^{\dagger}B_{j+}+B_{j+1}^{\dagger}B_{j+1}= B_{(j+1)+}^{\dagger}B_{(j+1)+}+B_{(j+1)-}^{\dagger}B_{(j+1)-}.
\end{equation}
Then the Hamiltonian (\ref{H1j1j1}) can be expressed as
\begin{equation}
H^{[1,(j+1)]} =  \Delta_{1}A_{1}^{\dagger}A_{1}+\Omega\sum_{s=2}^{j+1} B_{s-}^{\dagger}B_{s-}+\Omega B^{\dagger}_{(j+1)+}B_{(j+1)+}+G_{1(j+1)+}(A_{1}^{\dagger}B_{(j+1)+}+B_{(j+1)+}^{\dagger}A_{1}).~\label{H1JI}
\end{equation}
Equation (\ref{H1JI}) indicates that there are one type-$b$ bright mode $B_{(j+1)+}$ and $j$ type-$b$ dark modes: $B_{2-},B_{3-},\dotsb,$ and $B_{(j+1)-}$. Therefore, the statement in Theorem 1(ii)  is valid for $l=j+1$. Based on the above proof, we can conclude that Theorem 1(ii) is valid for an arbitrary positive integer $l$.

We can summarize the statement of Theorem 1(ii) as follows. For the Hamiltonian
\begin{equation}
H=\Delta_1 A_1^{\dagger }A_1+\Omega \sum_{j=1}^{l}B_{j}^{\dagger}B_{j}+\sum_{j=1}^{l}( G_{1j}A_1^{\dagger }B_{j}+G_{1j}^{\ast
}B_{j}^{\dagger }A_1),
\end{equation}
there are one type-$b$ bright mode
\begin{equation}
B_{l+}=\frac{1}{G_{1l+}}(G_{1\left( l-1\right)
+}B_{\left( l-1\right) +}+G_{1l}B_{l})=\frac{1}{\sqrt{\sum_{j=1}^{l}\vert G_{1j}\vert ^{2}}}
\sum_{j=1}^{l}G_{1j}B_{j},
\end{equation}
and $l-1$ type-$b$ dark modes: $B_{2-},B_{3-},B_{4-},\dotsb$, and $B_{l-}$, which have been defined in Eq. (\ref{BJBJ}).

\subsection{Dark-mode theorems for the ($M+N$)-mode linear quantum networks \label{mnmechan}}

In this section, we propose and prove the theorems for determining the number of dark modes in the two-component $(M+N)$-mode linear quantum networks, which consist of $M$ ($M\geq2$) type-$a$ modes coupled to $N$ type-$b$ modes. According to the above analyses, we know that the $(M+N)$-mode  linear quantum networks can be transformed into bipartite-graph networks, which can be described by a thick arrowhead matrix.  Below we present the detailed calculations concerning the dark-mode theorems and their proofs for this $(M+N)$-mode linear network by analyzing the properties of the arrowhead matrix $\mathbf{H}_{AB}$.

\textbf{Theorem~2.}---\emph{Consider a $(M+N)$-mode linear quantum network described by the Hamiltonian $H=(\mathbf{A}^{\dag},\mathbf{B}^{\dag})\boldsymbol{H}_{AB}(\mathbf{A},\mathbf{B})^{T}$, the coefficient matrix  is given by a thick arrowhead matrix $\textbf{H}_{AB}=\{\{\textbf{H}_{A},\textbf{C}_{AB}\},\{\textbf{C}_{AB}^{\dag},\textbf{H}_{B}\}\}$, where $\textbf{C}_{AB}=[\textbf{G}_{1},\textbf{G}_{2},\cdots,\textbf{G}_{N}]$ with column vectors $\textbf{G}_{j}=[G_{1j},G_{2j},\cdots,G_{Mj}]^{T}$}, $\mathbf{H}_{A}=\text{diag}\{\Delta_{1},\Delta_{2},\ldots ,\Delta_{M}\}$, and $\textbf{H} _{B}=\text{diag}\{\Omega_{1},\Omega_{2},\ldots,\Omega_{N}\}$.

(i) \emph{If the $j$th column vector $\textbf{G}_{j}=\textbf{0}$, then the corresponding mode $B_{j}$ is a type-$b$ dark mode.}

(ii) \emph{If $l$ (for $l=2-N$) type-$b$ modes are degenerate (for example $\Omega_{j=1\text{-}l}=\Omega$), and all these corresponding column vectors in $\textbf{C}_{AB}$ are linearly dependent, then in the degenerate-mode subspace, there are one type-$b$ bright mode and ($l-1$) type-$b$ dark modes.}

(iii) \emph{In the case of $M$ type-$a$ normal modes and $N$ degenerate type-$b$ normal modes ($\Omega_{j=1\text{-}N}=\Omega$),  there are at least $N-M$ type-$b$ dark modes when $N>M$.}

(iv) \emph{When all the $N$ type-$b$ normal modes can be divided into different groups with the same resonance frequency (each group corresponds to a coupling sub-matrix formed by the corresponding coupling column vectors $\textbf{G}_{j}$, the dimension of the sub-matrix could be $M\times1$ for a single non-degenerate mode). The total number of type-$b$ bright modes in the network is $R=\sum_{s} R_{s}$, where $R_{s}$ is the rank of the $s$th submatrix in the coupling matrix $\textbf{C}_{AB}$, and  the number of the type-$b$ dark modes is $N-R$.}

(v) \emph{The change of the coupling strengths $\xi_{kk^{\prime}}$  between these type-$a$ modes $a_{k}$ and $a_{k^{\prime}}$ will not change the number of the type-$b$ dark modes, but will change the forms of these type-$b$ dark modes.}

(vi) \emph{If $\textbf{G}_{j}\neq\textbf{0}$ for $j=1\text{-}N$ and $\Omega_{j}\neq\Omega_{j^{\prime}}$ for all  $j\neq j^{\prime}$, then there is no type-$b$ dark mode in the system.}

The proofs of  Theorems 2(i) and 2(vi) have been given in the main text. Below we present the detailed proofs of Theorems 1(ii-v) .

\subsubsection{Proof of Theorem 2(ii)}
For the ($M+l$)-mode linear quantum network with $M$ type-$a$ modes and $l$ degenerate type-$b$ normal modes (i.e., $\Omega_{j=1\text{-}l}=\Omega$), the coefficient matrix of the network in the normal-mode ($A_k$,$B_j$) representation reads
\begin{equation}
\textbf{H}_{AB}=	\left(
	\begin{array}{ccccc|cccccc}
		\Delta _{1} & \cdots  & 0 & \cdots  & 0 & G_{11} & G_{12} & \cdots  & G_{1j}
		& \cdots  & G_{1l} \\
		\cdots  & \cdots  & \cdots  & \cdots  & \cdots  & \cdots  & \cdots  & \cdots
		& \cdots  & \cdots  & \cdots  \\
		0 & \cdots  & \Delta _{k} & \cdots  & 0 & G_{k1} & G_{k2} & \cdots  & G_{kj}
		& \cdots  & G_{kl} \\
		\cdots  & \cdots  & \cdots  & \cdots  & \cdots  & \cdots  & \cdots  & \cdots
		& \cdots  & \cdots  & \cdots  \\
		0 & \cdots  & 0 & \cdots  & \Delta _{M} & G_{M1} & G_{M2} & \cdots  & G_{Mj}
		& \cdots  & G_{Ml} \\\hline
		G_{11}^{\ast } & \cdots  & G_{k1}^{\ast } & \cdots  & G_{M1}^{\ast } &
		\Omega  & 0 & \cdots  & 0 & \cdots  & 0 \\
		G_{12}^{\ast } & \cdots  & G_{k2}^{\ast } & \cdots  & G_{M2}^{\ast } & 0 &
		\Omega  & \cdots  & 0 & \cdots  & 0 \\
		\ \cdots  & \cdots  & \cdots  & \cdots  & \cdots  & \cdots  & \cdots  &
		\cdots  & \cdots  & \cdots  & \cdots  \\
		G_{1j}^{\ast } & \cdots  & G_{kj}^{\ast } & \cdots  & G_{Mj}^{\ast } & 0 & 0
		& \cdots  & \Omega  & \cdots  & 0 \\
		\cdots  & \cdots  & \cdots \  & \cdots  & \cdots  & \cdots  & \cdots  &
		\cdots  & \cdots  & \cdots  & \cdots  \\
		G_{1l}^{\ast } & \cdots  & G_{kl}^{\ast } & \cdots  & G_{Ml}^{\ast } & 0 & 0
		& \cdots  & 0 & \cdots  & \Omega
	\end{array}%
	\right).~\label{HABDELTA}
\end{equation}
By defining the following matrix $\mathbf{\Delta}$ and column vectors $\textbf{G}_{j=1\text{-}l}$,
\begin{equation}
	\mathbf{\Delta } \mathbf{=}\left(
	\begin{array}{ccccc}
		\Delta _{1} & \cdots  & 0 & \cdots  & 0 \\
		\cdots  & \cdots  & \cdots  & \cdots  & \cdots  \\
		0 & \cdots  & \Delta _{k} & \cdots  & 0 \\
		\cdots  & \cdots  & \cdots  & \cdots  & \cdots  \\
		0 & \cdots  & 0 & \cdots  & \Delta _{M}%
	\end{array}%
	\right),\hspace{1cm}
	\mathbf{G}_{j=1-l} =\left(
	\begin{array}{c}
		G_{1j} \\
		\cdots  \\
		G_{kj} \\
		\cdots  \\
		G_{Mj}%
	\end{array}%
	\right) ,
\end{equation}
the Hamiltonian of this ($M+l$)-mode linear quantum network can be expressed as
\begin{eqnarray}
	H^{[M,l]} &=&(
	\begin{array}{ccccccc}
		\mathbf{A}^{\dagger } & B_{1}^{\dagger } & B_{2}^{\dagger } & \cdots  &
		B_{j}^{\dagger } & \cdots  & B_{l}^{\dagger }%
	\end{array}%
	) \left(
	\begin{array}{ccccccc}
		\mathbf{\Delta } & \mathbf{G}_{1} & \mathbf{G}_{2} & \cdots  & \mathbf{G}_{j}
		& \cdots  & \mathbf{G}_{l} \\
		\mathbf{G}_{1}^{\dagger } & \Omega  & 0 & \cdots  & 0 & \cdots  & 0 \\
		\mathbf{G}_{2}^{\dagger } & 0 & \Omega  & \cdots  & 0 & \cdots  & 0 \\
		\cdots  & \cdots  & \cdots  & \cdots  & \cdots  & \cdots  & \cdots  \\
		\mathbf{G}_{j}^{\dagger } & 0 & 0 & \cdots  & \Omega  & \cdots  & 0 \\
		\cdots  & \cdots  & \cdots  & \cdots  & \cdots  & \cdots  & \cdots  \\
		\mathbf{G}_{l}^{\dagger } & 0 & 0 & \cdots  & 0 & \cdots  & \Omega
	\end{array}%
	\right) \left(
	\begin{array}{c}
		\mathbf{A} \\
		B_{1} \\
		B_{2} \\
		\cdots  \\
		B_{j} \\
		\cdots  \\
		B_{l}%
	\end{array}%
	\right)  \nonumber\\
	&=&\mathbf{A}^{\dagger }\mathbf{\Delta A\mathbf{+}}\Omega
	\sum_{j=1}^{l}B_{j}^{\dagger }B_{j}\mathbf{+}\sum_{j=1}^{l}\mathbf{A}%
	^{\dagger }\mathbf{G}_{j}B_{j}+\sum_{j=1}^{l}B_{j}^{\dagger }\mathbf{G}%
	_{j}^{\dagger }\mathbf{A},~\label{Htextb}
\end{eqnarray}
where the superscript ``[$M,l$]'' denotes the mode numbers of type-$a$ and -$b$ modes.
Below, we show that, in the linear quantum network governed by Hamiltonian (\ref{Htextb}), there are one type-$b$ bright mode and $l-1$ type-$b$ dark modes.

To prove Theorem 2(ii), we adopt the mathematical induction method.

(i) Step 1: When $l=1,$ we have%
\begin{eqnarray}
	H^{\left[ M,1\right] }&=&\mathbf{A}^{\dagger }\mathbf{\Delta A\mathbf{+}}\Omega
	B_{1}^{\dagger }B_{1}+\mathbf{A}
	^{\dagger }\mathbf{G}_{1}B_{1}+B_{1}^{\dagger }\mathbf{G}%
	_{1}^{\dagger }\mathbf{A}\nonumber\\
	&=&
	\sum_{k=1}^{M}\Delta _{k}A_{k}^{\dagger }A_{k}+\Omega
	B_{1}^{\dagger }B_{1}+\sum_{k=1}^{M}( G_{k1}A_{k}^{\dagger
	}B_{1}+G_{k1}^{\ast }B_{1}^{\dagger }A_{k}).
\end{eqnarray}
Obviously, there is one type-$b$ bright mode $B_{1}$ and no type-$b$
dark mode.

(ii) Step 2: When $l=2,$ we assume that the corresponding column vectors $\textbf{G}_{2}$ and $\textbf{G}_{1}$ are linearly dependent, i.e.,
\begin{equation}
	\mathbf{G}_{2}=\left(
	\begin{array}{c}
		G_{12} \\
		\cdots  \\
		G_{k2} \\
		\cdots  \\
		G_{M2}%
	\end{array}%
	\right) =\lambda _{2}\mathbf{G}_{1}=\left(
	\begin{array}{c}
		\lambda _{2}G_{11} \\
		\cdots  \\
		\lambda _{2}G_{k1} \\
		\cdots  \\
		\lambda _{2}G_{M1}%
	\end{array}%
	\right).
\end{equation}
In this case, the Hamiltonian becomes
\begin{eqnarray}
	H^{\left[ M,2\right] } &=&H^{[M,1]}+\Omega B^{\dagger}_{2}B_{2}+\mathbf{A}
	^{\dagger }\mathbf{G}_{2}B_{2}+B_{2}^{\dagger }\mathbf{G}
	_{2}^{\dagger }\mathbf{A} \nonumber \\
	&=&\sum_{k=1}^{M}\Delta _{k}A_{k}^{\dagger }A_{k}+\Omega (B_{1}^{\dagger
	}B_{1}+B_{2}^{\dagger }B_{2})+\sum_{k=1}^{M}G_{k1}A_{k}^{\dagger }\left(
	B_{1}+\lambda _{2}B_{2}\right) +\sum_{k=1}^{M}G_{k1}^{\ast }(
	B_{1}^{\dagger }+\lambda _{2}^{\ast }B_{2}^{\dagger }) A_{k}.
\end{eqnarray}%
We further introduce the operators for the two new modes $B_{2\pm}$ as
\begin{eqnarray}
	B_{2+} &=&\frac{1}{\lambda _{2+}}\left( B_{1}+\lambda
	_{2}B_{2}\right) ,\hspace{1cm}B_{2+}^{\dagger } = \frac{
		1}{\lambda _{2+}}( B_{1}^{\dagger }+\lambda _{2}^{\ast }B_{2}^{\dagger
	}),\nonumber\\
	B_{2-} &=&\frac{1}{\lambda _{2+}}\left(
	\lambda _{2}^{\ast }B_{1}-B_{2}\right)
	, \hspace{1cm}	B_{2-}^{\dagger } =\frac{1}{
		\lambda _{2+}}( \lambda _{2}B_{1}^{\dagger }-B_{2}^{\dagger }),
\end{eqnarray}%
with $\lambda _{2+}=\sqrt{1+\left\vert \lambda _{2}\right\vert ^{2}}>0$. Then we can show
\begin{equation}
B_{1}^{\dagger }B_{1}+B_{2}^{\dagger }B_{2}=B_{2+}^{\dagger }B_{2+}+B_{2-}^{\dagger }B_{2-} ,
\end{equation}%
and the Hamiltonian $H^{\left[ M,2\right] }$ can be expressed as
\begin{equation}
	H^{\left[ M,2\right] } 	=\sum_{k=1}^{M}\Delta _{k}A_{k}^{\dagger }A_{k}+\Omega B_{2-}^{\dagger
	}B_{2-}+\Omega B_{2+}^{\dagger }B_{2+}+\sum_{k=1}^{M}(
	G_{k2+}A_{k}^{\dagger }B_{2+}+G_{k2+}^{\ast }B_{2+}^{\dagger }A_{k}),
\end{equation}%
with $G_{k2+}=\lambda _{2+}G_{k1}$. Here we can see that there are one type-$b$ bright mode $B_{2+}$ and one
type-$b$ dark mode $B_{2-}$.

(iii) Step 3: We assume that the statement in Theorem 2(ii) is valid  for $l=j$, namely the Hamiltonian can be expressed as
\begin{eqnarray}
	H^{[ M,j] }&=&H^{[M,(j-1)]}+\Omega B^{\dagger}_{j}B_{j}+\mathbf{A}
	^{\dagger }\mathbf{G}_{j}B_{j}+B_{j}^{\dagger }\mathbf{G}%
	_{j}^{\dagger }\mathbf{A} \nonumber \\
	&=&\sum_{k=1}^{M}\Delta _{k}A_{k}^{\dagger }A_{k}+\Omega\sum_{s=2}^{j}
	B_{s-}^{\dagger }B_{s-}+\Omega B_{j+}^{\dagger }B_{j+}+\sum_{k=1}^{M}(
	G_{kj+}A_{k}^{\dagger }B_{j+}+G_{kj+}^{\ast }B_{j+}^{\dagger }A_{k}),
\end{eqnarray}
where $G_{kj+}=\lambda _{j+}G_{k1}$ with $\lambda_{j+}=\sqrt{1+\sum_{j^{\prime}=2}^{j}\vert\lambda_{j^{\prime}}\vert^{2}}>0$, and the modes $B_{j\pm}$ are defined by
\begin{eqnarray}
	B_{j+} &=&
	\frac{1}{\lambda _{j+}}( \lambda _{\left( j-1\right) +}B_{\left(
		j-1\right) +}+\lambda _{j}B_{j}) ,\hspace{1cm}B_{j+}^{\dagger } =\frac{1}{\lambda _{j+}}( \lambda _{\left(
		j-1\right) +}B_{\left( j-1\right) +}^{\dagger }+\lambda _{j}^{\ast
	}B_{j}^{\dagger }),\nonumber \\
	B_{j-}& =&\frac{1}{\lambda _{j+}}( \lambda _{j}^{\ast }B_{\left(
		j-1\right) +}-\lambda _{\left( j-1\right) +}B_{j}) ,\hspace{1cm}B_{j-}^{\dagger } =\frac{1}{\lambda _{j+}}( \lambda
	_{j}B_{\left( j-1\right) +}^{\dagger }-\lambda _{\left( j-1\right)
		+}B_{j}^{\dagger }).~\label{HMJD}
\end{eqnarray}
Here, $B_{j+}$ is the type-$b$ bright mode, and these  $j-1$ modes
\{$B_{2-},B_{3-},B_{4-},...B_{j-}$\} are  type-$b$ dark modes.

(iv) Step 4: We show that the statement in Theorem 2(ii) is valid for $l=j+1$. For the
Hamiltonian
\begin{eqnarray}
	H^{[ M,(j+1)] } &=&H^{[M,j]}+\Omega B^{\dagger}_{j+1}B_{j+1}+\mathbf{A}
	^{\dagger }\mathbf{G}_{j+1}B_{j+1}+B_{j+1}^{\dagger }\mathbf{G}%
	_{j+1}^{\dagger }\mathbf{A} \nonumber \\
	&=&\sum_{k=1}^{M}\Delta _{k}A_{k}^{\dagger
	}A_{k}+\Omega B_{2-}^{\dagger }B_{2-}+\Omega B_{3-}^{\dagger
	}B_{3-}+...+\Omega B_{j-}^{\dagger }B_{j-}+\Omega B_{j+}^{\dagger
	}B_{j+} +\Omega B_{j+1}^{\dagger }B_{j+1} \nonumber\\
	&&+\sum_{k=1}^{M}( G_{kj+}A_{k}^{\dagger }B_{j+}+G_{kj+}^{\ast
	}B_{j+}^{\dagger }A_{k})+\sum_{k=1}^{M}( G_{k\left(
		j+1\right) }A_{k}^{\dagger }B_{j+1}+G_{k\left( j+1\right) }^{\ast
	}B_{j+1}^{\dagger }A_{k}),
\end{eqnarray}
with
\begin{equation}
	\mathbf{G}_{j+1}=\left(
	\begin{array}{c}
		G_{1\left( j+1\right) } \\
		\cdots  \\
		G_{k\left( j+1\right) } \\
		\cdots  \\
		G_{M\left( j+1\right) }%
	\end{array}%
	\right) =\lambda _{j+1}\mathbf{G}_{1}=\left(
	\begin{array}{c}
		\lambda _{j+1}G_{11} \\
		\cdots  \\
		\lambda _{j+1}G_{k1} \\
		\cdots  \\
		\lambda _{j+1}G_{M1}%
	\end{array}%
	\right).
\end{equation}%
We introduce the operators
\begin{eqnarray}
	B_{\left( j+1\right) +} &=&\frac{1}{\lambda _{(
			j+1) +}}( \lambda _{j+}B_{j+}+\lambda _{j+1}B_{j+1}),\hspace{1cm}	B_{\left( j+1\right) +}^{\dagger } =\frac{1}{\lambda _{\left( j+1\right) +}}( \lambda
		_{j+}B_{j+}^{\dagger }+\lambda _{j+1}^{\ast }B_{j+1}^{\dagger }) , \nonumber\\
	B_{\left( j+1\right) -} &=&\frac{1}{\lambda
		_{\left( j+1\right) +}}( \lambda _{j+1}^{\ast }B_{j+}-\lambda
	_{j+}B_{j+1}),\hspace{1cm}
	B_{\left( j+1\right) -}^{\dagger } =\frac{1}{\lambda _{\left( j+1\right) +}}( \lambda
	_{j+1}B_{j+}^{\dagger }-\lambda _{j+}B_{j+1}^{\dagger }),
\end{eqnarray}%
with $\lambda _{\left( j+1\right) +}=\sqrt{\lambda _{j+}^{2}+\vert \lambda
	_{j+1}\vert ^{2}}=\sqrt{1+\vert \lambda _{2}\vert
	^{2}+\vert \lambda _{3}\vert ^{2}+...+\vert \lambda
	_{j+1}\vert ^{2}}>0$. It can be shown that
\begin{equation}
B_{l+}^{\dagger }B_{l+}+B_{l+1}^{\dagger }B_{l+1}=B_{\left( j+1\right) +}^{\dagger }B_{\left( j+1\right) +}+B_{\left(
		j+1\right) -}^{\dagger }B_{\left( j+1\right) -} .
\end{equation}%
 Then, the Hamiltonian can be expressed as%
\begin{equation}
	H^{[ M,(j+1)] }=\sum_{k=1}^{M}\Delta _{k}A_{k}^{\dagger }A_{k}+\Omega \sum_{s=2}^{j+1}B_{s-}^{\dagger
	}B_{s-}+\Omega B_{\left( j+1\right) +}^{\dagger }B_{\left( j+1\right) +} +\sum_{k=1}^{M}( G_{k\left( j+1\right) +}A_{k}^{\dagger }B_{\left(
		j+1\right) +}+G_{k\left( j+1\right) +}^{\ast }B_{\left( j+1\right)
		+}^{\dagger }A_{k}),
\end{equation}
where  $G_{k\left( j+1\right) +} =\lambda _{\left( j+1\right) +}G_{k1}$. Here, there are one type-$b$ bright mode $B_{(j+1)+}$ and $j$ type-$b$ dark modes \{$B_{2-}$, $B_{3-}$,$\dotsb$, $B_{(j+1)-}$\}. Therefore,
the statement in Theorem 2(ii)  is valid for $l=j+1.$ Based on the above analyses, we can conclude that Theorem 2(ii) is valid for an arbitrary positive integer $l$.

Based on the above proof, we know that for the ($M+l$)-mode linear quantum network, when these $l$ modes are degenerate,  and the corresponding column vectors in the coupling matrix are linearly dependent, then there are one type-$b$ bright mode
	$B_{l+} =\lambda _{l+}^{-1}\left( \lambda _{\left( l-1\right) +}B_{\left(
		l-1\right) +}+\lambda _{l}B_{l}\right)$,
and $l-1$ type-$b$ dark modes $\left\{B_{2-},B_{3-},B_{4-},...,B_{l-}\right\} $, which have been defined in Eq.~(\ref{HMJD}).

\subsubsection{Proof of Theorem 2(iii)}
 We consider the ($M+N$)-mode linear quantum network in which all the $N$ type-$b$ modes are degenerate, i.e., $\Omega_{j=1\text{-}N}=\Omega$. The Hamiltonian of this ($M+N$)-mode network can be written as
	$H=( \mathbf{A}^{\dag},\mathbf{B}^{\dag}) \mathbf{H
	}_{AB}(\textbf{A},\textbf{B})^{{T}}$,
where the coefficient matrix in the normal-mode ($A_k,B_j$) representation is given by Eq. (\ref{HABMNMN}) under the condition $\Omega_{1\text{-}N}=\Omega$.
In this network, the coupling Hamiltonian describing the interactions between the modes $A_{k}$ and $B_{j}$ can be written as
\begin{eqnarray}
	H_{I} &=&\sum_{k=1}^{M}A_{k}^{\dagger
	}(G_{k1}B_{1}+G_{k2}B_{2}+G_{k3}B_{3}+...+G_{kN}B_{N}) \nonumber\\
	&=&A_{1}^{\dagger }(G_{11}B_{1}+G_{12}B_{2}+G_{13}B_{3}+...+G_{1N}B_{N}) \nonumber\\
	&&+A_{2}^{\dagger }(G_{21}B_{1}+G_{22}B_{2}+G_{23}B_{3}+...+G_{2N}B_{N}) \nonumber\\
	&&+\dotsb\nonumber\\
	&&+A_{k}^{\dagger }(G_{k1}B_{1}+G_{k2}B_{2}+G_{k3}B_{3}+...+G_{kN}B_{N}) \nonumber\\
	&&+\dotsb\nonumber\\
	&&+A_{M}^{\dagger }(G_{M1}B_{1}+G_{M2}B_{2}+G_{M3}B_{3}+...+G_{MN}B_{N}).
\end{eqnarray}

In this network, there are $N$ type-$b$ normal modes. Therefore, we can map these $N$ type-$b$ normal modes into the bases of an inner-product space
with dimension $N.$ To this end, we assume that each type-$b$ normal mode $B_{j}$
corresponds to a basis $\mathbf{e}_{j}$, then the operator  $(G_{k1}B_{1}+G_{k2}B_{2}+G_{k3}B_{3}+...+G_{kN}B_{N})$ can be mapped to a
vector $\mathbf{x}_{k}=\sum_{j=1}^{N}G_{kj}\mathbf{e}_{j}$. We first assume the case where these vectors $\mathbf{x}_{k=1-M}$ are linearly independent.
Therefore, we can use the Gram-Schmidt orthogonalization~\cite{Franklin2012S} to find $M$
orthogonal bases $\mathbf{v}_{k=1-M}.$ According to the
Gram-Schmidt orthogonalization method, we can choose
\begin{eqnarray}
	\mathbf{v}_{1} &=&\mathbf{x}_{1}, \nonumber\\
	\mathbf{v}_{2} &=&\mathbf{x}_{2}-\frac{\left\langle \mathbf{x}_{2},\mathbf{v}
		_{1}\right\rangle }{\left\langle \mathbf{v}_{1},\mathbf{v}_{1}\right\rangle }
	\mathbf{v}_{1}, \nonumber\\
	\mathbf{v}_{3} &=&\mathbf{x}_{3}-\frac{\left\langle \mathbf{x}_{3},\mathbf{v}
		_{1}\right\rangle }{\left\langle \mathbf{v}_{1},\mathbf{v}_{1}\right\rangle }
	\mathbf{v}_{1}-\frac{\left\langle \mathbf{x}_{3},\mathbf{v}_{2}\right\rangle
	}{\left\langle \mathbf{v}_{2},\mathbf{v}_{2}\right\rangle }\mathbf{v}_{2}, \nonumber\\
	&&\dotsb \nonumber\\
	\mathbf{v}_{M} &=&\mathbf{x}_{M}-\frac{\left\langle \mathbf{x}_{M},\mathbf{v}
		_{1}\right\rangle }{\left\langle \mathbf{v}_{1},\mathbf{v}_{1}\right\rangle }
	\mathbf{v}_{1}-...-\frac{\left\langle \mathbf{x}_{M},\mathbf{v}
		_{M-1}\right\rangle }{\left\langle \mathbf{v}_{M-1},\mathbf{v}
		_{M-1}\right\rangle }\mathbf{v}_{M-1},
\end{eqnarray}
where the inner product is defined by
\begin{equation}
	\left\langle \mathbf{x}_{k},\mathbf{x}_{k^{\prime }}\right\rangle
	=G_{k1}^{\ast }G_{k^{\prime }1}+G_{k2}^{\ast }G_{k^{\prime }2}+G_{k3}^{\ast
	}G_{k^{\prime }3}+...+G_{kM}^{\ast }G_{k^{\prime }M}.
\end{equation}
We can show that all these bases $\textbf{v}_{k=1\text{-}M}$ are orthogonal, i.e., $\langle \textbf{v}_{k},\textbf{v}_{k^{\prime}\neq k}\rangle=0$. In addition, the normalization of these orthogonal bases $\textbf{v}_{k}$ can be realized by introducing $\tilde{\textbf{v}}_{k}=\textbf{v}_{k}/\vert\vert \textbf{v}_{k}\vert\vert$, where $\vert\vert \textbf{v}_{k}\vert\vert $ is the norm of the vector $\textbf{v}_{k}$, and then we have $\langle \tilde{\textbf{v}}_{k},\tilde{\textbf{v}}_{k^{\prime}}\rangle=\delta_{k,k^{\prime}}$.
In the corresponding mode space, the orthonormalization of the bases $\tilde{\textbf{v}}_k$ implies that the bosonic communication relation is satisfied.

Here we can see that these type-$a$ modes only couple to the type-$b$ modes corresponding to the vectors in the subspace spanned by these orthogonal vectors $\mathbf{v}_{k=1-M}.$ This means that there are $N-M$ type-$b$ dark modes which decouple from these type-$a$ modes. Namely, there are $M$ type-$b$ bright modes coupled to these type-$a$ modes. In the above analyses, we have assumed that all the vectors $\textbf{x}_{l=1-M}$ are linearly independent, then there are $N-M$ type-$b$ dark modes. When some of these vectors $\textbf{x}_{l=1-M}$  are linearly dependent, then the number of the orthogonal bases coupled to type-$a$ modes will be smaller than $M$. Therefore, there are at least $N-M$ type-$b$ dark modes in the network when $N>M$.

\subsubsection{Proof of Theorem 2(iv)}
For generality, we assume that in the ($M+N$)-mode linear quantum network, there are some groups of degenerate type-$b$ normal modes and some groups of non-degenerate type-$b$ normal modes. In the normal-mode ($A_{k},B_{j}$) representation, the coefficient matrix becomes
\begin{eqnarray}
		\mathbf{H}_{AB}&=&\left(\begin{array}{c c}\textbf{H}_{A} &\textbf{C}_{AB}\\\textbf{C}_{AB}^{\dagger} &\textbf{H}_{B}
		\end{array}\right)\nonumber\\
		&=&\left(\begin{array}{cccc|cccc|cccc|c|c|c}
			\Delta_{1} & 0 & \cdots & 0 & G_{11} & G_{12} & \cdots & G_{1l_{1}} & G_{1(l_{1}+1)} & G_{1(l_{1}+2)} & \cdots & G_{1(l_{1}+l_{2})} & \cdots & G_{1(N-1)} & G_{1N}\\
			0 & \Delta_{2} & \cdots & 0 & G_{21} & G_{22} & \cdots & G_{2l_{1}} & G_{2(l_{1}+1)} & G_{2(l_{1}+2)} & \cdots & G_{2(l_{1}+l_{2})} & \cdots & G_{2(N-1)} & G_{2N}\\
			\cdots & \cdots & \cdots & \cdots & \cdots & \cdots & \cdots & \cdots & \cdots & \cdots & \cdots & \cdots & \cdots & \cdots & \cdots\\
			0 & 0 & \cdots & \Delta_{M} & G_{M1} & G_{M2} & \cdots & G_{Ml_{1}} & G_{M(l_{1}+1)} & G_{M(l_{1}+2)} & \cdots & G_{M(l_{1}+l_{2})} & \cdots & G_{M(N-1)} & G_{MN}\\
			\hline G_{11}^{\ast} & G_{21}^{\ast} & \cdots & G_{M1}^{\ast} & \text{\ensuremath{\Omega_{1}}} & 0 & \cdots & 0 & 0 & 0 & \cdots & 0 & \cdots & 0 & 0\\
			G_{12}^{\ast} & G_{22}^{\ast} & \cdots & G_{M2}^{\ast} & 0 & \text{\ensuremath{\Omega_{1}}} & \cdots & 0 & 0 & 0 & \cdots & 0 & \cdots & 0 & 0\\
			\cdots & \cdots & \cdots & \cdots & \cdots & \cdots & \cdots & \cdots & \cdots & \cdots & \cdots & \cdots & \cdots & \cdots & \cdots\\
			G_{1l_{1}}^{\ast} & G_{2l_{1}}^{\ast} & \cdots & G_{Ml_{1}}^{\ast} & 0 & 0 & \cdots & \text{\ensuremath{\Omega_{1}}} & 0 & 0 & \cdots & 0 & \cdots & 0 & 0\\
			\hline G_{1(l_{1}+1)}^{\ast} & G_{2(l_{1}+1)}^{\ast} & \cdots & G_{M(l_{1}+1)}^{\ast} & 0 & 0 & \cdots & 0 & \text{\ensuremath{\Omega_{l_{1}+1}}} & 0 & \cdots & 0 & \cdots & 0 & 0\\
			G_{1(l_{1}+2)}^{\ast} & G_{2(l_{1}+2)}^{\ast} & \cdots & G_{M(l_{1}+2)}^{\ast} & 0 & 0 & \cdots & 0 & 0 & \text{\text{\ensuremath{\Omega_{l_{1}+1}}}} & \cdots & 0 & \cdots & 0 & 0\\
			\cdots & \cdots & \cdots & \cdots & \cdots & \cdots & \cdots & \cdots & \cdots & \cdots & \cdots & \cdots & \cdots & \cdots & \cdots\\
			G_{1(l_{1}+l_{2})}^{\ast} & G_{2(l_{1}+l_{2})}^{\ast} & \cdots & G_{M(l_{1}+l_{2})}^{\ast} & 0 & 0 & \cdots & 0 & 0 & 0 & \cdots & \text{\ensuremath{\Omega_{l_{1}+1}}} & \cdots & 0 & 0\\
			\hline \cdots & \cdots & \cdots & \cdots & \cdots & \cdots & \cdots & \cdots & \cdots & \cdots & \cdots & \cdots & \cdots & \cdots & \cdots\\
			\hline G_{1(N-1)}^{\ast} & G_{2(N-1)}^{\ast} & \cdots & G_{M(N-1)}^{\ast} & 0 & 0 & \cdots & 0 & 0 & 0 & \cdots & 0 & \cdots & \Omega_{(N-1)} & 0\\
			\hline G_{1N}^{\ast} & G_{2N}^{\ast} & \cdots & G_{MN}^{\ast} & 0 & 0 & \cdots & 0 & 0 & 0 & \cdots & 0 & \cdots & \cdots & \Omega_{N}
		\end{array}\right)\nonumber\\
		&=&\left(\begin{array}{cccccc}
			\mathbf{H}_{A}^{\{M\}} & \mathbf{C}_{AB}^{[M,\{1\text{-} l_{1}\}]} & \mathbf{C}_{AB}^{[M,\{(l_{1}+1)\text{-}(l_{1}+l_{2})\}]} & \cdots & \mathbf{C}_{AB}^{[M,\{N-1\}]} & \mathbf{C}_{AB}^{[M,\{N\}]}\\
			\left(\mathbf{C}_{AB}^{[M,\{1\text{-} l_{1}\}]}\right)^{\dagger} & \mathbf{H}_{B}^{\{1\text{-} l_{1}\}} & 0 & \cdots & 0 & 0\\
			\left(\mathbf{C}_{AB}^{[M,\{(l_{1}+1)\text{-}(l_{1}+l_{2})\}]}\right)^{\dagger} & 0 & \mathbf{H}_{B}^{\{(l_{1}+1)\text{-} (l_1+l_{2})\}} & \cdots & 0 & 0\\
			\cdots & \cdots & \cdots & \cdots & \cdots & \cdots\\
			\left(\mathbf{C}_{AB}^{[M,\{N-1\}]}\right)^{\dagger} & 0 & 0 & \cdots & \mathbf{H}_{B}^{\{N-1\}} & 0\\
			\left(\mathbf{C}_{AB}^{[M,\{N\}]}\right)^{\dagger} & 0 & 0 & \cdots & 0 & \mathbf{H}_{B}^{\{N\}}
		\end{array}\right).~\label{HABLMNB}
\end{eqnarray}%
As shown in Eq.~(\ref{HABLMNB}), here we assume that these type-$b$ normal modes $B_{j=1\text{-}l_1}$ are degenerate, with the resonance frequency $\Omega_{1}$. In addition, these type-$b$ normal modes $B_{j=[(l_1+1)\text{-} (l_1+l_2)]}$ (i.e., $B_{l_1+1},B_{l_1+2},\dotsb,B_{l_1+l_2}$) are also degenerate, with the resonance frequency $\Omega_{l_1+1}$ (we use the index of the first type-$b$ mode in this degenerate-mode subspace to denote the resonance frequency of these degenerate normal modes).  In addition, we assume that the type-$b$ normal modes $B_{N-1}$ and $B_{N}$ are non-degenerate, i.e., their resonance frequencies are $\Omega_{N-1}$ ($\Omega_{N-1}\neq \Omega_{j\neq N-1}$) and $\Omega_N$ ($\Omega_N\neq \Omega_{j\neq N}$), respectively.
In principle, all other modes can also be divided into either other degenerate-mode subspaces or non-degenerate-mode subspaces.
Therefore, the coupling matrix $\textbf{C}_{AB}$ can be divided into many coupling sub-matrices: $\textbf{C}_{AB}^{[M,\{1\text{-} l_1\}]}$, $\textbf{C}_{AB}^{[M,\{(l_1+1)\text{-}(l_1+l_2)\}]}$,$\dotsb$, $\textbf{C}_{AB}^{[M,\{N-1\}]}$, $\textbf{C}_{AB}^{[M,\{N\}]}$, where the superscript $\{1\text{-} l_1\}$ denotes that these type-$b$ normal modes $B_1, B_2,\dotsb, B_l$ are degenerate, the superscript $\{(l_1+1)\text{-}(l_1+l_2)\}$ indicates that these type-$b$ normal modes $B_{l_1+1}, B_{l_1+2},\dotsb, B_{l_1+l_2}$ are degenerate, and the superscripts $\{N-1\}$ and $\{N\}$ denote the two non-degenerate type-$b$ normal modes $B_{N-1}$ and $B_N$, respectively. Since the dark modes only exist in the same degenerate-mode subspace, therefore, we can know the number of type-$b$ bright and dark modes in this network by analyzing the dark-mode effect in all these degenerate-mode subspaces. Below, we analyze the number of type-$b$ bright and dark modes in a given degenerate-mode subspace.

For example, in the degenerate type-$b$ normal-mode subspace (spanned by the modes $B_{j=1-l_1}$ with the resonance frequency $\Omega_1$), the number of type-$b$ bright modes can be determined by analyzing the rank of the following coupling sub-matrix associated with the type-$a$ normal modes $A_{k=1\text{-}M}$ and the type-$b$ normal modes $B_{j=1\text{-}l_1}$,
\begin{equation}
	\textbf{C}_{AB}^{[M,\{1\text{-} l_1\}]}=\left(
	\begin{array}{ccccc}
		G_{11} & G_{12} & G_{13} & \cdots  & G_{1l_1} \\
		G_{21} & G_{22} & G_{23} & \cdots  & G_{2l_1} \\
		G_{31} & G_{32} & G_{33} & \cdots  & G_{3l_1} \\
		\cdots  & \cdots  & \cdots  & \cdots  & \cdots\\
		G_{M1} & G_{M2} & G_{M3} & \cdots  & G_{Ml_1}
	\end{array}
	\right).
\end{equation}
Below we will show that the number of type-$b$ bright modes within this degenerate-mode subspace is equal to the rank of the  sub-matrix $\textbf{C}_{AB}^{[M,\{1\text{-} l_1\}]}$.

(i) When $l_1>M$ and there is no zero column vector in the matrix $\textbf{C}_{AB}^{[M,\{1\text{-} l_1\}]}$, as well as all the column vectors are linearly independent, then the rank of $\textbf{C}_{AB}^{[M,\{1\text{-} l_1\}]}$ is $M$. According to Theorem 2(iii), we know that the number of type-$b$ bright modes in this degenerate type-$b$ normal-mode subspace is $M$.

(ii) When $l_1>M$ and there are $s$ ($1 \leq s\leq l_1$) column vectors which can be transformed to zero column vectors by matrix elementary transformation, for example $(\textbf{C}_{AB}^{[M,\{1\text{-} l_1\}]})_{j2}=\eta_{2}(\textbf{C}_{AB}^{[M,\{1\text{-} l_1\}]})_{j1}$, $\cdots$,  $(\textbf{C}_{AB}^{[M,\{1\text{-} l_1\}]})_{j(s+1)}=\eta_{s+1}(\textbf{C}_{AB}^{[M,\{1\text{-} l_1\}]})_{j1}$ for $j=1\text{-}M$, then
the rank of $\textbf{C}_{AB}^{[M,\{1\text{-} l_1\}]}$ is $M$ when $l_1-s\geq M$ (In this case, the number of type-$b$ bright modes in this degenerate type-$b$ normal-mode subspace is $M$), and the rank of $\textbf{C}_{AB}^{[M,\{1\text{-} l_1\}]}$ is $l_1-s$ when $l_1-s\leq M$. In this case, the number of type-$b$ bright modes in this degenerate type-$b$ normal-mode subspace is $l_1-s$.

(iii) When $l_1<M$ and there is no zero column vector in the matrix $\textbf{C}_{AB}^{[M,\{1\text{-} l_1\}]}$, as well as all the column vectors are linearly independent, then the rank of $\textbf{C}_{AB}^{[M,\{1\text{-} l_1\}]}$ is $l_1$. In this case, the number of type-$b$ bright modes in this degenerate type-$b$ normal-mode subspace is $l_1$.

(iv) When $l_1<M$ and there are $s$ ($1 \leq s\leq l_1$) column vectors can be transformed to zero column vectors by matrix elementary transformation, then the rank of $\textbf{C}_{AB}^{[M,\{1\text{-} l_1\}]}$ is $l_1-s$. The number of type-$b$ bright modes in this degenerate type-$b$ normal-mode subspace is $l_1-s$.

We can see that in all these cases, the number of type-$b$ bright modes in this degenerate type-$b$ normal-mode subspace is equal to the rank of the sub-matrix $\textbf{C}_{AB}^{[M,\{1\text{-} l_1\}]}$. Similarly, in the degenerate type-$b$ normal-mode subspace spanned by the modes $B_{j=(l_1+1)\text{-}(l_1+l_2)}$ with the degenerate frequency $\Omega_{l_1+1}$, the number of type-$b$ bright modes is equal to the rank of $\textbf{C}_{AB}^{[M,\{(l_1+1)\text{-} (l_1+l_2)\}]}$. Based on the above analyses, we know that the number of type-$b$ bright modes associated with all degenerate subspaces is a sum of the ranks of all these coupling sub-matrices corresponding to these degenerate-mode subspaces. In addition, for a non-degenerate type-$b$ mode, it will always couple to these type-$a$ modes, namely, the number of type-$b$ bright mode is one, which is also right the rank of the corresponding column vector. Therefore, the total number of type-$b$ bright modes in this network is equal to the sum of the ranks of all these coupling sub-matrices associated with degenerate type-$b$ normal-mode subspaces and the number of non-degenerate type-$b$ modes.  In other words, the number of type-$b$ bright modes in the network is equal to the sum $\sum_{s}R_{s}$ of the ranks of all the coupling sub-matrices, where $R_s$ is the rank of the $s$th coupling sub-matrix. As a result, the number of type-$b$ dark modes in the linear quantum network is the total number of type-$b$ modes minus the number of type-$b$ bright modes, i.e., $N-\sum_{s}R_{s}$.

\subsubsection{Proof of Theorem 2(v)}
We consider a ($M+N$)-mode linear quantum network consisting of $M$ ($M\geq2$) type-$a$
modes and $N$ type-$b$ modes. Without loss of generality, we assume that there are some degenerate and non-degenerate mode subspaces in the network. In particular, we consider the degenerate-mode distribution as the same as the case considered in the proof of Theorem 2(iv). Below, we prove Theorem 2(v) with two steps: (a) We show the dark-mode effect for the network when $\xi_{kk^{\prime}}= 0$, i.e., in the absence of the couplings among these type-$a$ modes; (b) We show the dark-mode effect in the presence of excitation-hopping interactions among all these type-$a$ modes. In these two cases, both the resonance frequencies of type-$a$ and type-$b$ modes and the coupling distribution among these type-$b$ modes keep unchanged.

(a) We first consider the case $\xi_{kk^{\prime}}=0$, i.e., there are no couplings among all these type-$a$ modes. When $\xi_{kk^{\prime}}=0$,
the Hamiltonian of this ($M+N$)-mode linear quantum network can be expressed
as $H=(\boldsymbol{a}^{\dagger},\boldsymbol{b}^{\dagger})\textbf{H}_{ab}(\boldsymbol{a},\boldsymbol{b})^{T}$,
where the coefficient matrix $\textbf{H}_{ab}$ is defined in Eq.~(\ref{HabMNMN}) under the assumption of degenerate frequencies and $\xi_{kk^{\prime}}=0$.
To prove Theorem 2(v), we introduce the unitary operators
$\textbf{U}_{b}$ and $\textbf{U}_{b}^{\dagger}$ to diagonalize the
Hamiltonian $\textbf{H}_{b}$ of the type-$b$ mode sub-network
as $\mathbf{U}_{b}\mathbf{H}_{b}\mathbf{U}_{b}^{\dagger}=\mathbf{H}_{B}=\textrm{diag}\{\Omega_{1},\Omega_{2},\dotsb,\Omega_{N}\}$.
Then the Hamiltonian becomes
\begin{equation}
	H=(\boldsymbol{a}^{\dagger},\textbf{B}^{\dagger})\textbf{H}_{aB}\left(\begin{array}{c}
		\boldsymbol{a}\\
		\textbf{B}
	\end{array}\right),~\label{HHHHHHHH}
\end{equation}
where $\textbf{B}=(B_{1},B_{2},\dotsb,B_{N})^{T}$ with $B_{j}=\sum_{j^{\prime}=1}^{N}(\textbf{U}_{b})_{jj^{\prime}}b_{j^{\prime}}$. The coefficient matrix in Eq. (\ref{HHHHHHHH}) is defined by
\begin{eqnarray}
	\textbf{H}_{aB}&=&\left(\begin{array}{cc}
	\textbf{H}_{a} & \textbf{C}_{aB}\\
	\textbf{C}_{aB}^{\dagger} & \textbf{H}_{B}
    \end{array}\right)\nonumber\\&=&\left(\begin{array}{cccc|cccc|cccc|c|c|c}
    \delta_{1} & 0 & \cdots & 0 & g_{11}^{[aB]} & g_{12}^{[aB]} & \cdots & g_{1l_{1}}^{[aB]} & g_{1(l_{1}+1)}^{[aB]} & g_{1(l_{1}+2)}^{[aB]} & \cdots & g_{1(l_{1}+l_{2})}^{[aB]} & \cdots & g_{1(N-1)}^{[aB]} & g_{1N}^{[aB]}\\
    0 & \delta_{2} & \cdots & 0 & g_{21}^{[aB]} & g_{22}^{[aB]} & \cdots & g_{2l_{1}}^{[aB]} & g_{2(l_{1}+1)}^{[aB]} & g_{2(l_{1}+2)}^{[aB]} & \cdots & g_{2(l_{1}+l_{2})}^{[aB]} & \cdots & g_{2(N-1)}^{[aB]} & g_{2N}^{[aB]}\\
    \cdots & \cdots & \cdots & \cdots & \cdots & \cdots & \cdots & \cdots & \cdots & \cdots & \cdots & \cdots & \cdots & \cdots & \cdots\\
    0 & 0 & \cdots & \delta_{M} & g_{M1}^{[aB]} & g_{M2}^{[aB]} & \cdots & g_{Ml_{1}}^{[aB]} & g_{M(l_{1}+1)}^{[aB]} & g_{M(l_{1}+2)}^{[aB]} & \cdots & g_{M(l_{1}+l_{2})}^{[aB]} & \cdots & g_{M(N-1)}^{[aB]} & g_{MN}^{[aB]}\\
    \hline \left(g_{11}^{[aB]}\right)^{\ast} & \left(g_{21}^{[aB]}\right)^{\ast} & \cdots & \left(g_{M1}^{[aB]}\right)^{\ast} & \text{\ensuremath{\Omega_{1}}} & 0 & \cdots & 0 & 0 & 0 & \cdots & 0 & \cdots & 0 & 0\\
    \left(g_{12}^{[aB]}\right)^{\ast} & \left(g_{22}^{[aB]}\right)^{\ast} & \cdots & \left(g_{M2}^{[aB]}\right)^{\ast} & 0 & \text{\ensuremath{\Omega_{1}}} & \cdots & 0 & 0 & 0 & \cdots & 0 & \cdots & 0 & 0\\
    \cdots & \cdots & \cdots & \cdots & \cdots & \cdots & \cdots & \cdots & \cdots & \cdots & \cdots & \cdots & \cdots & \cdots & \cdots\\
    \left(g_{1l_{1}}^{[aB]}\right)^{\ast} & \left(g_{2l_{1}}^{[aB]}\right)^{\ast} & \cdots & \left(g_{Ml_{1}}^{[aB]}\right)^{\ast} & 0 & 0 & \cdots & \text{\ensuremath{\Omega_{1}}} & 0 & 0 & \cdots & 0 & \cdots & 0 & 0\\
    \hline \left(g_{1(l_{1}+1)}^{[aB]}\right)^{\ast} & \left(g_{2(l_{1}+1)}^{[aB]}\right)^{\ast} & \cdots & \left(g_{M(l_{1}+1)}^{[aB]}\right)^{\ast} & 0 & 0 & \cdots & 0 & \text{\ensuremath{\Omega_{l_{1}+1}}} & 0 & \cdots & 0 & \cdots & 0 & 0\\
    \left(g_{1(l_{1}+2)}^{[aB]}\right)^{\ast} & \left(g_{2(l_{1}+2)}^{[aB]}\right)^{\ast} & \cdots & \left(g_{M(l_{1}+2)}^{[aB]}\right)^{\ast} & 0 & 0 & \cdots & 0 & 0 & \text{\text{\ensuremath{\Omega_{l_{1}+1}}}} & \cdots & 0 & \cdots & 0 & 0\\
    \cdots & \cdots & \cdots & \cdots & \cdots & \cdots & \cdots & \cdots & \cdots & \cdots & \cdots & \cdots & \cdots & \cdots & \cdots\\
    \left(g_{1(l_{1}+l_{2})}^{[aB]}\right)^{\ast} & \left(g_{2(l_{1}+l_{2})}^{[aB]}\right)^{\ast} & \cdots & \left(g_{M(l_{1}+l_{2})}^{[aB]}\right)^{\ast} & 0 & 0 & \cdots & 0 & 0 & 0 & \cdots & \text{\ensuremath{\Omega_{l_{1}+1}}} & \cdots & 0 & 0\\
    \hline \cdots & \cdots & \cdots & \cdots & \cdots & \cdots & \cdots & \cdots & \cdots & \cdots & \cdots & \cdots & \cdots & \cdots & \cdots\\
    \hline \left(g_{1(N-1)}^{[aB]}\right)^{\ast} & \left(g_{2(N-1)}^{[aB]}\right)^{\ast} & \cdots & \left(g_{M(N-1)}^{[aB]}\right)^{\ast} & 0 & 0 & \cdots & 0 & 0 & 0 & \cdots & 0 & \cdots & \Omega_{(N-1)} & 0\\
    \hline \left(g_{1N}^{[aB]}\right)^{\ast} & \left(g_{2N}^{[aB]}\right)^{\ast} & \cdots & \left(g_{MN}^{[aB]}\right)^{\ast} & 0 & 0 & \cdots & 0 & 0 & 0 & \cdots & 0 & \cdots & \cdots & \Omega_{N}
\end{array}\right)\nonumber\\&=&\left(\begin{array}{cccccc}
	\mathbf{H}_{a}^{\{M\}} & \mathbf{C}_{aB}^{[M,\{1\text{-} l_{1}\}]} & \mathbf{C}_{aB}^{[M,\{(l_{1}+1)\text{-}(l_{1}+l_{2})\}]} & \cdots & \mathbf{C}_{aB}^{[M,\{N-1\}]} & \mathbf{C}_{aB}^{[M,\{N\}]}\\
	\left(\mathbf{C}_{aB}^{[M,\{1\text{-} l_{1}\}]}\right)^{\dagger} & \mathbf{H}_{B}^{\{1\text{-} l_{1}\}} & 0 & \cdots & 0 & 0\\
	\left(\mathbf{C}_{aB}^{[M,\{(l_{1}+1)\text{-}(l_{1}+l_{2})\}]}\right)^{\dagger} & 0 & \mathbf{H}_{B}^{\{(l_{1}+1)\text{-}(l_{1}+l_{2})\}} & \cdots & 0 & 0\\
	\cdots & \cdots & \cdots & \cdots & \cdots & \cdots\\
	\left(\mathbf{C}_{aB}^{[M,\{N-1\}]}\right)^{\dagger} & 0 & 0 & \cdots & \mathbf{H}_{B}^{\{N-1\}} & 0\\
	\left(\mathbf{C}_{aB}^{[M,\{N\}]}\right)^{\dagger} & 0 & 0 & \cdots & 0 & \mathbf{H}_{B}^{\{N\}}
\end{array}\right),\label{Ha1111-1}
\end{eqnarray}
where $g_{kj}^{[aB]}=\sum_{j^{\prime}=1}^{N}g_{kj^{\prime}}(\textbf{U}_{b}^{\dagger})_{j^{\prime}j}$
is the effective coupling strength between the $k$th type-$a$ bare mode $a_k$ and the
$j$th type-$b$ normal mode $B_j$. Note that Eq. (\ref{Ha1111-1}) can be obtained from Eq. (\ref{HABLMNB}) by setting $\xi_{kk^{\prime}}=0$ and making the replacement $\delta_{k}=\Delta_{k}$ (i.e., $\textbf{U}_{a}=\textbf{I}$). In this case, we can divide $\textbf{C}_{aB}$ into many coupling sub-matrices corresponding to different degenerate-mode subspaces (we understand the non-degenerate subspace as a special case of degenerate subspace). According
to Theorem 2(iv), the number of type-$b$ bright modes is the sum $\sum_{s}R_{s}$ of the ranks
of all these coupling sub-matrices, and
the number of type-$b$ dark modes is $N-\sum_{s}R_{s}$.

(b) When $\xi_{kk^{\prime}}\neq0$, we further introduce the unitary operators $\textbf{U}_{a}$
and $\textbf{U}_{a}^{\dagger}$ to diagonalize the Hamiltonian $\textbf{H}_{a}$
of the type-$a$ mode sub-network as $\mathbf{U}_{a}\mathbf{H}_{a}\mathbf{U}_{a}^{\dagger}=\mathbf{H}_{A}=\textrm{diag}\{\Delta_{1},\Delta_{2},\dotsb,\Delta_{M}\}$.
In the normal-mode ($A_{k},B_{j}$) representation, the Hamiltonian of this network can be expressed as $H=(\mathbf{A}^{\dagger},\mathbf{B}^{\dagger})\mathbf{H}_{AB}(\textbf{A},\textbf{B})^{T}$, where the coefficient matrix $\mathbf{H}_{AB}$ is given by Eq. (\ref{HABLMNB}).
Here, we introduce the vector $\textbf{A}=(A_{1},A_{2},\dotsb,A_{M})^{T}$
with $A_{k}=\sum_{k^{\prime}=1}^{\textsc{\textsc{M}}}(\textbf{U}_{a})_{kk^{\prime}}a_{k^{\prime}}$, and $G_{kj}=\sum_{k^{\prime}=1}^{M}(\textbf{U}_{a})_{kk^{\prime}}g_{k^{\prime}j}^{[aB]}$
is the effective coupling strength between the $k$th type-$a$ normal
mode $A_k$ and the $j$th type-$b$ normal mode $B_j$. In this case, we know that $\text{\textbf{C}}_{AB}=\textbf{U}_{a}$$\text{\textbf{C}}_{aB}$. These sub-matrices of the coupling matrix $\textbf{C}_{AB}$ can be obtained as $\mathbf{C}_{AB}^{[M,\{1\text{-} l_{1}\}]}=\textbf{U}_a\mathbf{C}_{aB}^{[M,\{1\text{-} l_{1}\}]}  $, $\mathbf{C}_{AB}^{[M,\{(l_1+1)\text{-} (l_{1}+l_2)\}]}=\textbf{U}_a\mathbf{C}_{aB}^{[M,\{(l_1+1)\text{-} (l_{1}+l_2)\}]}  $,$\dotsb$, $\mathbf{C}_{AB}^{[M,\{N-1\}]}=\textbf{U}_a\mathbf{C}_{aB}^{[M,\{N-1\}]}  $, $\mathbf{C}_{AB}^{[M,\{N\}]}=\textbf{U}_a\mathbf{C}_{aB}^{[M,\{N\}]}  $.  Based on the fact that \emph{the rank of the coupling sub-matrix will not change when it multiples a unitary matrix}, we know that the ranks of the two corresponding coupling matrices $\textbf{C}_{aB}$ and $\textbf{C}_{AB}$ are equal.
Based on Theorem 2(iv), we know that the number of type-$b$ bright modes in the network
is $\sum_{s}R_s$,
and the number of type-$b$ dark modes is $N-\sum_{s}R_s$.  However, we should point out that the forms of the dark modes will be changed due to the different effective coupling strengths.
Therefore, we can conclude that the change of the coupling strengths $\xi_{kk^{\prime}}$  between these type-$a$ modes $a_{k}$ and $a_{k^{\prime}}$ will not change the number of type-$b$ dark modes, but it will change the forms of these type-$b$ dark modes.

\section{Implementation of the type-($a, b$) ($M+N$)-mode linear quantum network with linearized optomechanical network~\label{modelmn}}
In this section, we present the detailed derivations concerning the implementation of the ($M+N$)-mode linear network with  linearized optomechanical network. For generality, we consider a ($M+N$)-mode optomechanical network consisting of $M$ optical modes optomechanically coupled to $N$ mechanical modes. Here, these optical (mechanical) modes are coupled with each other via the photon  (phonon)-hopping interactions. In addition, each optical mode is driven by a monochromatic field.
The Hamiltonian of the optomechanical network reads
\begin{eqnarray}
	\label{HamiltMNS}
	H^{[M,N]}_{\text{omn}}(t)&=&\sum_{k=1}^{M}\omega _{c,k}\tilde{a}_{k}^{\dagger}\tilde{a}_{k}+\sum_{j=1}^{N}\omega _{j}\tilde{b}_{j}^{\dagger}\tilde{b}_{j}+\sum_{\textcolor{black}{k,k^{\prime }=1, k<k^{\prime }}}^{M}(\xi _{kk^{\prime}}\tilde{a}_{k}^{\dagger }\tilde{a}_{k^{\prime }}+\xi^{\ast} _{kk^{\prime}}\tilde{a}_{k^{\prime }}^{\dagger }\tilde{a}_{k})+\sum_{\textcolor{black}{j,j^{\prime}=1,j<j^{\prime
	}}}^{N}(\eta _{jj^{\prime }}\tilde{b}_{j}^{\dagger }\tilde{b}_{j^{\prime }}+\eta^{\ast} _{jj^{\prime }}\tilde{b}_{j^{\prime }}^{\dagger }\tilde{b}_{j})\nonumber \\
	&&+\sum_{k=1}^{M}\sum_{j=1}^{N}\tilde{g}_{kj}\tilde{a}_{k}^{\dagger
}\tilde{a}_{k}(\tilde{b}_{j}^{\dagger }+\tilde{b}_{j})+\sum_{k=1}^{M}(\Lambda_{k}\tilde{a}_{k}^{\dagger }e^{-i\omega_{L,k}t}+\Lambda^{\ast}_{k}\tilde{a}_{k}e^{i\omega_{L,k}t}),
\end{eqnarray}
where $\tilde{a}_{k=1\text{-}M}$ $(\tilde{a}^{\dagger}_{k=1\text{-}M})$ and $\tilde{b}_{j=1\text{-}N}$ $(\tilde{b}^{\dagger}_{j=1\text{-}N})$ are, respectively, the annihilation (creation) operators of the $k$th optical mode and the $j$th mechanical mode, with the corresponding resonance frequencies $\omega_{c,k}$ and $\omega_{j}$. The $\tilde{g}_{kj}$ terms describe the optomechanical interactions between the $k$th optical mode $\tilde{a}_{k}$ and the $j$th mechanical mode $\tilde{b}_{j}$, with $\tilde{g}_{kj}$ being the single-photon optomechanical coupling strengths.   The $\xi _{kk^{\prime }}$ ($\eta_{jj^{\prime}}$) terms describe the photon (phonon)-hopping interactions [with coupling strengths $\xi_{kk^{\prime}}$ ($\eta_{jj^{\prime}}$)] between the two optical (mechanical) modes  $\tilde{a}_{k}$ and $\tilde{a}_{k^{\prime}}$ ($\tilde{b}_{j}$ and $\tilde{b}_{j^{\prime}}$). Moreover, $\Lambda_{k}$ and $\omega_{L,k}$ are, respectively,  the driving amplitude and  driving frequency of the monochromatic field associated with  the $k$th optical mode.  In our following discussions, we assume that all the driving fields have the same resonance frequencies, i.e., $\omega_{L,k}=\omega_{L}$.

In a rotating frame defined by $\exp(-i\omega_{L}t\sum_{k=1}^{M}\tilde{a}_{k}^{\dagger}\tilde{a}_{k})$, the Hamiltonian~(\ref{HamiltMNS})  becomes
\begin{eqnarray}
	\label{HamiltrotMNS}
	H^{[M,N]}_{I}&=&\sum_{k=1}^{M}\tilde{\delta} _{k}\tilde{a}_{k}^{\dagger}\tilde{a}_{k}+\sum_{j=1}^{N}\omega _{j}\tilde{b}_{j}^{\dagger}\tilde{b}_{j}+\sum_{\textcolor{black}{k,k^{\prime}=1, k<k^{\prime }}}^{M}(\xi _{kk^{\prime}}\tilde{a}_{k}^{\dagger }\tilde{a}_{k^{\prime }}+\xi^{\ast} _{kk^{\prime}}\tilde{a}_{k^{\prime }}^{\dagger }\tilde{a}_{k})+\sum_{\textcolor{black}{j,j^{\prime}=1,j<j^{\prime
	}}}^{N}(\eta _{jj^{\prime }}\tilde{b}_{j}^{\dagger }\tilde{b}_{j^{\prime }}+\eta^{\ast} _{jj^{\prime }}\tilde{b}_{j^{\prime }}^{\dagger }\tilde{b}_{j})\nonumber \\
	&&+\sum_{k=1}^{M}\sum_{j=1}^{N}\tilde{g}_{kj}\tilde{a}_{k}^{\dagger}\tilde{a}_{k}(\tilde{b}_{j}^{\dagger }+\tilde{b}_{j})+\sum_{k=1}^{M}(\Lambda_{k}\tilde{a}_{k}^{\dagger }+\Lambda^{\ast}_{k}\tilde{a}_{k}),
\end{eqnarray}
where $\tilde{\delta}_{k}=\omega_{c,k}-\omega_{L}$ is the driving detuning of the resonance frequency $\omega_{c,k}$ of the $k$th optical mode with respect to the driving frequency $\omega_{L}$.

To include the dissipations in this optomechanical network, we assume that the optical modes and the mechanical modes are, respectively, connected to vacuum baths and heat baths. Within the Markovian-dissipation framework, we can obtain the quantum Langevin equations for the system operators as
\begin{eqnarray}
	\dot{\tilde{a}}_{1} &=&-(\kappa _{1}+i\tilde{\delta}_{1})\tilde{a}_{1}-i\sum_{j=1}^{N}\tilde{g}_{1j}\tilde{a}_{1}(\tilde{b}_{j}^{\dagger }+\tilde{b}_{j})-i\xi_{12}\tilde{a}_{2}-i\xi _{13}\tilde{a}_{3}-...-i\xi _{1M}\tilde{a}_{M}-i\Lambda _{1}+\sqrt{2\kappa_{1}}\tilde{a}_{1\text{,in}}, \nonumber\\
	\dot{\tilde{a}}_{2} &=&-(\kappa _{2}+i\tilde{\delta}_{2})\tilde{a}_{2}-i\sum_{j=1}^{N}\tilde{g}_{2j}\tilde{a}_{2}(\tilde{b}_{j}^{\dagger }+\tilde{b}_{j})-i\xi^{\ast}_{12}\tilde{a}_{1}-i\xi _{23}\tilde{a}_{3}-...-i\xi _{2M}\tilde{a}_{M}-i\Lambda _{2}+\sqrt{2\kappa_{2}}\tilde{a}_{2\text{,in}}, \nonumber\\
	&&\vdots  \nonumber\\
	\dot{\tilde{a}}_{M}&=&-(\kappa _{M}+i\tilde{\delta}_{M})\tilde{a}_{M}-i\sum_{j=1}^{N}\tilde{g}_{Mj}\tilde{a}_{M}(\tilde{b}_{j}^{\dagger }+\tilde{b}_{j})-i\xi^{\ast}
	_{1M}\tilde{a}_{1}-i\xi^{\ast} _{2M}\tilde{a}_{2}-...-i\xi^{\ast} _{M-1,M}\tilde{a}_{M-1}-i\Lambda_{M}+\sqrt{2\kappa _{M}}\tilde{a}_{M\text{,in}}, \nonumber\\
	\dot{\tilde{b}}_{1}&=&-(\gamma_{1}+i\omega_{1})\tilde{b}_{1}-i\sum_{k=1}^{M}\tilde{g}_{k1}\tilde{a}_{k}^{\dagger }\tilde{a}_{k}-i\eta _{12}\tilde{b}_{2}-i\eta_{13}\tilde{b}_{3}-...-i\eta _{1N}\tilde{b}_{N}+\sqrt{2\gamma _{1}}\tilde{b}_{1,\text{in}}, \nonumber\\
	\dot{\tilde{b}}_{2} &=&-(\gamma _{2}+i\omega_{2})\tilde{b}_{2}-i\sum_{k=1}^{M}\tilde{g}_{k2}\tilde{a}_{k}^{\dagger }\tilde{a}_{k}-i\eta^{\ast} _{12}\tilde{b}_{1}-i\eta_{23}\tilde{b}_{3}-...-i\eta _{2N}\tilde{b}_{N}+\sqrt{2\gamma _{2}}\tilde{b}_{2,\text{in}}, \nonumber\\
	&&\vdots  \nonumber\\
	\dot{\tilde{b}}_{N} &=&-(\gamma _{N}+i\omega_{N})\tilde{b}_{N}-i\sum_{k=1}^{M}\tilde{g}_{kN}\tilde{a}_{k}^{\dagger }\tilde{a}_{k}-i\eta^{\ast} _{1N}\tilde{b}_{1}-i\eta^{\ast}_{2N}\tilde{b}_{2}-...-i\eta^{\ast} _{N-1,N}\tilde{b}_{N-1}+\sqrt{2\gamma _{N}}\tilde{b}_{N,\text{in}},
\end{eqnarray}
where $\kappa_{k=1\text{-}M}$ and $\gamma_{j=1\text{-}N}$ denote the decay rates of the $k$th optical mode and the $j$th mechanical mode, respectively. The operators $\tilde{a}_{k,\textrm{in}}$ and $\tilde{b}_{j,\textrm{in}}$ are, respectively, the noise operators associated to the $k$th optical mode and the $j$th mechanical mode. These noise operators are characterized by the zero average values and nonzero correlation functions~\cite{Gardiner2000S}
\begin{eqnarray}
	\langle \tilde{a}_{k,\textrm{in}}(t) \tilde{a}_{k,\textrm{in}}^{\dagger}(t^{\prime})\rangle&=&\delta(t-t^{\prime}), \nonumber \\
	\langle \tilde{b}_{j,\textrm{in}}(t) \tilde{b}_{j,\textrm{in}}^{\dagger}(t^{\prime})\rangle&=&(\bar{n}_{j}+1)\delta(t-t^{\prime}), \nonumber \\
	\langle \tilde{b}_{j,\textrm{in}}^{\dagger}(t)\tilde{b}_{j,\textrm{in}}(t^{\prime})\rangle&=&\bar{n}_{j}\delta(t-t^{\prime}),
\end{eqnarray}
where $\bar{n}_{j=1\text{-}N}$ denotes the mean thermal phonon number associated to the heat bath of the $j$th mechanical mode.

In the strong-driving case, both the optical modes and mechanical modes will be largely populated, then we can express these system operators as a sum of average values and fluctuation operators. We focus on the steady-state properties of the network and consider the strong-driving case, then we can linearize the optomechanical interactions. In this case, the dynamics of this network is governed by the linearized optomechanical Hamiltonian. To perform the linearization, we expand the operators as a sum of their mean values and fluctuation operators
\begin{equation}
\tilde{o}=\langle \tilde{o}\rangle+o, \hspace{0.5cm}\text{for}\hspace{0.2cm}o\in\{a_{k}, a_{k}^{\dagger}, b_{j}, b^{\dagger}_{j}\}.\label{oooo}
\end{equation}
By separating the equations of motion for the average values and fluctuation operators, we can obtain the equations of motion for the average values of operators as
\begin{eqnarray}
	\frac{d\langle\tilde{a}_{1}\rangle}{dt} & = & -(\kappa_{1}+i\delta_{1})\langle\tilde{a}_{1}\rangle-i\xi_{12}\left\langle \tilde{a}_{2}\right\rangle -i\xi_{13}\left\langle \tilde{a}_{3}\right\rangle -...-i\xi_{1M}\left\langle \tilde{a}_{M}\right\rangle -i\Lambda_{1},\nonumber \\
	\frac{d\langle\tilde{a}_{2}\rangle}{dt} & = & -(\kappa_{2}+i\delta_{2})\langle\tilde{a}_{2}\rangle-i\xi_{12}^{\ast}\left\langle \tilde{a}_{1}\right\rangle -i\xi_{23}\left\langle \tilde{a}_{3}\right\rangle -...-i\xi_{2M}\left\langle \tilde{a}_{M}\right\rangle -i\Lambda_{2},\nonumber \\
	& \vdots\nonumber \\
	\frac{d\langle\tilde{a}_{M}\rangle}{dt} & = & -(\kappa_{M}+i\delta_{M})\langle\tilde{a}_{M}\rangle-i\xi_{1M}^{\ast}\left\langle \tilde{a}_{1}\right\rangle -i\xi_{2M}^{\ast}\left\langle \tilde{a}_{2}\right\rangle -...-i\xi_{M-1,M}^{\ast}\left\langle \tilde{a}_{M-1}\right\rangle -i\Lambda_{M},\nonumber \\
	\frac{d\langle\tilde{b}_{1}\rangle}{dt} & = & -(\gamma_{1}+i\omega_{1})\langle\tilde{b}_{1}\rangle-i\sum_{k=1}^{M}\tilde{g}_{k1}|\langle\tilde{a}_{k}\rangle|^{2}-i\eta_{12}\langle\tilde{b}_{2}\rangle-i\eta_{13}\langle\tilde{b}_{3}\rangle-...-i\eta_{1N}\langle\tilde{b}_{N}\rangle,\nonumber \\
	\frac{d\langle\tilde{b}_{2}\rangle}{dt} & = & -(\gamma_{2}+i\omega_{2})\langle\tilde{b}_{2}\rangle-i\sum_{k=1}^{M}\tilde{g}_{k2}|\langle\tilde{a}_{k}\rangle|^{2}-i\eta_{12}^{\ast}\langle\tilde{b}_{1}\rangle-i\eta_{23}\langle\tilde{b}_{3}\rangle-...-i\eta_{2N}\langle\tilde{b}_{N}\rangle,\nonumber \\
	& \vdots\nonumber \\
	\frac{d\langle\tilde{b}_{N}\rangle}{dt} & = & -(\gamma_{N}+i\omega_{N})\langle\tilde{b}_{N}\rangle-i\sum_{k=1}^{M}\tilde{g}_{kN}|\langle\tilde{a}_{k}\rangle|^{2}-i\eta_{1N}^{\ast}\langle\tilde{b}_{1}\rangle-i\eta_{2N}^{\ast}\langle\tilde{b}_{2}\rangle-...-i\eta_{N-1,N}^{\ast}\langle\tilde{b}_{N-1}\rangle,\label{average}
\end{eqnarray}
where $\delta _{k}=\tilde{\delta}_{k}+\sum_{j=1}^{N}\tilde{g}_{kj}(\langle \tilde{b}_{j}\rangle _{\text{ss}}+\langle \tilde{b}_{j}\rangle _{\text{ss}}^{\ast })$ is the detuning modified by the optomechanical couplings.
Since we consider the linearization around the steady state of the network, then we set the left-hand  side of Eq.~(\ref{average}) to be zero, and obtain the equations determining these steady-state mean values as
\begin{eqnarray}
	\langle \tilde{a}_{1}\rangle _{\text{ss}} &=&\frac{-i\Lambda _{1}-i\xi_{12}\left\langle \tilde{a}_{2}\right\rangle _{\text{ss}}-i\xi _{13}\left\langle \tilde{a}_{3}\right\rangle _{\text{ss}}-...-i\xi _{1M}\left\langle \tilde{a}_{M}\right\rangle _{\text{ss}}}{\kappa _{1}+i\delta _{1}}, \notag \\
	\langle \tilde{a}_{2}\rangle _{\text{ss}} &=&\frac{-i\Lambda _{2}-i\xi^{\ast}_{12}\left\langle \tilde{a}_{1}\right\rangle _{\text{ss}}-i\xi _{23}\left\langle \tilde{a}_{3}\right\rangle _{\text{ss}}-...-i\xi _{2M}\left\langle \tilde{a}_{M}\right\rangle _{\text{ss}}}{\kappa _{2}+i\delta _{2}},\notag \\
	&&\vdots   \notag \\
	\langle \tilde{a}_{M}\rangle _{\text{ss}} &=&\frac{-i\Lambda _{M}-i\xi^{\ast}_{1M}\left\langle \tilde{a}_{1}\right\rangle _{\text{ss}}-i\xi^{\ast} _{2M}\left\langle \tilde{a}_{2}\right\rangle _{\text{ss}}-...-i\xi^{\ast} _{M-1,M}\left\langle \tilde{a}_{M-1}\right\rangle _{\text{ss}}}{\kappa _{M}+i\delta _{M}}, \notag \\
	\langle \tilde{b}_{1}\rangle _{\text{ss}} &=&\frac{-i\sum_{k=1}^{M}\tilde{g}_{k1}|\langle \tilde{a}_{k}\rangle _{\text{ss}}|^{2}-i\eta _{12}\langle \tilde{b}_{2}\rangle _{\text{ss}}-i\eta _{13}\langle \tilde{b}_{3}\rangle _{\text{ss}}-...-i\eta _{1N}\langle \tilde{b}_{N}\rangle _{\text{ss}}}{\gamma _{1}+i\omega _{1}},  \notag \\
	\langle \tilde{b}_{2}\rangle _{\text{ss}} &=&\frac{-i\sum_{k=1}^{M}\tilde{g}_{k2}|\langle \tilde{a}_{k}\rangle _{\text{ss}}|^{2}-i\eta^{\ast} _{12}\langle \tilde{b}_{1}\rangle _{\text{ss}}-i\eta _{23}\langle \tilde{b}_{3}\rangle _{\text{ss}}-...-i\eta _{2N}\langle \tilde{b}_{N}\rangle _{\text{ss}}}{\gamma _{2}+i\omega _{2}},  \notag \\
	&&\vdots   \notag \\
	\langle \tilde{b}_{N}\rangle _{\text{ss}} &=&\frac{-i\sum_{k=1}^{M}\tilde{g}_{kN}|\langle \tilde{a}_{k}\rangle _{\text{ss}}|^{2}-i\eta^{\ast} _{1N}\langle \tilde{b}_{1}\rangle _{\text{ss}}-i\eta^{\ast} _{2N}\langle \tilde{b}_{2}\rangle _{\text{ss}}-...-i\eta^{\ast} _{N-1,N}\langle \tilde{b}_{N-1}\rangle _{\text{ss}}}{\gamma _{N}+i\omega _{N}}.
	\label{steadyMNS}
\end{eqnarray}
Around the steady state, we can obtain the equations of motion for the fluctuation operators. In particular, we consider the case where the optomechanical network works in the linearization regime. Then we can perform the linearization by discarding all the second-order fluctuation terms, and obtain the linearized Langevin equations as
\begin{eqnarray}
	\label{linearlanMMS}
	\dot{a}_{1} &=&-(\kappa _{1}+i\delta _{1}) a_{1}-i\sum_{j=1}^{N}g_{1j}( b_{j}^{\dagger }+ b_{j})-i\xi
	_{12} a_{2}-i\xi _{13} a_{3}-...-i\xi _{1M} a_{M}+\sqrt{2\kappa _{1}}\tilde{a}_{\text{1,in}}, \notag \\
	\dot{a}_{2} &=&-(\kappa _{2}+i\delta _{2}) a_{2}-i\sum_{j=1}^{N}g_{2j}( b_{j}^{\dagger }+ b_{j})-i\xi^{\ast}
	_{12}a_{1}-i\xi _{23} a_{3}-...-i\xi _{2M} a_{M}+\sqrt{2\kappa _{2}}\tilde{a}_{2\text{,in}}, \notag \\
	&&\vdots  \notag \\
	\dot{a}_{M} &=&-(\kappa _{M}+i\delta _{M}) a_{M}-i\sum_{j=1}^{N}g_{Mj}( b_{j}^{\dagger}+ b_{j})-i\xi^{\ast}
	_{1M} a_{1}-i\xi^{\ast} _{2M} a_{2}-...-i\xi^{\ast} _{M-1,M} a_{M-1}+\sqrt{2\kappa _{M}}\tilde{a}_{M\text{,in}}, \notag \\
	\dot{b}_{1} &=&-(\gamma _{1}+i\omega _{1}) b_{1}-i\sum_{k=1}^{M}(g^{\ast}_{k1} a_{k}+g_{k1} a_{k}^{\dagger }) -i\eta _{12} b_{2}-i\eta _{13} b_{3}-\ldots -i\eta _{1N} b_{N}+\sqrt{2\gamma _{1}}\tilde{b}_{1,\text{in}}, \notag \\
	\dot{b}_{2} &=&-(\gamma _{2}+i\omega _{2})b_{2}-i\sum_{k=1}^{M}(g^{\ast}_{k2}  a_{k}+g_{k2}
	a_{k}^{\dagger}) -i\eta^{\ast}_{12} b_{1}-i\eta_{23}b_{3}-\ldots -i\eta _{2N} b_{N}+\sqrt{2\gamma _{2}}\tilde{b}_{2,\text{in}}, \notag \\
	&&\vdots  \notag \\
	\dot{b}_{N} &=&-(\gamma _{N}+i\omega _{N}) b_{N}-i\sum_{k=1}^{M} (g^{\ast}_{kN} a_{k}+ g_{kN}a_{k}^{\dagger }) -i\eta^{\ast} _{1N} b_{1}-i\eta^{\ast} _{2N} b_{2}-\ldots -i\eta^{\ast} _{N-1,N} b_{N-1}+\sqrt{2\gamma _{N}}\tilde{b}_{N,\text{in}},
\end{eqnarray}
where the parameter $g_{kj}=\tilde{g}_{kj}\langle \tilde{a}_{k}\rangle _{\text{ss}}$ is the linearized optomechanical coupling strength.

Based on Eq.~(\ref{linearlanMMS}) and ignoring the damping and noise terms, we can derive an approximate linearized Hamiltonian for this network as
\begin{equation}
	H^{[M,N]}_{\text{lin}}=\sum_{k=1}^{M}\delta_{k}a_{k}^{\dagger}a_{k}+\sum_{j=1}^{N}\omega_{j}b_{j}^{\dagger}b_{j}+\sum_{\textcolor{black}{k,k^{\prime}=1,k<k^{\prime}}}^{M}(\xi_{kk^{\prime}}a_{k}^{\dagger}a_{k^{\prime}}+\xi_{kk^{\prime}}^{\ast}a_{k^{\prime}}^{\dagger}a_{k})+\sum_{\textcolor{black}{j,j^{\prime}=1,j<j^{\prime}}}^{N}(\eta_{jj^{\prime}}b_{j}^{\dagger}b_{j^{\prime}}+\eta_{jj^{\prime}}^{\ast}b_{j^{\prime}}^{\dagger}b_{j})+\sum_{k=1}^{M}\sum_{j=1}^{N}(g_{kj}a_{k}^{\dagger}+g^{\ast}_{kj}a_{k})(b_{j}^{\dagger}+b_{j}),~\label{HMNLIN}
\end{equation}
where the operators and variables have been defined in Eqs.~(\ref{HamiltMNS}) and (\ref{oooo}).
We consider the red-sideband-resonance dominating regime and the weak-coupling case, then we can perform the rotating-wave approximation (RWA) to obtain the approximate Hamiltonian as
\begin{equation}
	H_{\text{RWA}}^{[M,N]}=\sum_{k=1}^{M}\delta _{k}a_{k}^{\dagger}a_{k}+\sum_{j=1}^{N}\omega _{j} b_{j}^{\dagger
	} b_{j}+\sum_{\textcolor{black}{k,k^{\prime}=1,k<k^{\prime}}}^{M}(\xi_{kk^{\prime}}a_{k}^{\dagger}a_{k^{\prime}}+\xi_{kk^{\prime}}^{\ast}a_{k^{\prime}}^{\dagger}a_{k})+\sum_{\textcolor{black}{j,j^{\prime}=1,j<j^{\prime}}}^{N}(\eta_{jj^{\prime}}b_{j}^{\dagger}b_{j^{\prime}}+\eta_{jj^{\prime}}^{\ast}b_{j^{\prime}}^{\dagger}b_{j})+\sum_{k=1}^{M}\sum_{j=1}^{N}(g_{kj} a_{k}^{\dagger}b_{j}+g^{\ast}_{kj}b_{j}^{\dagger} a_{k} ).~\label{MNRWA}
\end{equation}
 The Hamiltonian (\ref{MNRWA}) is the starting point for our study on the dark-mode theorems in the main text.

\section{Cooling of multiple mechanical modes in the ($1+N$)-mode optomechanical networks including one optical mode and $N$ mechanical modes\label{CooloneN}}
In optomechanical networks, the cooling performance of these mechanical modes is significantly determined by the mechanical dark modes.  In the presence of dark modes, the mechanical modes involved into the dark normal modes cannot be cooled into their ground states. If the simultaneous ground-state cooling of all these mechanical modes can be realized, then there are no dark modes. Therefore,  we can examine the dark-mode effect by evaluating the simultaneous ground-state cooling of these mechanical modes. To examine the validity of  the dark-mode theorems, in this section, we evaluate the simultaneous ground-state cooling of multiple mechanical modes in the ($1+N$)-mode optomechanical network. Below, we first present the detailed calculations concerning the final mean phonon numbers in these mechanical modes, and then we study the simultaneous ground-state cooling of mechanical modes in the ($1+N$)-mode optomechanical networks for $N=2$, 3, and 4.

\subsection{Final mean phonon numbers}
To study quantum cooling of these mechanical modes in optomechanical networks, we introduce both the vector of fluctuation operators
\begin{eqnarray}
	\mathbf{u}(t)=(a_{1},a_{2},\cdots,a_{M},b_{1},b_{2},\cdots,b_{N},a^{\dagger}_{1}, a^{\dagger}_{2},\cdots, a^{\dagger}_{M},b^{\dagger}_{1}, b^{\dagger}_{2},\cdots, b^{\dagger}_{N})^{T},
\end{eqnarray}
and the vector of noise operators
\begin{eqnarray}
	\mathbf{N}(t)&=&\sqrt{2}(\sqrt{\kappa_{1}}\tilde{a}_{1,\text{in}},\sqrt{\kappa_{2}}\tilde{a}_{2,\text{in}},\cdots,\sqrt{\kappa_{M}}\tilde{a}_{M,\text{in}},\sqrt{\gamma_{1}}\tilde{b}_{1,\text{in}},\sqrt{\gamma_{2}}\tilde{b}_{2,\text{in}},\cdots,\sqrt{\gamma_{N}}\tilde{b}_{N,\text{in}},\sqrt{\kappa_{1}}\tilde{a}^{\dagger}_{1,\text{in}}, \nonumber\\ &&\sqrt{\kappa_{2}}\tilde{a}^{\dagger}_{2,\text{in}},\cdots,\sqrt{\kappa_{M}}\tilde{a}^{\dagger}_{M,\text{in}},\sqrt{\gamma_{1}}\tilde{b}^{\dagger}_{1,\text{in}},\sqrt{\gamma_{2}}\tilde{b}^{\dagger}_{2,\text{in}},\cdots, \sqrt{\gamma_{N}}\tilde{b}^{\dagger}_{N,\text{in}})^{T}.
\end{eqnarray}
Then the linearized Langevin equations~(\ref{linearlanMMS}) can be written as a compact form
\begin{eqnarray}
	\mathbf{\dot{u}}(t)=\mathbf{A}\mathbf{u}(t)+\mathbf{N}(t),\label{compactMNS}
\end{eqnarray}
where we introduce the drift matrix
$\mathbf{A}=\left(
\begin{array}{cc}
-\mathbf{E} & -\mathbf{F} \\
-\mathbf{F}^{\ast} & -\mathbf{E}^{\ast}
\end{array}
\right)$,
with
\begin{equation}
	\mathbf{E}=\left(
	\begin{array}{cccccccc}
		i\delta_{1}+\kappa_{1} & i\xi _{12} & \cdots  & i\xi _{1M} & ig_{11} & ig_{12} & \cdots  & ig_{1N} \\
		i\xi^{\ast} _{12} & i\delta_{2}+\kappa_{2} & \cdots  & i\xi _{2M} & ig_{21} & ig_{22} & \cdots  & ig_{2N} \\
		\vdots  & \vdots  & \ddots  & \vdots  & \vdots  & \vdots  & \ddots  & \vdots \\
		i\xi^{\ast} _{1M} & i\xi^{\ast} _{2M} & \cdots  & i\delta_{M}+\kappa_{M} & ig_{M1} & ig_{M2} & \cdots  & ig_{MN} \\
		ig^{\ast}_{11} & ig^{\ast}_{21} & \cdots  & ig^{\ast}_{M1} &  i\omega_{1}+\gamma_{1} & i\eta _{12} & \cdots  & i\eta _{1N} \\
		ig^{\ast}_{12} & ig^{\ast}_{22} & \cdots  & ig^{\ast}_{M2} & i\eta^{\ast} _{12} &  i\omega_{2}+\gamma_{2} & \cdots  & i\eta _{2N} \\
		\vdots  & \vdots  & \ddots  & \vdots & \vdots  & \vdots & \ddots  & \vdots  \\
		ig^{\ast}_{1N} & ig^{\ast}_{2N} & \cdots  & ig^{\ast}_{MN} & i\eta^{\ast} _{1N} & i\eta^{\ast} _{2N} & \cdots  &  i\omega_{N}+\omega _{N}
	\end{array}\right)
\end{equation}
and
\begin{equation}
	\mathbf{F}=\left(
	\begin{array}{cccccccc}
		0 & 0 & \cdots  &0 & ig_{11} & ig_{12} & \cdots  & ig_{1N} \\
		0 &0& \cdots  & 0 & ig_{21} & ig_{22} & \cdots  & ig_{2N} \\
		\vdots  & \vdots  & \ddots  & \vdots  & \vdots  & \vdots  & \ddots  & \vdots \\
		0 & 0 & \cdots  &0 & ig_{M1} & ig_{M2} & \cdots  & ig_{MN} \\
		ig^{\ast}_{11} & ig^{\ast}_{21} & \cdots  & ig^{\ast}_{M1} & 0 & 0 & \cdots  &0 \\
		ig^{\ast}_{12} & ig^{\ast}_{22} & \cdots  & ig^{\ast}_{M2} & 0 & 0 & \cdots  & 0 \\
		\vdots  & \vdots  & \ddots  & \vdots & \vdots  & \vdots & \ddots  & \vdots  \\
		ig^{\ast}_{1N} & ig^{\ast}_{2N} & \cdots  & ig^{\ast}_{MN} & 0 & 0 & \cdots  &  0
	\end{array}\right).
\end{equation}
The solution of this system is stable when all the eigenvalues of the matrix $\mathbf{A}$ have negative real parts. This stability conditions can be analyzed by using the Routh-Hurwitz criterion~\cite{Gradshteyn2014S}. In all simulations in this work, the parameters used satisfy the stability conditions.

The final mean phonon numbers of these mechanical modes $b_{j=1\text{-}N}$ can be obtained by solving the steady state of the network. For this purpose, we introduce the covariance matrix $\mathbf{V}$ defined  by the elements
\begin{equation}
	\mathbf{V}_{ij}=\frac{1}{2}[\langle \mathbf{u}_{i}(\infty) \mathbf{u}_{j}(\infty ) \rangle +\langle \mathbf{u}_{j}( \infty) \mathbf{u}_{i}(\infty )\rangle], \hspace{0.3 cm}i,j=1\text{-}2(M\!+\!N). \label{covarianceMNS}
\end{equation}
The covariance matrix $\mathbf{V}$ satisfies the Lyapunov equation~\cite{Vitali2007S}
\begin{equation}
	\mathbf{A}\mathbf{V}+\mathbf{V}\mathbf{A}^{T}=-\mathbf{Q}, \label{LyapunovMNS}
\end{equation}
where the diffusion matrix is introduced as
$\mathbf{Q}=\left(
\begin{array}{cc}
\mathbf{0} & \mathbf{P} \\
\mathbf{P} & \mathbf{0}
\end{array}
\right)$,
with
\begin{equation}
	\mathbf{P}=\left(
	\begin{array}{cccccccc}
		\kappa _{1} & 0 & \cdots  & 0 & 0 & 0 & \cdots  & 0 \\
		0 & \kappa _{2} & \cdots  & 0 & 0 & 0 & \cdots  & 0 \\
		\vdots  & \vdots  & \ddots  & \vdots  & \vdots  & \vdots  & \ddots  & \vdots\\
		0 & 0 & \cdots  & \kappa _{M} & 0 & 0 & \cdots  & 0 \\
		0 & 0 & \cdots  & 0 & \gamma _{1}(2\bar{n}_{1}+1) & 0 & \cdots  & 0 \\
		0 & 0 & \cdots  & 0 & 0 & \gamma _{2}(2\bar{n}_{2}+1) & \cdots  & 0 \\
		\vdots  & \vdots  & \ddots  & \vdots  & \vdots  & \vdots  & \ddots  & \vdots\\
		0 & 0 & \cdots  & 0 & 0 & 0 & \cdots  & \gamma _{N}(2\bar{n}_{N}+1)
	\end{array}
	\right) .
\end{equation}
Based on the Lyapunov equation and the diffusion matrix, we can obtain the expression of the covariance matrix $\mathbf{V}$, namely, all the matrix elements defined in Eq.~(\ref{covarianceMNS}), then the final mean phonon numbers in the $j$th mechanical mode can be obtained as
\begin{equation}
	n^{f}_{j}=\langle  b_{j}^{\dagger} b_{j}\rangle=\mathbf{V}_{2M+N+j,M+j}-\frac{1}{2},\label{phononNS}
\end{equation}
for $j=1\text{-}N$, where $\mathbf{V}_{2M+N+j,M+j}$ is the matrix element defined in Eq.~(\ref{covarianceMNS}).

For a given network, we obtain the final mean phonon numbers in these mechanical modes by calculating the covariance matrix. Under the red-sideband resonance and in the resolved-sideband regime, the simultaneous ground-state cooling of all these mechanical modes will be realized when there are no dark modes. In the presence of the dark modes, the thermal noise excitations stored in these dark modes cannot be extracted through the optomechanical-cooling channel associated with the optical mode,  then the simultaneous ground-state cooling of these mechanical modes cannot be realized. Therefore, by evaluating the simultaneous ground-state cooling performance of all these mechanical modes, we can examine the dark-mode theorems in the ($1+N$)-mode optomechanical networks.

\subsection{One-optical-mode and two-mechanical-mode optomechanical networks  \label{modeltwo}}

\begin{figure}[tbp]
\center
\includegraphics[width=0.6 \textwidth]{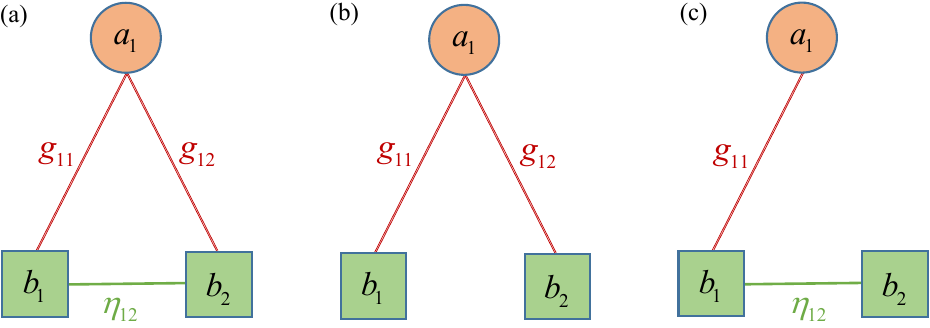}
\caption{Three coupling configurations of the one-optical-mode and two-mechanical-mode optomechanical networks. Panel (a) represents the case where all couplings are connected. Panels (b) and (c) describe the cases where the couplings $\eta_{12}$ and $g_{12}$ are disconnected, respectively.}
\label{FigS2}
\end{figure}

To verify the validity of Theorem 1 in the main text, we first consider the one-optical-mode and two-mechanical-mode optomechanical network [Fig.~\ref{FigS2}(a)]. This network is a special case ($N=2$) of the ($1+N$)-mode optomechanical networks. In this case, the linearized Hamiltonian of this ($1+2$)-mode optomechanical network reads

\begin{equation}
	H_{\text{lin}}^{[1,2]}=\delta_{1}a_{1}^{\dagger}a_{1}+\sum_{j=1}^{2}\omega_{j}b_{j}^{\dagger}b_{j}+(\eta_{12}b_{1}^{\dagger}b_{2}+\eta_{12}^{\ast}b_{2}^{\dagger}b_{1})+\sum_{j=1}^{2}(g_{1j}a_{1}^{\dagger}+g_{1j}^{\ast}a_{1})(b_{j}^{\dagger}+b_{j}),~\label{H12LIN}
\end{equation}
where the phonon-hopping coupling strength between the two mechanical modes $b_{1}$ and $b_{2}$ is denoted by $\eta_{12}$, other parameters and operators have been defined in Eq.~(\ref{HMNLIN}). Hereafter, we add the superscript ``[1,2]" to denote the involved mode number in this network.

For this ($1+2$)-mode optomechanical network, the linearized Langevin equations are given by
\begin{equation}
\mathbf{\dot{u}}^{[1,2]}(t)=\mathbf{A}^{[1,2]}\mathbf{u}^{[1,2]}(t)+\mathbf{N}^{[1,2]}(t).~\label{dotu12}
\end{equation}
 Here, the vectors of fluctuation operators and noise operators are, respectively, defined by
 \begin{equation}
 \mathbf{u}^{[1,2]}(t)=( a_{1}, b_{1},b_{2}, a_{1}^{\dagger}, b^{\dagger}_{1}, b^{\dagger}_{2})^{T}
 \end{equation}
 and
 \begin{equation}
 \mathbf{N}^{[1,2]}(t)=\sqrt{2}(\sqrt{\kappa_{1}}\tilde{a}_{\text{1,\text{in}}},\sqrt{\gamma_{1}}\tilde{b}_{1,\text{in}},\sqrt{\gamma_{2}}\tilde{b}_{2,\text{in}},\sqrt{\kappa_{1}}\tilde{a}^{\dagger}_{1,\text{in}}, \sqrt{\gamma_{1}}\tilde{b}^{\dagger}_{1,\text{in}},\sqrt{\gamma_{2}}\tilde{b}^{\dagger}_{2,\text{in}})^{T}.
  \end{equation}
The drift matrix $\mathbf{A}^{[1,2]}$ in Eq. (\ref{dotu12}) is given by
\begin{equation}
\label{driftN}
\mathbf{A}^{[1,2]}=-\left(
\begin{array}{cccccc}
i\delta_{1}+\kappa_{1} & ig_{11} & ig_{12} & 0 & ig_{11} & ig_{12}\\
ig^{\ast}_{11} & i\omega_{1}+\gamma_{1} & i\eta_{12} & ig_{11} & 0 & 0\\
ig^{\ast}_{12} & i\eta^{\ast}_{12} & i\omega_{2}+\gamma_{2} & ig_{12} & 0 & 0\\
0 & -ig^{\ast}_{11} & -ig^{\ast}_{12} & \kappa_{1}-i\delta_{1} & -ig^{\ast}_{11} & -ig^{\ast}_{12} \\
-ig^{\ast}_{11} & 0 & 0 &  -ig_{11} & \gamma_{1}-i\omega_{1} & -i\eta^{\ast}_{12}\\
-ig^{\ast}_{12} & 0 & 0 &  -ig_{12} & -i\eta_{12} & \gamma_{2}-i\omega_{2}
\end{array}
\right).
\end{equation}
Based on the covariance matrix~(\ref{covarianceMNS}) and the Lyapunov equation~(\ref{LyapunovMNS}), we can obtain the final mean phonon numbers in the two mechanical modes.

In practice, the optomechanical and phonon-hopping couplings are crucial factors for realization of simultaneous ground-state cooling in this (1+2)-mode optomechanical network, so it is an important topic to analyze the influence of these couplings on ground-state cooling of two mechanical modes. To be consistent with the case of  $N$-mechanical-mode optomechanical network, we consider the case where the two mechanical modes have the same resonance frequencies, i.e.,  $\omega_{1}=\omega_{2}=\omega_{m}$. To analyze the ground-state cooling and dark-mode effect in this network, we study several cases of coupling configurations by controlling the three couplings $g_{11}$, $g_{12}$, and $\eta_{12}$. Note that the two mechanical modes must be either directly or indirectly coupled to the optical mode, thus we can only cut off at most one coupling, then there are three different coupling configurations, as shown in Fig.~\ref{FigS2}. In panel (a), none of the couplings are cut off. In panels (b) and (c), the couplings $\eta_{12}$ and $g_{12}$ are cut off, respectively.

\begin{figure}[tbp]
\center
\includegraphics[width=0.7 \textwidth]{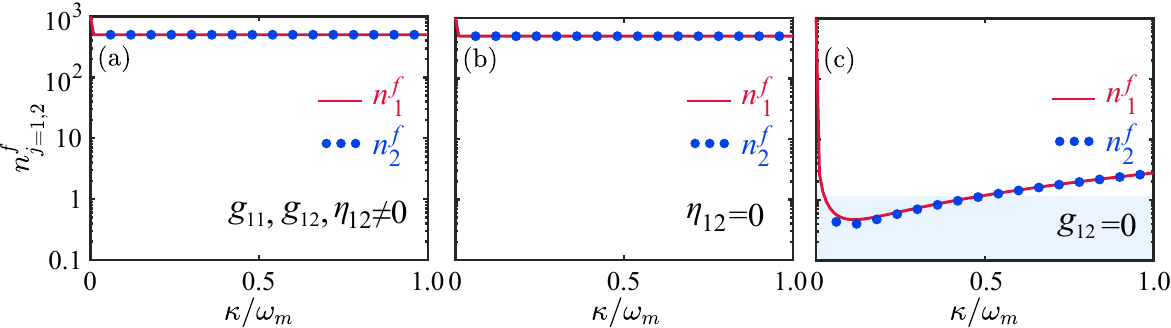}
\caption{Final mean phonon numbers $n^{f}_{1}$ and $n^{f}_{2}$ versus the scaled decay rate $\kappa/\omega_{m}$ for three different coupling configurations depicted in Fig.~\ref{FigS2}. The parameters used are given by  $\gamma_{1}/\omega_{m}=\gamma_{2}/\omega_{m}=10^{-5}$, $\delta_{1}/\omega_{m}=1$, $g_{11}/\omega_{m}=g_{12}/\omega_{m}=0.1$, $\eta_{12}/\omega_{m}=0.09$, and $\bar{n}_{1}=\bar{n}_{2}=1000$ when all couplings are connected (a). Note that the couplings $\eta_{12}$ and $g_{12}$ are, respectively, disconnected when (b) $\eta_{12}=0$ and (c) $g_{12}=0$. The optomechanical-coupling strength after the linearization is denoted by $g_{12}=\tilde{g}_{12}\langle \tilde{a}_{1}\rangle_{\text{ss}}$.}
\label{FigS3}
\end{figure}

Corresponding to the three cases illustrated in Fig.~\ref{FigS2}, we plot the final mean phonon numbers $n^{f}_{1}$ and $n^{f}_{2}$ as functions of the decay rate $\kappa/\omega_{m}$, as shown in Fig.~\ref{FigS3}. Figures~\ref{FigS3}(a) and~\ref{FigS3}(b) show that the two mechanical modes cannot be cooled into their ground states when \{$g_{11}$, $g_{12}$, $\eta_{12}\!\neq\!0$\} or \{$g_{11}$, $g_{12}\!\neq\!0$, $\eta_{12}\!=\!0$\}, which imply that mechanical dark modes appear in the coupling configurations depicted in Figs.~\ref{FigS2}(a) and ~\ref{FigS2}(b). In Fig.~\ref{FigS3}(c) corresponding to \{$g_{11}$, $\eta_{12}\!\neq\!0$, $g_{12}\!=\!0$\}, the simultaneous ground-state cooling of the two mechanical modes can be realized ($n^{f}_{1}$ and $n^{f}_{2}<1$) in the resolved-sideband regime,  which means that the mechanical dark mode does not exist in the coupling configuration depicted in Fig.~\ref{FigS2}(c).

The parameter conditions for the appearance of mechanical dark modes in this loop-coupled optomechanical network can be derived by using arrowhead-matrix method. Based on Eq.~(\ref{H12LIN}) and ignoring the damping and noise terms, we can derive an approximate linearized Hamiltonian for this network as
\begin{eqnarray}
\label{Hamitrwa3}
H_{\text{RWA}}^{[1,2]}=\delta_1a_1^{\dagger} a_1+\omega_{m}\sum_{j=1}^{2}b_{j}^{\dagger} b_{j} +(\eta_{12}b_{1}^{\dagger} b_{2}+ \eta^{\ast}_{12}b_{2}^{\dagger}b_{1})+\sum_{j=1}^{2}(g_{1j} a_1^{\dagger}b_{j}+g_{1j}^{\ast}b_{j}^{\dagger}a_1).
\end{eqnarray}
The Hamiltonian~(\ref{Hamitrwa3}) can be further expressed as
\begin{equation}
H_{\text{RWA}}^{[1,2]}=( a_1^{\dagger}, b_{1}^{\dagger}, b_{2}^{\dagger})\textbf{H}_{ab}^{[1,2]}(a_1,b_{1},b_{2})^{T},~\label{HRWA121}
\end{equation}
where we introduce the coefficient matrix
\begin{eqnarray}
\label{matrixm2}
\textbf{H}^{[1,2]}_{ab}&=&\left(
\begin{array}{c|cc}
\delta_1 & g_{11} & g_{12} \\ \hline
g_{11}^{\ast} & \omega_{m} & \eta_{12} \\
g_{12}^{\ast} & \eta^{\ast}_{12} & \omega_{m}
\end{array}
\right).
\end{eqnarray}
To clearly clarify the parameter conditions for the existence of mechanical dark modes, we first diagonalize the two coupled mechanical modes by introducing the matrix $\textbf{U}_{b}$. Then in the normal-mode ($A_1,B_j$) representation, the Hamiltonian (\ref{HRWA121}) can be expressed as
\begin{equation}
	H_{\text{RWA}}^{[1,2]}=( A_1^{\dagger}, B_{1}^{\dagger}, B_{2}^{\dagger})\textbf{H}_{AB}^{[1,2]}(A_1,B_{1},B_{2})^{T},
\end{equation}
where the coefficient matrix in the normal-mode representation is expressed as
\begin{eqnarray}
	\label{matrixAB}
	\textbf{H}^{[1,2]}_{AB}&=&\left(
	\begin{array}{c|cc}
		\Delta_1 & G_{11} & G_{12} \\ \hline
		G^{\ast}_{11} & \Omega_1 & 0 \\
		G^{\ast}_{12} & 0 & \Omega_2
	\end{array}
	\right).\label{H12AB}
\end{eqnarray}
Here, $\Delta_1=\delta_1$ is the effective resonance frequency of the optical mode $a_1=A_1$,  $\Omega_j$ is the effective resonance frequency of the $j$th mechanical normal mode $B_{j=1,2}=\sum_{j^{\prime}}^{2}(\textbf{U}_{b})_{jj^{\prime}}b_{j^{\prime}}$, and $G_{1j}=\sum_{j^{\prime}=1}^{2}g_{1j^{\prime}}(\textbf{U}_{b}^{\dagger})_{j^{\prime}j}$.

Based on Eq. (\ref{H12AB}), we find that, when $G_{1j}=0$ for an arbitrary $\Omega_{j}$, then according to Theorem 1(i), the corresponding mode $B_j$ becomes a mechanical dark mode. In this case, the simultaneous ground-state cooling of the two mechanical modes is unfeasible. When $G_{1j}\neq 0$ and $\Omega_{1}=\Omega_{2}$, based on Theorem 1(ii), there are one mechanical bright mode $B_{2+}$ and one mechanical dark mode $B_{2-}$ defined by Eq. (\ref{b222}), then the simultaneous ground-state cooling of the two mechanical modes cannot be realized.
When $G_{1j}\neq 0$ and $\Omega_{1}\neq\Omega_{2}$, based on Theorem 1(iii), we find that there is no mechanical dark mode, which means that the simultaneous ground-state cooling of the two mechanical modes can be realized.

In this ($1+2$)-mode optomechanical network, there are three coupling configurations depicted in Fig.~\ref{FigS2}. Below, we assume that the optomechanical and phonon-hopping coupling strengths are real. We also analyze the influence of these couplings on ground-state cooling of two mechanical modes case by case.

(i) In the case of $\eta_{12}\neq0$ and $g_{11}\!=\!g_{12}\!=\!g\!\neq\!0$, the coefficient matrix (\ref{H12AB}) becomes
\begin{eqnarray}
	\label{matrixAB}
	\textbf{H}^{[1,2]}_{AB}&=&\left(
	\begin{array}{c|cc}
		\Delta_1 & 0 & \sqrt{2}g \\ \hline
		0& \omega_{m}-\eta & 0 \\
		 \sqrt{2}g & 0 &  \omega_{m}+\eta
	\end{array}
	\right).~\label{eq88}
\end{eqnarray}
We can see from Eq.~(\ref{eq88}) that the mechanical normal mode $B_1$ is decoupled from both the optical normal mode $A_1=a_1$ and the mechanical normal mode $B_2$. Therefore, the mode $B_1$ becomes a mechanical dark mode, which means that the simultaneous ground-state cooling of the two mechanical modes is unfeasible. This result is consistent with the numerical result depicted in Fig. \ref{FigS3}(a).

(ii) In the case of $\eta_{12}=0$ and $g_{11}=g_{12}=g\neq0$, the coefficient matrix (\ref{H12AB}) becomes
\begin{eqnarray}
	\label{matrixAB}
	\textbf{H}^{[1,2]}_{AB}&=&\left(
	\begin{array}{c|cc}
		\Delta_1 & g & g \\ \hline
		g& \omega_{m} & 0 \\
		g & 0 &  \omega_{m}
	\end{array}
	\right).~\label{eq89}
\end{eqnarray}
Equation (\ref{eq89}) indicates that the two degenerate normal modes $B_1$ and $B_2$ are coupled to the optical normal mode $A_1$. Based on Theorem 1(ii), we know that there is one mechanical
dark mode in this ($1+2$)-mode optomechanical network, and the simultaneous ground-state cooling of the two mechanical modes cannot be realized. This result agrees well with the numerical
result depicted in Fig. \ref{FigS3}(b).

(iii) In the case of $\eta_{12}\!\neq\!0$, $g_{11}\!=\!g\!\neq\!0$, and $ g_{12}=0$, the coefficient matrix (\ref{H12AB}) becomes
\begin{eqnarray}
	\label{matrixAB}
	\textbf{H}^{[1,2]}_{AB}&=&\left(
	\begin{array}{c|cc}
		\Delta_1 & -g/\sqrt{2} & g/\sqrt{2} \\ \hline
		-g/\sqrt{2}& \omega_{m}-\eta & 0 \\
		g/\sqrt{2} & 0 &  \omega_{m}+\eta
	\end{array}
	\right).~\label{eq82}
\end{eqnarray}
It can be seen from Eq. (\ref{eq82}) that the two mechanical normal modes $B_1$ and $B_2$ are coupled to the optical normal mode $A_1$, and the frequencies of the
two normal modes $B_1$ and $B_2$ are non-degenerate. Based on Theorem 1(iii), we know that there is no mechanical dark mode and hence the simultaneous ground-state cooling of the two mechanical modes can be realized in the resolved-sideband regime, which is consistent with the numerical result depicted in Fig. \ref{FigS3}(c).

\subsection{One-optical-mode and three-mechanical-mode optomechanical networks \label{modelthree}}

\begin{figure}[tbp]
\center
\includegraphics[width=0.75 \textwidth]{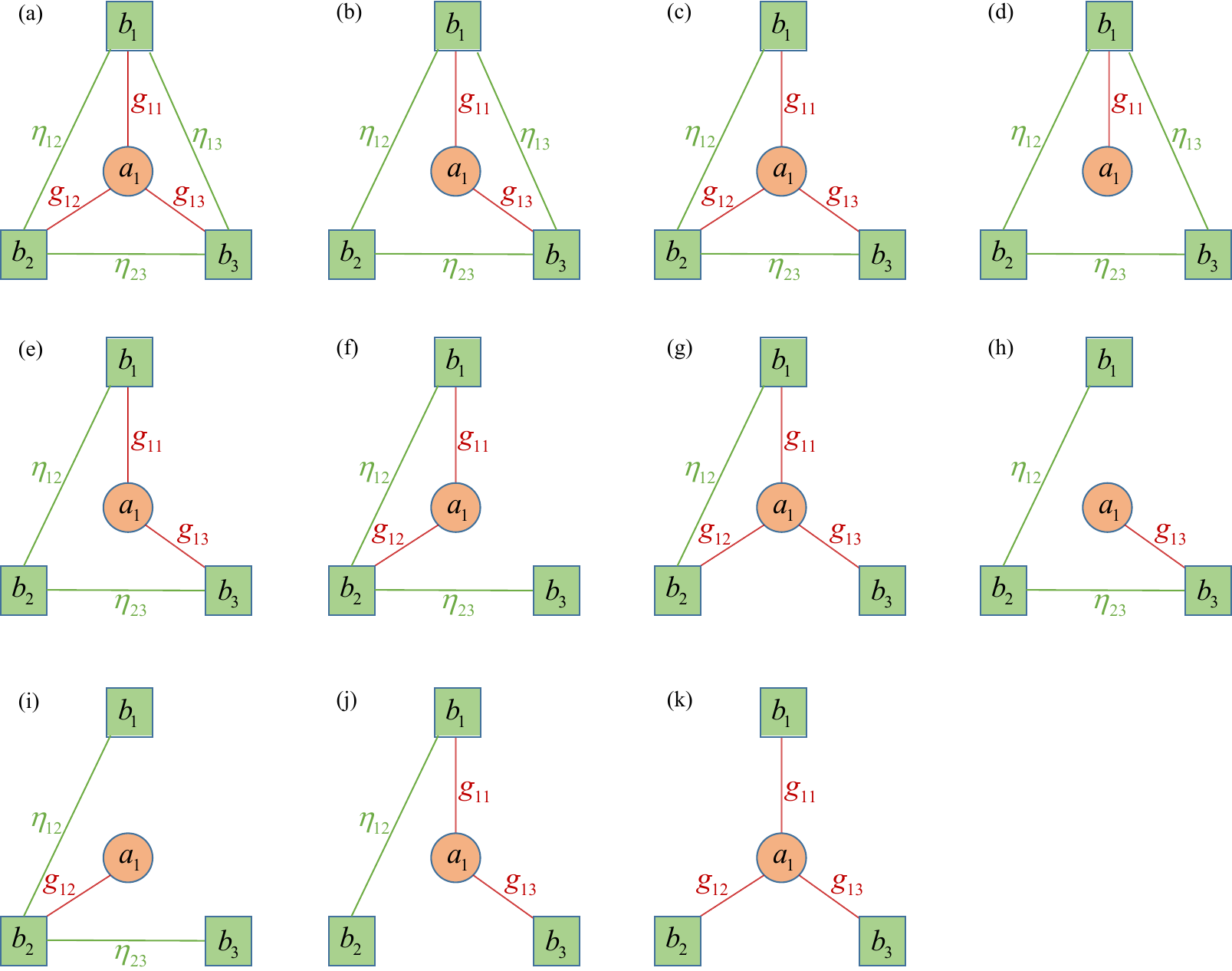}
\caption{Eleven coupling configurations of the one-optical-mode and three-mechanical-mode optomechanical networks. Here, the three identical mechanical modes are optomechanically coupled to the optical mode. To ensure that the four modes can be directly or indirectly coupled together, we keep at least three couplings in the network, and the other three couplings are switchable on demand. Panel (a) shows the case where all couplings are connected. Panels (b) and (c) represent these cases where one coupling is cut off. Panels (d)-(g) correspond to the two-coupling-disconnected cases. Panels (h)-(k) indicate these cases where three couplings are cut off.}
\label{FigS4}
\end{figure}

In this section, we examine the dark-mode theorem by evaluating the simultaneous ground-state cooling of mechanical modes in the one-optical-mode and three-mechanical-mode optomechanical networks. This system is a special case ($N=3$) of  the ($1 + N$)-mode optomechanical networks. For this optomechanical network, its  linearized optomechanical Hamiltonian reads
\begin{equation}
	H_{\text{lin}}^{[1,3]}=\delta_{1}a_{1}^{\dagger}a_{1}+\sum_{j=1}^{3}\omega_{j}b_{j}^{\dagger}b_{j}+\sum_{\textcolor{black}{j,j^{\prime}=1,j<j^{\prime}}}^{3}(\eta_{jj^{\prime}}b_{j}^{\dagger}b_{j^{\prime}}+\eta_{jj^{\prime}}^{\ast}b_{j^{\prime}}^{\dagger}b_{j})+\sum_{j=1}^{3}(g_{1j}a_{1}^{\dagger}+g_{1j}^{\ast}a_{1})(b_{j}^{\dagger}+b_{j}),~\label{H13LIN}
\end{equation}
where the notations are consistent with those defined in Eq.~(\ref{HMNLIN}).
Based on the linearized optomechanical Hamiltonian, we can obtain the covariance matrix and the final mean phonon numbers in the three mechanical modes.

To analyze the ground-state cooling and the dark-mode conditions in this four-mode optomechanical network, we consider different coupling configurations by controlling these six couplings $g_{1(j=1\text{-}3)}$, $\eta_{12}$, $\eta_{13}$, and $\eta_{23}$. For convenience, we consider the three-identical-mechanical-mode case, thus the mechanical modes have the same resonance frequencies, i.e., $\omega_{1}=\omega_{2}=\omega_{3}=\omega_{m}$.
In addition, we consider different coupling configurations of the optomechanical network. Namely, the couplings in the network could be turned on or cut off on demand. In particular, we assume that if the couplings are present, then these couplings have the same coupling strengths. To ensure that these three mechanical modes can be either directly or indirectly coupled to the optical mode, then there exist at least three couplings in the network. Thus, there are $11$ different coupling configurations. As shown in Fig.~\ref{FigS4}(a), none of the couplings is cut off. Figures~\ref{FigS4}(b)--\ref{FigS4}(c) and Figures~\ref{FigS4}(d)--\ref{FigS4}(g) show that one and two of the six couplings are cut off, respectively. Figures~\ref{FigS4}(h)--\ref{FigS4}(k) represent these cases where three couplings in $g_{1(j=1\text{-}3)}$, $\eta_{12}$, $\eta_{13}$, and $\eta_{23}$ are cut off.

\begin{figure}[tbp]
\center
\includegraphics[width=0.97 \textwidth]{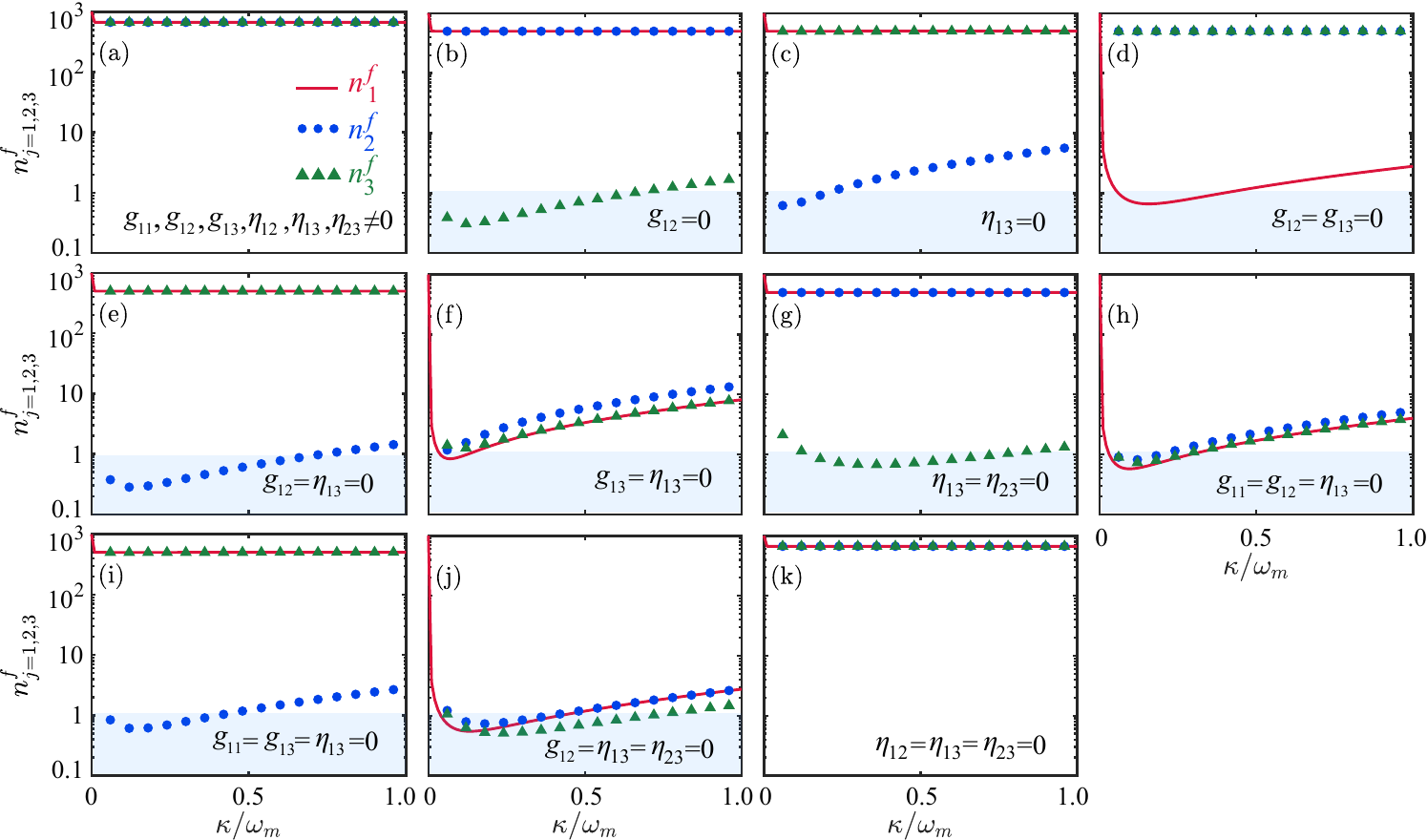}
\caption{(Color online) Final mean phonon numbers $n^{f}_{1}$, $n^{f}_{2}$, and $n^{f}_{3}$ versus $\kappa/\omega_{m}$ corresponding to eleven coupling configurations depicted in Fig.~\ref{FigS4}. The parameters are taken as  $\gamma_{j=1\text{-}3}/\omega_{m}=10^{-5}$, $\delta_{1}/\omega_{m}=1.01$, $g_{11}/\omega_{m}=g_{12}/\omega_{m}=g_{13}/\omega_{m}=0.1$, $\eta_{12}/\omega_{m}=\eta_{13}/\omega_{m}=\eta_{23}/\omega_{m}=0.09$, and $\bar{n}_{j=1\text{-}3}=1000$. The labels of $g_{1(j=1\text{-}3)}=0$, $\eta_{12}=0$, $\eta_{13}=0$, and $\eta_{23}=0$ indicate that the corresponding couplings are disconnected, i.e., the couplings are cut off.}
\label{FigS5}
\end{figure}

Corresponding to these $11$ cases illustrated in Fig.~\ref{FigS4}, we plot the final mean phonon numbers $n^{f}_{1}$, $n^{f}_{2}$, and $n^{f}_{3}$ as functions of the decay rate $\kappa/\omega_{m}$, as shown in Fig.~\ref{FigS5}. Figure~\ref{FigS5}(a) shows that the simultaneous ground-state cooling of the three mechanical modes cannot be realized when none of the couplings is cut off, which indicates that the mechanical dark modes exist for the coupling configuration depicted in Fig.~\ref{FigS4}(a). From Figs.~\ref{FigS5}(b) and~\ref{FigS5}(c), we find that the three mechanical modes cannot be cooled into their ground states when either $g_{12}=0$ or $\eta_{13}=0$, which implies that there are mechanical dark modes in the coupling configurations depicted in Figs.~\ref{FigS4}(b) and \ref{FigS4}(c). In Figs.~\ref{FigS5}(d-g), the simultaneous ground-state cooling of the three mechanical modes can only be realized ($n^{f}_{1}$, $n^{f}_{2}$, and $n^{f}_{3}<1$) in the case of $g_{13}=\eta_{13}=0$. We know that the mechanical dark modes do not exist in the coupling configuration depicted in Fig.~\ref{FigS4}(f), but exist in the coupling configurations depicted in Figs.~\ref{FigS4}(d), \ref{FigS4}(e), and \ref{FigS4}(g). Figures~\ref{FigS5}(h-k) show that the three mechanical modes can be cooled into their ground states when either $g_{11}=g_{12}=\eta_{13}=0$ or $g_{12}=\eta_{13}=\eta_{23}=0$, corresponding to the cases shown in Figs.~\ref{FigS4}(h) and \ref{FigS4}(j), respectively. However, in the other two cases of $g_{11}=g_{13}=\eta_{13}=0$ and $\eta_{12}=\eta_{13}=\eta_{23}=0$, the simultaneous ground-state cooling of the three mechanical modes cannot be realized, which means that the mechanical dark modes exist in the coupling configurations depicted in Figs.~\ref{FigS4}(i) and \ref{FigS4}(k).

The cooling results of the three mechanical modes in these eleven coupling configurations can be explained based on the dark-mode conditions derived with the arrowhead-matrix method. Below we analyze the arrowhead matrices of the linearized Hamiltonians corresponding to these eleven coupling configurations. Based on Eq.~(\ref{H13LIN}) and ignoring the counter-rotating wave terms, we obtain the approximate Hamiltonian
\begin{equation}
H_{\text{RWA}}^{[1,3]}=\delta_1a_1^{\dagger} a_1+\omega_{m}\sum_{j=1}^{3}b_{j}^{\dagger} b_{j}+\sum_{\textcolor{black}{j,j^{\prime}=1,j<j^{\prime}}}^{3}(\eta_{jj^{\prime}}b_{j}^{\dagger}b_{j^{\prime}}+\eta_{jj^{\prime}}^{\ast}b_{j^{\prime}}^{\dagger}b_{j})+\sum_{j=1}^{3}(g_{1j} a_1^{\dagger} b_{j}+g_{1j}^{\ast}b_{j}^{\dagger} a_1 ).
\label{Hamitrwafour}
\end{equation}

To clearly analyze the dark-mode conditions in these optomechanical networks, we rewrite Hamiltonian~(\ref{Hamitrwafour}) as
\begin{equation}
H_{\text{RWA}}^{[1,3]}=(a_1^{\dagger},b_{1}^{\dagger}, b_{2}^{\dagger}, b_{3}^{\dagger})\textbf{H}^{[1,3]}_{ab}(a_1,b_{1},b_{2},b_{3})^{T},
\end{equation}
where we introduce the coefficient matrix
\begin{eqnarray}
\label{matrixm3}
\textbf{H}^{[1,3]}_{ab}&=&\left(
\begin{array}{c|ccc}
\delta_{1} & g_{11} & g_{12} & g_{13}\\     \hline
g_{11}^{\ast} & \omega_{m} & \eta_{12} & \eta_{13}\\
g_{12}^{\ast} & \eta^{\ast}_{12}  & \omega_{m} & \eta_{23}\\
g_{13}^{\ast} & \eta^{\ast}_{13}  & \eta^{\ast}_{23}  & \omega_{m}
\end{array}
\right).
\end{eqnarray}
In particular, we will consider the identical coupling case when the corresponding optomechanical couplings and phonon-hopping couplings exist. To clearly clarify the parameter conditions for the mechanical dark modes, we first diagonalize the three mechanical modes. Then in the normal-mode  ($A_1,B_j$) representation, the coefficient matrix becomes
\begin{eqnarray}
	\label{matrixAB}
	\textbf{H}^{[1,3]}_{AB}&=&\left(
	\begin{array}{c|ccc}
		\Delta_{1} & G_{11} & G_{12} & G_{13}\\     \hline
		G_{11}^{\ast} & \Omega_1 &0&0\\
		G_{12}^{\ast} & 0  & \Omega_2 &0\\
		G_{13}^{\ast} & 0  & 0 & \Omega_3
	\end{array}
	\right),~\label{H13ABB}
\end{eqnarray}
where $\Delta_1=\delta_1$, $\Omega_j$ is the effective frequency of the $j$th mechanical normal mode $B_{j}=\sum_{j^{\prime}}^{3}(\textbf{U}_{b})_{jj^{\prime}}b_{j^{\prime}}$, and $G_{1j}=\sum_{j^{\prime}=1}^{3}g_{1j^{\prime}}(\textbf{U}_{b}^{\dagger})_{j^{\prime}j}$.

Below, we consider the real optomechanical and phonon-hopping coupling strengths, and analyze the influence of these couplings on the simultaneous ground-state cooling of the three mechanical modes for these eleven configurations.

(i) In the case of $\eta_{12}=\eta_{13}=\eta_{23}=\eta\neq0$ and $g_{11}=g_{12}=g_{13}=g\neq0$, as depicted in Fig. \ref{FigS4}(a), the coefficient matrix (\ref{H13ABB}) becomes
\begin{eqnarray}
	\textbf{H}_{AB}^{[1,3]}&=&\left(
	\begin{array}{c|ccc}
		\delta_{1} & 0 & 0 & \sqrt{3}g\\   \hline
		0 & \omega_{m}-\eta & 0 & 0\\
		0 & 0  & \omega_{m}-\eta & 0\\
		\sqrt{3}g & 0  & 0  & \omega_{m}+2\eta
	\end{array}
	\right).~\label{HAB131}
\end{eqnarray}
According to Theorem 1(i), we know that the two mechanical normal modes $B_{1}$ and $B_{2}$ are decoupled from the optical normal mode $A_1$. As a result, the simultaneous ground-state cooling of these mechanical modes cannot be realized, which is consistent with the numerical result depicted in Fig.~\ref{FigS5}(a). In this case, the two mechanical dark modes can be expressed as $B_{1}=(-b_{1}+b_{3})/\sqrt{2}$ and $B_{2}=(-b_{1}+ 2b_{2}-b_3)/\sqrt{6}$.

(ii) For the coupling configuration depicted in Fig. \ref{FigS4}(b), we have $\eta_{12}=\eta_{13}=\eta_{23}=\eta \neq0$, $g_{11}=g_{13}=g\neq0$, and  $g_{12}=0$. In this case, the coefficient matrix (\ref{H13ABB}) becomes
\begin{eqnarray}
	\textbf{H}_{AB}^{[1,3]}&=&\left(
	\begin{array}{c|ccc}
		\delta_{1} &0&  -\sqrt{2}g/\sqrt{3}  & 2g/\sqrt{3}\\   \hline
		0 & \omega_{m}-\eta & 0 & 0\\
	 -\sqrt{2}g/\sqrt{3} & 0  & \omega_{m}-\eta & 0\\
		2g/\sqrt{3} & 0  & 0  & \omega_{m}+2\eta
	\end{array}
	\right).
\end{eqnarray}
According to Theorem 1(i), we can see that the mechanical normal mode $B_1=(-b_{1}+ b_{3})/\sqrt{2}$ is decoupled from the optical normal mode $A_1$. Therefore, the simultaneous ground-state cooling of these mechanical modes cannot be realized,  which is consistent with the numerical result depicted in Fig.~\ref{FigS5}(b).

(iii) In the case of $\eta_{12}=\eta_{23}=\eta\neq0$, $g_{11}=g_{12}=g_{13}=g\neq0$ and $\eta_{13}=0$, as depicted in Fig. \ref{FigS4}(c), the coefficient matrix (\ref{H13ABB}) is reduced to
\begin{eqnarray}
	\textbf{H}_{\textsc{AB}}^{[1,3]}&=&\left(
	\begin{array}{c|ccc}
		\delta_{1} & 0 & (2-\sqrt{2})g/2 & (2+\sqrt{2})g/2\\   \hline
		0 & \omega_{m} & 0 & 0\\
		(2-\sqrt{2})g/2& 0  & \omega_{m}-\sqrt{2}\eta & 0\\
		(2+\sqrt{2})g/2 & 0  & 0  & \omega_{m}+\sqrt{2}\eta
	\end{array}
	\right).
\end{eqnarray}
In this case, we find that the mode $B_1=(- b_{1}+ b_{3})/\sqrt{2}$ becomes a mechanical dark mode according to Theorem 1(i). Then the simultaneous ground-state cooling of these mechanical modes is suppressed, which is consistent with the numerical result depicted in Fig.~\ref{FigS5}(c).

(iv) For the coupling configuration described by Fig. \ref{FigS4}(d), we have $\eta_{12}=\eta_{13}=\eta_{23}=\eta\neq0$,  $g_{11}=g\neq0$, and $g_{12}=g_{13}=0$. In this case, the coefficient matrix (\ref{H13ABB}) is expressed as
\begin{eqnarray}
	\textbf{H}_{AB}^{[1,3]}&=&\left(
	\begin{array}{c|ccc}
		\delta_{1} &-g/\sqrt{2}&-g/\sqrt{6} & g/\sqrt{3}\\   \hline
		-g/\sqrt{2} & \omega_{m}-\eta & 0 & 0\\
	-g/\sqrt{6} & 0  & \omega_{m}-\eta & 0\\
		g/\sqrt{3} & 0  & 0  & \omega_{m}+2\eta
	\end{array}
	\right).
\end{eqnarray}
According to Theorem 1(ii), we see that the mode $B_{2-}=(b_{2}- b_{3})/\sqrt{2}$ becomes a mechanical dark mode. Then the simultaneous ground-state cooling of these mechanical modes is suppressed, which is consistent with the numerical result depicted in Fig.~\ref{FigS5}(d).

(v) In the case of $\eta_{12}=\eta_{23}=\eta\neq0$, $g_{11}=g_{13}=g\neq0$, and $g_{12}=\eta_{13}=0$, as shown in Fig. \ref{FigS4}(e), the coefficient matrix (\ref{H13ABB}) is reduced to
\begin{eqnarray}
	\textbf{H}_{AB}^{[1,3]}&=&\left(
	\begin{array}{c|ccc}
		\delta_{1} & 0 & g & g\\   \hline
		0 & \omega_{m} & 0 & 0\\
		g& 0  & \omega_{m}-\sqrt{2}\eta & 0\\
		g & 0  & 0  & \omega_{m}+\sqrt{2}\eta
	\end{array}
	\right).
\end{eqnarray}
In this case, we know based on Theorem 1(i) that the mode $B_1=(- b_{1}+ b_{3})/\sqrt{2}$ becomes a dark mode. Then the simultaneous ground-state cooling of these mechanical modes is suppressed, which is consistent with the numerical result depicted in Fig.~\ref{FigS5}(e).

(vi) For the coupling configuration shown in Fig. \ref{FigS4}(f), we have $\eta_{12}=\eta_{23}=\eta\neq0$, $g_{11}=g_{12}=g\neq0$, and $g_{13}=\eta_{13}=0$. In this case, the coefficient matrix (\ref{H13ABB}) is expressed as
\begin{eqnarray}
	\textbf{H}_{AB}^{[1,3]}&=&\left(
	\begin{array}{c|ccc}
		\delta_{1} & -g/\sqrt{2} & (1-\sqrt{2})g/2 & (1+\sqrt{2})g/2\\   \hline
		-g/\sqrt{2} & \omega_{m} & 0 & 0\\
		(1-\sqrt{2})g/2& 0  & \omega_{m}-\sqrt{2}\eta & 0\\
		(1+\sqrt{2})g/2 & 0  & 0  & \omega_{m}+\sqrt{2}\eta
	\end{array}
	\right).
\end{eqnarray}
Based on Theorem 1(iii), we know that there is no mechanical dark mode. Then the simultaneous ground-state cooling of these mechanical modes can be realized, which is confirmed by the numerical result depicted in Fig.~\ref{FigS5}(f).

(vii) For the coupling configuration described by Fig. \ref{FigS4}(g), the couplings are given by $\eta_{12}=\eta\neq0$, $g_{11}=g_{12}=g_{13}=g\neq0$, and $\eta_{13}=\eta_{23}=0$. In this case, the coefficient matrix (\ref{H13ABB}) becomes
\begin{eqnarray}
	\textbf{H}_{AB}^{[1,3]}&=&\left(
	\begin{array}{c|ccc}
		\delta_1 & g & 0 & \sqrt{2}g\\   \hline
		g & \omega_{m} & 0 & 0\\
		0& 0  & \omega_{m}-\eta & 0\\
		\sqrt{2}g& 0  & 0  & \omega_{m}+\eta
	\end{array}
	\right).
\end{eqnarray}
According to Theorem 1(i), we know that the mode $B_2=(- b_{1}+b_{2})/\sqrt{2}$ is decoupled from the optical normal mode $A_1$ and becomes a mechanical dark mode. Therefore, the simultaneous ground-state cooling of these mechanical modes is unfeasible, which is consistent with the numerical result depicted in Fig.~\ref{FigS5}(g).

(viii) For the coupling configuration depicted in Fig. \ref{FigS4}(h), we have $\eta_{12}=\eta_{23}=\eta \neq0$, $g_{13}=g\neq0$, and  $g_{11}=g_{12}=\eta_{13}=0$. In this case, the coefficient matrix (\ref{H12AB}) is expressed as
\begin{eqnarray}
	\textbf{H}_{AB}^{[1,3]}&=&\left(
	\begin{array}{c|ccc}
		\delta_1 & g/\sqrt{2} & g/2 & g/2\\   \hline
		g/\sqrt{2} & \omega_{m} & 0 & 0\\
		g/2& 0  & \omega_{m}-\sqrt{2}\eta & 0\\
		g/2 & 0  & 0  & \omega_{m}+\sqrt{2}\eta
	\end{array}
	\right).
\end{eqnarray}
We can see that the three effective coupling strengths are nonzero and the three effective frequencies are non-degenerate. Based on Theorem 1(iii), we know that there is no mechanical dark mode, which means that the simultaneous ground-state cooling of these mechanical modes is feasible. This result is consistent with the numerical simulation depicted in Fig.~\ref{FigS5}(h).

(ix) For the coupling configuration described by Fig. \ref{FigS4}(i), we have $\eta_{12}=\eta_{23}=\eta \neq0$, $g_{12}=g\neq0$, and  $g_{11}=g_{13}=\eta_{13}=0$. In this case, the coefficient matrix (\ref{H13ABB}) can be expressed as
\begin{eqnarray}
	\textbf{H}_{AB}^{[1,3]}&=&\left(
	\begin{array}{c|ccc}
		\delta_1 & 0 & -g/\sqrt{2} & g/\sqrt{2}\\   \hline
		0 & \omega_{m} & 0 & 0\\
		-g/\sqrt{2}& 0  & \omega_{m}-\sqrt{2}\eta & 0\\
		g/\sqrt{2} & 0  & 0  & \omega_{m}+\sqrt{2}\eta
	\end{array}
	\right).
\end{eqnarray}
Based on Theorem 1(i), we know that the mode $B_1=(- b_{1}+b_{3})/\sqrt{2}$ becomes a mechanical dark mode. Then the simultaneous ground-state cooling of these mechanical modes cannot be realized, which is consistent with the numerical result depicted in Fig.~\ref{FigS5}(i).

(x) For the coupling configuration shown in Fig. \ref{FigS4}(j), the couplings are given by $\eta_{12}=\eta \neq0$, $g_{11}=g_{13}=g\neq0$, and  $g_{12}=\eta_{13}=\eta_{23}=0$. In this case, the coefficient matrix (\ref{H13ABB}) becomes
\begin{eqnarray}
	\textbf{H}_{AB}^{[1,3]}&=&\left(
	\begin{array}{c|ccc}
		\delta_1 & g &  -g/\sqrt{2} & g/\sqrt{2}\\   \hline
		g & \omega_{m} & 0 & 0\\
		-g/\sqrt{2}& 0  & \omega_{m}-\eta & 0\\
		g/\sqrt{2} & 0  & 0  & \omega_{m}+\eta
	\end{array}
	\right).
\end{eqnarray}
Based on Theorem 1(iii), we know that there is no mechanical dark mode.  Then the simultaneous ground-state cooling of these mechanical modes can be realized, which is confirmed by the numerical simulation depicted in Fig.~\ref{FigS5}(j).

\begin{figure}[tbp]
	\center
	\includegraphics[width=1.0 \textwidth]{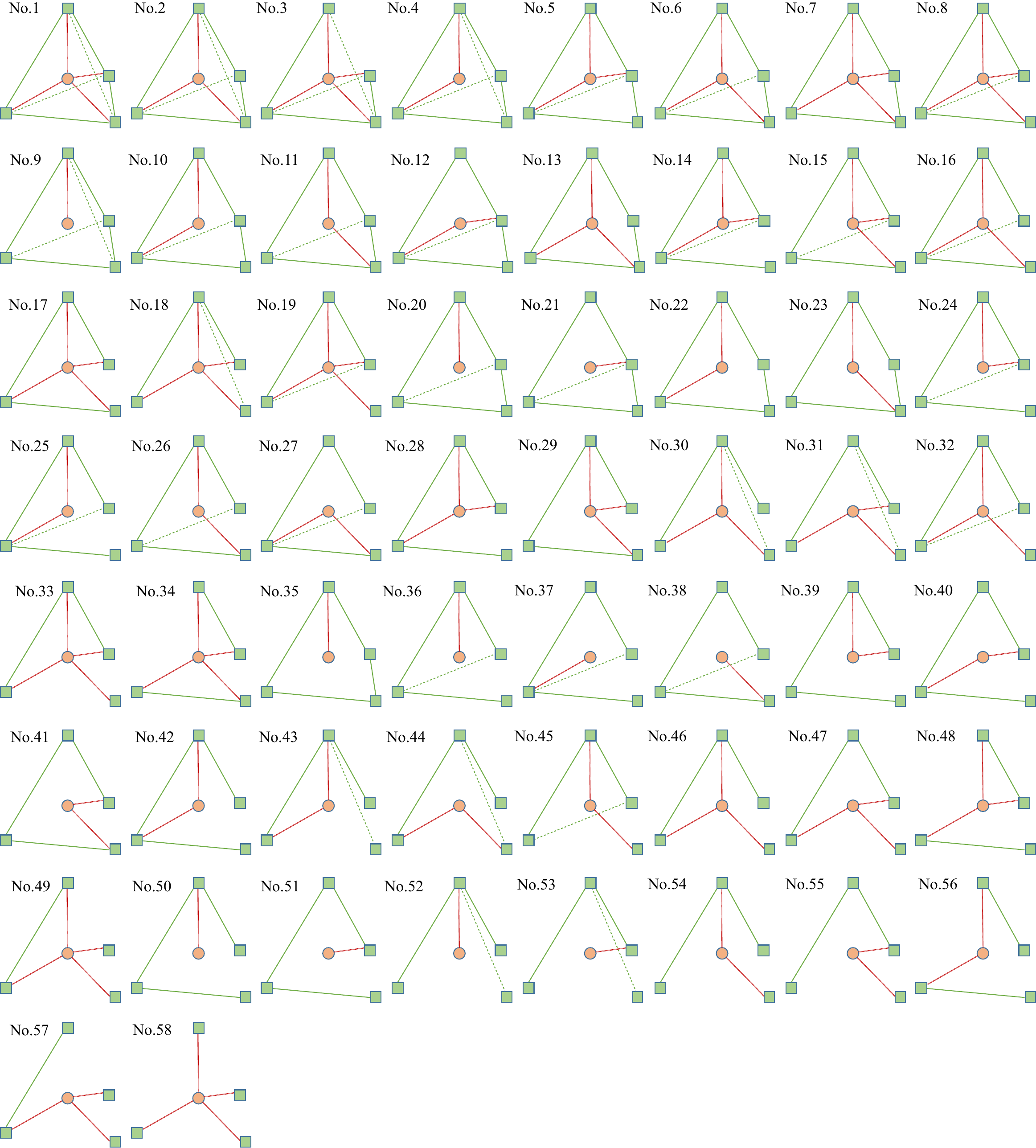}
	\caption{Fifty-eight coupling configurations for the five-mode optomechanical networks consisting of one optical mode (circle) and four mechanical modes (square).  Here, the four mechanical modes are identical, thus we have  $\omega_{1}=\omega_{2}=\omega_{3}=\omega_{4}=\omega_{m}$, $g_{1}=g_{2}=g_{3}=g_{4}=g$, and $\eta_{12}=\eta_{13}=\eta_{14}=\eta_{23}=\eta_{24}=\eta_{34}=\eta$. To ensure that these four mechanical modes can be either directly or indirectly coupled to the cavity mode, we keep at least four couplings in the network, and the other six couplings can be switched on demand. In this case, there are fifty-eight coupling configurations. Concretely, panel No.1 describes the case where all the couplings are present. Panels No.2-No.3 describe the cases where one coupling is cut off. The two-coupling-disconnected cases are described by panels No.4 - No.8. The cases corresponding to three-coupling-disconnected are shown in panels No.9 - No.19. The cases where four couplings are cut off are shown in panels No.20 - No.34. The five-coupling-disconnected cases are shown in panels No.35 - No.49. In addition, the cases corresponding to six-coupling-disconnected are shown in panels No.50 - No.58.}
	\label{FigS6}
\end{figure}

(xi) For the coupling configuration described by Fig. \ref{FigS4}(k), we have $\eta_{12}=\eta_{13}=\eta_{23}=0$, and $g_{11}=g_{12}=g_{13}=g\neq0$. In this case, the coefficient matrix (\ref{H12AB}) can be expressed as
\begin{eqnarray}
	\textbf{H}_{AB}^{[1,3]}&=&\left(
	\begin{array}{c|ccc}
		\delta_1 & g &  g & g\\   \hline
		g & \omega_{m} & 0 & 0\\
		g& 0  & \omega_{m} & 0\\
		g & 0  & 0  & \omega_{m}
	\end{array}
	\right).
\end{eqnarray}
We can see that the three degenerate normal modes are coupled to the optical normal mode $A_1$, which means that there are two dark modes, i.e., $B_{2-}=( b_{1}-b_{2})/\sqrt{2}$ and $B_{3-}=(b_{1}+ b_{2}-2b_3)/\sqrt{6}$. As a result, the simultaneous ground-state cooling of these mechanical modes cannot be realized. This result is consistent with the numerical result depicted in Fig.~\ref{FigS5}(k).

By analyzing the cooling results for these eleven coupling configurations, we find that the simultaneous ground-state cooling of the three mechanical modes cannot be realized when there exist mechanical dark modes in the network. These analyses are consistent with the statements given in Theorem 1. Generally speaking, to know whether the multiple mechanical modes can be cooled into their ground states in an arbitrary coupling configuration, an efficient way is to calculate the arrowhead matrix associated with the approximate linearized Hamiltonian. If the mechanical dark modes exist, then the mechanical modes cannot be simultaneously cooled into their ground states. Otherwise, the simultaneous ground-state cooling of these mechanical modes can be realized under proper parameters.

\subsection{One-optical-mode and four-mechanical-mode optomechanical networks \label{modelfour}}

To further examine the dark-mode theorem, we evaluate the simultaneous ground-state cooling of mechanical modes in the  one-optical-mode and four-mechanical-mode optomechanical networks, as shown in panel No.1 of Fig.~\ref{FigS6}. In this case, the linearized Hamiltonian reads
\begin{equation} H_{\text{lin}}^{[1,4]}=\delta_{1}a_{1}^{\dagger}a_{1}+\sum_{j=1}^{4}\omega_{j}b_{j}^{\dagger}b_{j}+\sum_{\textcolor{black}{j,j^{\prime}=1,j<j^{\prime}}}^{4}(\eta_{jj^{\prime}}b_{j}^{\dagger}b_{j^{\prime}}+\eta_{jj^{\prime}}^{\ast}b_{j^{\prime}}^{\dagger}b_{j})+\sum_{j=1}^{4}(g_{1j}a_{1}^{\dagger}+g_{1j}^{\ast}a_{1})(b_{j}^{\dagger}+b_{j}),\label{Hamitrot5}
\end{equation}
where the notations are consistent with those defined in Eq.~(\ref{HMNLIN}).
Based on Hamiltonian~(\ref{Hamitrot5}), we can obtain the final mean phonon numbers in the four mechanical modes by calculating the steady-state covariance matrix.

To clearly clarify the ground-state cooling and the dark-mode conditions in this five-mode optomechanical network, we consider different coupling configurations by switching the ten couplings $g_{1(l=1\text{-}4)}$, $\eta_{12}$, $\eta_{13}$, $\eta_{14}$, $\eta_{23}$, $\eta_{24}$, and $\eta_{34}$. For convenience, we consider that the four mechanical modes have the same resonance frequencies, i.e., $\omega_{1}=\omega_{2}=\omega_{3}=\omega_{4}=\omega_{m}$. To ensure that these four mechanical modes can be either directly or indirectly coupled with the cavity mode, we keep at least four couplings in the network. Therefore, there exist $58$ coupling configurations when zero, one, two, three, four, five, or six couplings are cut off.

	\begin{center}
		\begin{table}[ht!]
			\caption{The simultaneous ground-state cooling results of the four degenerate mechanical modes corresponding to these $58$ coupling configurations. We also present the existence of the dark mode and the number of the dark modes.}
			\label{Tab}
			\begin{tabular}{|c|c|c|c|c|c|c|c|c|c|c|c|}
				\hline
				\multicolumn{1}{|c|}{No.} & {\tabincell{c}{Simultaneous \\ground-state\\cooling}} & {\tabincell{c}{Existence\\ of\\ dark mode}}  & {\tabincell{c}{Dark\\ mode\\ (number)}} & No. & {\tabincell{c}{Ground\\state\\cooling}} & {\tabincell{c}{Existence\\ of\\ dark mode}} & {\tabincell{c}{Dark\\ mode\\ (number)}} & No. & {\tabincell{c}{Ground\\state\\cooling}} & {\tabincell{c}{Existence\\ of\\ dark mode}} & {\tabincell{c}{Dark\\ mode\\ (number)}}\\ \cline{1-12}
				\multicolumn{1}{|c|}{1} & no  & yes & 3 & {2} & no  & yes & 2 & {3}  & no  & yes & 2 \\ \cline{1-12}
				\multicolumn{1}{|c|}{4} & no  & yes & 2 &{5}  & no  & yes & 1  & {6} & no  & yes & 1 \\ \cline{1-12}
				\multicolumn{1}{|c|}{7} & no  & yes & 3 &{8}  & no  & yes & 1 & {9} & no  & yes & 2 \\ \cline{1-12}
				\multicolumn{1}{|c|}{10}& yes & no  & 0 &{11} & no  & yes & 2 & {12}& no  & yes & 2 \\ \cline{1-12}
				\multicolumn{1}{|c|}{13}& no  & yes & 1 &{14} & no  & yes & 1 & {15}& no  & yes & 1 \\ \cline{1-12}
				\multicolumn{1}{|c|} {16}& yes & no  & 0 &{17} & no  & yes & 2 & {18}& no  & yes & 2 \\ \cline{1-12}
				\multicolumn{1}{|c|} {19}& no  & yes & 2 &{20} & no  & yes & 1 & {21}& no  & yes & 1 \\ \cline{1-12}
				\multicolumn{1}{|c|} {22}& no  & yes & 2 &{23} & no  & yes & 2 & {24}& no  & yes & 1 \\ \cline{1-12}
				\multicolumn{1}{|c|} {25}& yes & no  & 0 &{26} & yes & no  & 0 & {27}& no  & yes & 1 \\ \cline{1-12}
				\multicolumn{1}{|c|} {28}& yes & no  & 0  &{29} & yes & no  & 0 & {30}& no  & yes & 1 \\ \cline{1-12}
				\multicolumn{1}{|c|} {31}& no  & yes & 2  &{32} & no  & yes & 1 & {33}& no  & yes & 1 \\ \cline{1-12}
				\multicolumn{1}{|c|} {34}& no  & yes & 3 &{35} & no  & yes & 1& {36}& yes & no  & 0 \\ \cline{1-12}
				\multicolumn{1}{|c|} {37}& no  & yes & 1 &{38} & no  & yes & 1 & {39}& yes & no  & 0 \\ \cline{1-12}
				\multicolumn{1}{|c|} {40}& yes & no  & 0 &{41} & no  & yes & 2 & {42}& no  & yes & 2 \\ \cline{1-12}
				\multicolumn{1}{|c|} {43}& no  & yes & 1  &{44} & no  & yes & 1 & {45}& no  & yes & 1  \\ \cline{1-12}
				\multicolumn{1}{|c|} {46}& no  & yes & 1  & {47} & no  & yes & 1 & {48}& no  & yes & 2 \\ \cline{1-12}
				\multicolumn{1}{|c|}{49}& no  & yes & 2   &{50} & yes & no  & 0 & {51}& yes & no  & 0  \\ \cline{1-12}
				\multicolumn{1}{|c|} {52}& no  & yes & 2 &{53} & no  & yes & 1 & {54}& no  & yes & 1 \\ \cline{1-12}
				\multicolumn{1}{|c|} {55}& no  & yes & 1  &{56} & no  & yes & 2 & {57}& no  & yes & 1 \\ \cline{1-12}
				\multicolumn{1}{|c|} {58}& no  & yes & 3   &        &         &      &     &      &     &     &    \\ \hline
			\end{tabular}
			\label{table1}
		\end{table}
	\end{center}

Based on these $58$ coupling configurations depicted in Fig. \ref{FigS6}, we can evaluate the simultaneous ground-state cooling of the four mechanical modes by analyzing the dark-mode conditions derived with the arrowhead-matrix method. Below we analyze the arrowhead matrices of the linearized Hamiltonian corresponding to these $58$ coupled configurations. Based on Eq.~(\ref{Hamitrot5}) and ignoring the counter-rotating wave terms, we can obtain the approximate Hamiltonian
\begin{equation}
	H_{\text{RWA}}^{[1,4]}=\delta_1a_1^{\dagger} a_1+\omega_{m}\sum_{j=1}^{4}b_{j}^{\dagger} b_{j}+\sum_{\textcolor{black}{j,j^{\prime}=1,j<j^{\prime}}}^{4}(\eta_{jj^{\prime}}b_{j}^{\dagger}b_{j^{\prime}}+\eta_{jj^{\prime}}^{\ast}b_{j^{\prime}}^{\dagger}b_{j})+\sum_{j=1}^{4}( g_{1j}a_1^{\dagger} b_{j}+g^{\ast}_{1j} b_{j}^{\dagger}a_1 ).~\label{Hamitrot104}
\end{equation}

To clearly analyze the dark-mode conditions in these optomechanical networks, we rewrite Hamiltonian~(\ref{Hamitrot104}) as
\begin{equation}
H_{\text{RWA}}^{[1,4]}=(a_1^{\dagger}, b_{1}^{\dagger}, b_{2}^{\dagger}, b_{3}^{\dagger},b_{4}^{\dagger})\textbf{H}_{ab}^{[1,4]}( a_1,b_{1}, b_{2}, b_{3},b_{4})^{T},
\end{equation}
where we introduce the coefficient matrix
\begin{eqnarray}
\textbf{H}_{ab}^{[1,4]}&=&\left(
\begin{array}{c|cccc}
\delta_{1} & g_{11} & g_{12} & g_{13} & g_{14}\\     \hline
g_{11}^{\ast} & \omega_{m} & \eta_{12} & \eta_{13} & \eta_{14}\\
g_{12}^{\ast} & \eta^{\ast}_{12}  & \omega_{m} & \eta_{23} & \eta_{24}\\
g_{13}^{\ast} & \eta^{\ast}_{13}  & \eta^{\ast}_{23}  & \omega_{m}& \eta_{34}\\
g_{14}^{\ast} & \eta^{\ast}_{14}  & \eta^{\ast}_{24}  & \eta^{\ast}_{34}& \omega_{m}
\end{array}
\right).
\end{eqnarray}
In particular,  we will consider the identical coupling case when the corresponding phonon-hopping and optomechanical couplings exist. To clearly clarify the parameter conditions for the dark modes, we first diagonalize the four mechanical modes. Then in the normal-mode ($A_1,B_j$) representation, the coefficient matrix becomes
\begin{eqnarray}
	\label{matrixAB}
	\textbf{H}^{[1,4]}_{AB}&=&\left(
	\begin{array}{c|cccc}
		\Delta_{1} & G_{11} & G_{12} & G_{13}& G_{14}\\     \hline
		G_{11}^{\ast} & \Omega_1 &0&0&0\\
		G_{12}^{\ast} & 0  & \Omega_2 &0&0\\
		G_{13}^{\ast} & 0  & 0 & \Omega_3&0\\
		G_{14}^{\ast} & 0  & 0 &0&\Omega_4
	\end{array}
	\right),~\label{H14ABA}
\end{eqnarray}
where $\Delta_1=\delta_1$, $\Omega_j$ is the effective frequency of the $j$th mechanical normal mode $B_{j}=\sum_{j^{\prime}}^{4}(\textbf{U}_{b})_{jj^{\prime}}b_{j^{\prime}}$, and $G_{1j}=\sum_{j^{\prime}=1}^{4}g_{1j^{\prime}}(\textbf{U}_{b}^{\dagger})_{j^{\prime}j}$.

Based on Eq.~(\ref{H14ABA}), we can obtain the number of dark modes and evaluate the simultaneous ground-state cooling of mechanical modes.
Since the corresponding results are too much to be presented here, we only exhibit the cooling results of these mechanical modes corresponding to these $58$ coupling configurations in Table~\ref{table1}.  Based on Table~\ref{table1}, we know that the performance of the simultaneous ground-state cooling is determined by the dark-mode effect. When there exist dark modes in the network, the simultaneous ground-state cooling of multiple mechanical modes cannot be realized. We emphasize that, in principle, the arrowhead-matrix theory is general, which can be used to study the simultaneous ground-state cooling of mechanical modes in any coupling configurations, regardless of whether the mechanical modes in the system are identical or not.

\subsection{One-optical-mode and two-mechanical-mode-chain optomechanical networks \label{modelchain}}

\begin{figure}[tbp]
	\center
	\includegraphics[width=0.95 \textwidth]{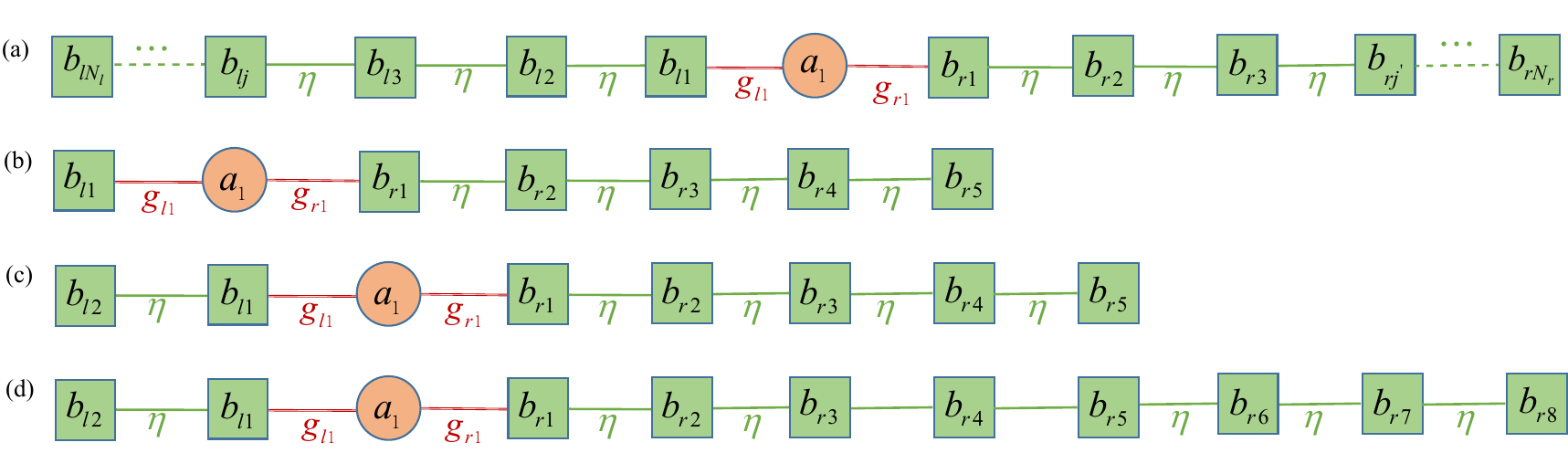}
	\caption{(a) Schematic of the optomechanical network consisting of an optical mode and two mechanical-mode chains formed by $N_{l}$ and $N_{r}$ mechanical modes. The optical mode is coupled to the first mechanical mode in the left (right)-mechanical-mode chain via radiation-pressure interaction with strength $g_{l1}$ ($g_{r1}$), and the neighboring mechanical modes are coupled to each other via phonon-hopping interactions with strength $\eta$. Schematic of the network for specific values of $N_{l}$ and $N_{r}$: (b) $N_{l}=1$ and $N_{r}=5$, (c) $N_{l}=2$ and $N_{r}=5$, and (d) $N_{l}=2$ and $N_{r}=8$.}
	\label{FigS7}
\end{figure}

In this section, we check the dark-mode theorem in the  one-optical-mode and two-mechanical-mode-chain optomechanical network (see Fig.~\ref{FigS7}). We assume that all the mechanical modes have the same resonance frequency $\omega_m$, and that all the coupling strengths between these neighboring mechanical modes are $\eta$. The linearized Hamiltonian of this network reads
\begin{eqnarray}
H & = & \delta_{1}a_{1}^{\dagger}a_{1}+\omega_{m}\sum_{j=1}^{N_{l}}b_{lj}^{\dagger}b_{lj}+\omega_{m}\sum_{j^{\prime}=1}^{N_{r}}b_{rj^{\prime}}^{\dagger}b_{rj^{\prime}}-(g_{l1}a_{1}^{\dagger}+g^{\ast}_{l1}a_{1})(b_{l1}^{\dagger}+b_{l1})-(g_{r1}a_{1}^{\dagger}+g^{\ast}_{r1}a_{1})(b_{r1}^{\dagger}+b_{r1}) \nonumber\\
&& +\eta\sum_{j=1}^{N_{l}-1}(b_{lj}b_{l(j+1)}^{\dagger}+b_{lj}^{\dagger}b_{l(j+1)})+\eta\sum_{j^{\prime}=1}^{N_{r}-1}(b_{rj^{\prime}}b_{r(j^{\prime}+1)}^{\dagger}+b_{rj^{\prime}}^{\dagger}b_{r(j^{\prime}+1)}),
\end{eqnarray}
where $b_{lj}$ $(b_{lj}^{\dagger})$ and $b_{rj^{\prime}}$ $(b^{\dagger}_{rj^{\prime}})$ are, respectively, the annihilation (creation) operators of the $j$th  mechanical mode in the left chain and the $j^{\prime}$th mechanical mode in the right chain. The parameter $g_{l1}$ ($g_{r1}$) is the linearized optomechanical-coupling strength between the optical mode $a_{1}$ and the mechanical mode  $b_{l1}$ ($b_{r1}$).

Similarly, we consider the case where the red-sideband-resonance terms dominate the physical processes, then we can perform the RWA to obtain the approximate Hamiltonian as
\begin{eqnarray}
\label{apprchain}
H & = & \delta_{1}a_{1}^{\dagger}a_{1}+\omega_{m}\sum_{j=1}^{N_{l}}b_{lj}^{\dagger}b_{lj}+\omega_{m}\sum_{j^{\prime}=1}^{N_{r}}b_{rj^{\prime}}^{\dagger}b_{rj^{\prime}}-(g_{l1}a_{1}^{\dagger}b_{l1}+g^{\ast}_{l1}a_{1}b_{l1}^{\dagger})-(g_{r1}a_{1}^{\dagger}b_{r1}+g^{\ast}_{r1}a_{1}b_{r1}^{\dagger})\nonumber\\
&&+\eta\sum_{j=1}^{N_{l}-1}(b_{lj}b_{l(j+1)}^{\dagger}+b_{lj}^{\dagger}b_{l(j+1)})+\eta\sum_{j^{\prime}=1}^{N_{r}-1}(b_{rj^{\prime}}b_{r(j^{\prime}+1)}^{\dagger}+b_{rj^{\prime}}^{\dagger}b_{r(j^{\prime}+1)}).
\end{eqnarray}
Below, we analyze the dark-mode effect in this system. To this end, we first diagonalize the two mechanical-mode chains. When $g_{l1}=0$, the Hamiltonian of the left mechanical-mode-chain can be diagonalized as
\begin{equation}
	H_{\text{lch}} =\omega_{m}\sum_{j=1}^{N_{l}}b_{lj}^{\dagger }b_{lj}+ \eta\sum_{j=1}^{N_{l}-1}( b_{lj}b_{l(j+1)}^{\dagger}+b_{lj}^{\dagger }b_{l(j+1)})=\sum_{n=1}^{N_{l}}\Omega_{n}B_{ln}^{\dagger}B_{ln},
\end{equation}
where the  resonance frequency of the $n$th mechanical normal mode $B_{ln}$ in the left mechanical-mode chain is introduced by
\begin{eqnarray}
	\Omega_{n}=\omega_{m}+2\eta\cos \left(\frac{n\pi}{N_{l}+1}\right),\hspace{0.5 cm}n=1\text{-}N_{l}.\label{omegan}
\end{eqnarray}
The relationship between  the mechanical mode $b_{lj}$ and the normal mode $B_{ln}$ is given by
\begin{equation}
	b_{lj}=\frac{1}{\sqrt{(N_{l}+1)/2}}\sum_{n=1}^{N_{l}}\sin \left(\frac{jn\pi}{N_{l}+1}\right)B_{ln}.
\end{equation}

\begin{figure}[tbp]
	\center
	\includegraphics[width=0.75 \textwidth]{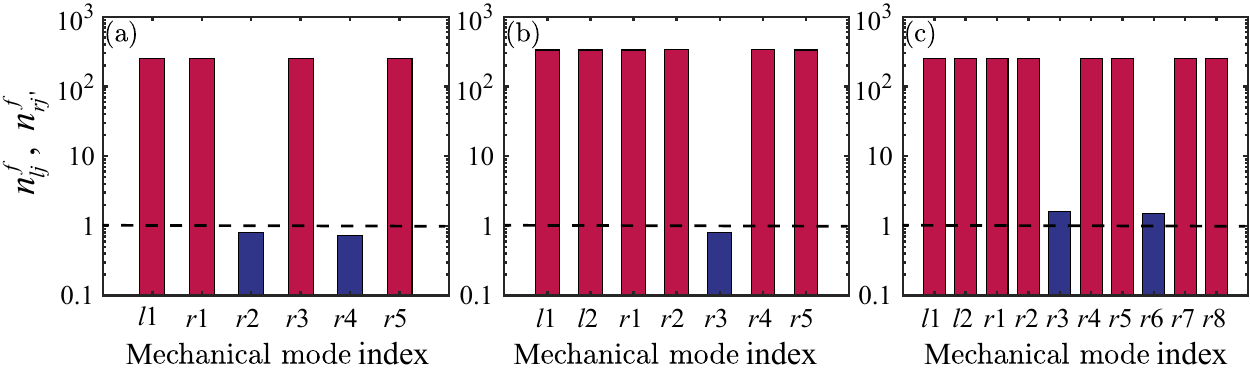}
	\caption{Final mean phonon numbers $n^{f}_{lj}$ and $n^{f}_{rj^{\prime}}$ in these mechanical modes for three different coupling configurations depicted in Figs.~\ref{FigS7}(b),~\ref{FigS7}(c), and~\ref{FigS7}(d), respectively. The parameters used are given by $\gamma_{lj}/\omega_{m}=\gamma_{rj^{\prime}}/\omega_{m}=10^{-5}$, $\delta_{1}/\omega_{m}=1$, $g_{l1}/\omega_{m}=g_{r1}/\omega_{m}=0.2$, $\eta/\omega_{m}=0.2$, $\kappa_{1}/\omega_{m}=0.2$, and $\bar{n}_{lj}=\bar{n}_{rj^{\prime}}=1000$. Note that $\gamma_{lj}$ ($\gamma_{rj^{\prime}}$) and $\bar{n}_{lj}$ ($\bar{n}_{rj^{\prime}}$) are the decay rate and the mean thermal phonon number of the $j$th ($j^{\prime}$th) mechanical mode on the left (right) of optical mode, respectively.}
	\label{Figchaincooling}
\end{figure}

Similarly, when $g_{r1}=0$,  the right mechanical-mode chain can be diagonalized as
\begin{equation}
	H_{\text{rch}} =\omega_{m}\sum_{j^{\prime}=1}^{N_{r}}b_{rj^{\prime}}^{\dagger }b_{rj^{\prime}}+ \eta\sum_{j^{\prime}=1}^{N_{r}-1}( b_{rj^{\prime}}b_{r(j^{\prime}+1)}^{\dagger}+b_{r(j^{\prime}+1)}b_{rj^{\prime}}^{\dagger })=\sum_{n^{\prime}=1}^{N_{r}}\Omega_{n^{\prime}}B_{rn^{\prime}}^{\dagger}B_{rn^{\prime}},
\end{equation}
where we introduce the resonance frequency of the $n^{\prime}$th mechanical normal mode $B_{rn^{\prime}}$ as
\begin{eqnarray}
	\Omega_{n^{\prime}}=\omega_{m}+2\eta\cos \left(\frac{n^{\prime}\pi}{N_{r}+1}\right),\hspace{0.5 cm}n^{\prime}=1\text{-}N_{r}.\label{omeganprime}
\end{eqnarray}
The relationship between the mechanical mode $b_{rj^{\prime}}$ and the normal mode $B_{rn^{\prime}}$ is determined by the following equation
\begin{equation}
	b_{rj^{\prime}}=\frac{1}{\sqrt{(N_{r}+1)/2}}\sum_{n^{\prime}=1}^{N_{r}}\sin \left(\frac{j^{\prime}n^{\prime}\pi}{N_{r}+1}\right)B_{rn^{\prime}}.
\end{equation}
Then the Hamiltonian~(\ref{apprchain})  of this optomechanical network can be written in the normal-mode ($A_1,B_{j}$) representation  as
$H=(A_1^{\dagger}, \textbf{B}^{\dagger})\textbf{H}_{AB}(A_1, \textbf{B})^{T}$, with the coefficient matrix
\small{
\begin{eqnarray}
	\label{matrileftrit}
		\setlength{\arraycolsep}{0.5pt}
	\textbf{H}_{AB}&=&\left(\begin{array}{c|cccccc}
		\delta_{1} & \frac{g_{l1}}{\sqrt{(N_l+1)/2}}\sin\left(\frac{\pi}{N_{l}+1}\right) & \ldots & \frac{g_{l1}}{\sqrt{(N_l+1)/2}}\sin\left(\frac{N_{l}\pi}{N_{l}+1}\right) & \frac{g_{r1}}{\sqrt{(N_r+1)/2}}\sin\left(\frac{\pi}{N_{r}+1}\right) & \ldots & \frac{g_{r1}}{\sqrt{(N_r+1)/2}}\sin\left(\frac{N_{r}\pi}{N_{r}+1}\right)\\
		\hline \frac{g_{l1}}{\sqrt{(N_l+1)/2}}\sin\left(\frac{\pi}{N_{l}+1}\right) & \Omega_{1} & 0 & 0 & 0 & 0 & 0\\
		\vdots & 0 & \ddots & 0 & 0 & 0 & 0\\
		\frac{g_{l1}}{\sqrt{(N_l+1)/2}}\sin\left(\frac{N_{l}\pi}{N_{l}+1}\right) & 0 & 0 & \Omega_{n} & 0 & 0 & 0\\
		\frac{g_{r1}}{\sqrt{(N_r+1)/2}}\sin\left(\frac{\pi}{N_{r}+1}\right) & 0 & 0 & 0 & \Omega_{1'} & 0 & 0\\
		\vdots & 0 & 0 & 0 & 0 & \ddots & 0\\
		\frac{g_{r1}}{\sqrt{(N_r+1)/2}}\sin\left(\frac{N_{r}\pi}{N_{r}+1}\right) & 0 & 0 & 0 & 0 & 0 & \Omega_{n^{\prime}}
	\end{array}\right).
\end{eqnarray}}
From matrix~(\ref{matrileftrit}),  we know that the coupling coefficients  are  nonzero for  $n=1-N_l$ and $n^{\prime}=1-N_r$. Therefore, the dark-mode existence condition is mainly determined by the resonance frequencies of these mechanical normal modes. If there exist degenerate modes among $\Omega_{n=1-N_l}$ and $\Omega_{n^{\prime}=1-N_r}$, then there exist mechanical dark modes. Equations (\ref{omegan}) and (\ref{omeganprime}) indicate that the mechanical-mode numbers $N_l$ and $N_r$ determines the resonance  frequencies  of these mechanical normal modes.
Below we analyze the dependence of these mechanical-mode frequencies on the variables $N_{l}$ and $N_{r}$ ($N_{l}\neq N_{r}$).

(i) When both $N_{l}$ and $N_{r}$ are odd numbers, we can always find an $n$ ($n^{\prime}$) to satisfy $\cos [n\pi /(N_{l}+1)]=0$ ($\cos [n^{\prime}\pi /(N_{r}+1)]=0$). In this case, we have $\Omega_{n}=\omega_{m}$ and  $\Omega_{n^{\prime}}=\omega_{m}$. Based on Theorem 1(i), there exist dark modes in this network.

(ii) When one of the numbers $N_{l}$ and $N_{r}$  is an even number and the other is an odd number, we choose $N_{l}=2$ and $N_{r}=5$ as an example,  and we can obtain $\textbf{H}_{2}=\text{diag}\{\omega_{m}+\eta,\omega_{m}-\eta\}$ and $\textbf{H}_{5}=\text{diag}\{\omega_ {m}+\sqrt{3}\eta,\omega_{m}+\eta,\omega_{m},\omega_{m}-\eta,\omega_{m}-\sqrt{3}\eta\}$. We see that there are two degenerate frequencies $\omega_{m}+\eta$ and $\omega_{m}-\eta$ in $\textbf{H}_{2}$ and $\textbf{H}_{5}$. Based on Theorem 1(ii), there exist dark modes in the network.

(iii) When both $N_{l}$ and $N_{r}$ are  even numbers, for example, $N_{l}=2$ and $N_{r}=8$, we have $\textbf{H}_{2}=\text{diag}\{\omega_{m}+\eta,\omega_{m}-\eta\}$ and $\textbf{H}_{8}=\text{diag}\{\omega_ {m}+2\eta\cos(\pi/9),\omega_{m}+2\eta\cos(2\pi/9),\omega_{m}+\eta,\omega_{m}+2\eta\cos(4\pi/9),\omega_{m}+2\eta\cos(5\pi/9),\omega_{m}-\eta,\omega_{m}+2\eta\cos(7\pi/9),\omega_{m}+2\eta\cos(8\pi/9)\}$. We find that there are two degenerate frequencies $\omega_{m}\pm\eta$ in $\textbf{H}_{2}$ and $\textbf{H}_{8}$. Based on Theorem 1(ii), there are two dark modes in this network.

To evaluate the dark-mode effect in the above three cases, we check the cooling performance of these mechanical modes. In Fig.~\ref{Figchaincooling}, we plot the final mean phonon numbers $n^{f}_{lj}$ and $n^{f}_{rj^{\prime}}$ in all these mechanical modes corresponding to the three cases shown in Figs.~\ref{FigS7}(b),~\ref{FigS7}(c), and~\ref{FigS7}(d). Here we can see that, though the mean phonon number for some mechanical modes could be smaller than or close to one, the simultaneous ground-state cooling of all these mechanical modes cannot be realized. These cooling results indicate that there exist mechanical dark modes in these three coupling configurations. These cooling results are consistent with our analytical discussions concerning the dark-mode effect in these coupling configurations.

\section{Cooling of multiple mechanical modes in the ($M+N$)-mode optomechanical networks including $M$ optical modes and $N$ mechanical modes \label{Nmodel}}

In this section, we evaluate the dark-mode effect in the ($M+N$)-mode optomechanical networks when $(M,N)=\{(2,2), (2,4), (3,2)$, and $(3,3)\}$. We also examine the dark-mode effect by checking the cooling of these mechanical modes.

\begin{figure}[tbp]
	\center
	\includegraphics[width=0.6 \textwidth]{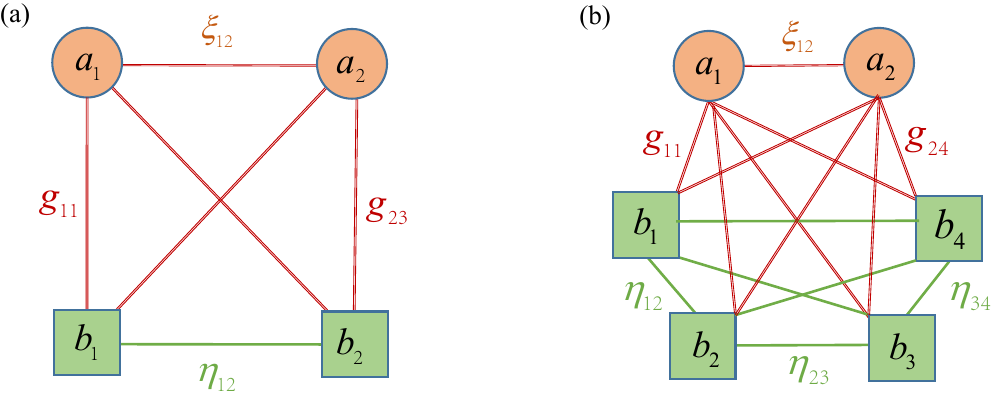}
	\caption{(a) Four-mode optomechanical network consisting of two optical modes and two mechanical modes.  The two optical modes are optomechanically coupled to the two mechanical modes, and the two optical (mechanical) modes are coupled via the photon-hopping (phonon-hopping) interactions. (b) Six-mode optomechanical network consisting of two optical modes optomechanically coupled to four mechanical modes.  The two optical (four mechanical) modes are coupled via the photon-hopping (phonon-hopping) interactions.}
	\label{FigS9}
\end{figure}

\subsection{Two-optical-mode and two-mechanical-mode optomechanical networks \label{modeltwotwo}}
We first consider a (2+2)-mode optomechanical network consisting of two optical modes and two mechanical modes  [see Fig.~\ref{FigS9}(a)]. In the case of $\delta_1=\delta_2=\delta$ and $\omega_{1}=\omega_{2}=\omega_{m}$,  based on Eq. (\ref{MNRWA}),
the approximate linearized Hamiltonian for this ($2+2$)-mode optomechanical network can be written as
\begin{equation}
	H_{\text{RWA}}^{[2,2]}=( a_{1}^{\dagger}, a_{2}^{\dagger}, b_{1}^{\dagger},b_{2}^{\dagger}) \mathbf{H}_{ab}^{[2,2]}(a_{1}, a_{2}, b_{1},b_{2})^{T},
\end{equation}
where the coefficient matrix in the bare-mode representation reads
\begin{equation}
	\mathbf{H}^{[2,2]}_{ab}=\left(
	\begin{array}{cccc}
		\delta  & \xi _{12} & g_{11} & g_{12}   \\
		\xi^{\ast}_{12} & \delta    & g_{21} & g_{22}  \\
		g^{\ast}_{11} & g^{\ast}_{21}   & \omega _{m} & \eta _{12} \\
		g^{\ast}_{12} & g^{\ast}_{22}& \eta^{\ast}_{12} & \omega _{m}
	\end{array}\right).
\end{equation}
In the normal-mode ($A_k,B_j$) representation, the Hamiltonian can be written as
\begin{equation}
	H_{\text{RWA}}^{[2,2]}=(A_{1}^{\dagger},A_{2}^{\dagger},B_{1}^{\dagger},B_{2}^{\dagger}) \mathbf{H}_{AB}^{[2,2]}( A_{1},A_{2},B_{1}, B_{2})^{T},
\end{equation}
where the coefficient matrix in the normal-mode representation becomes
\begin{equation}
	\label{matrix22}
	\mathbf{H}_{AB}^{[22]}=\left(\begin{array}{cc}
		\textbf{H}_{A} & \textbf{C}_{AB}\\
		\textbf{C}_{AB}^{\dagger} & \textbf{H}_{B}
	\end{array}\right)=\left(
	\begin{array}{ccccc}
		\delta-\xi_{12} & 0  &G_{11} &  G_{12}   \\
		0 & \delta+\xi_{12}   & G_{21}   & G_{22}   \\
	G^{\ast}_{11}  &G^{\ast}_{21}   & \omega _{m}-\eta _{12} & 0  \\
	G^{\ast}_{12} & G^{\ast}_{22} & 0 & \omega _{m}+\eta _{12}
	\end{array}\right).
\end{equation}
Here, $G_{kj}=\sum_{k^{\prime}=1}^{2}\sum_{j^{\prime}=1}^{2}(\textbf{U}_a)_{kk^{\prime}}g_{k^{\prime}j^{\prime}}(\textbf{U}_b^{\dagger})_{j^{\prime}j}$ is the coupling strength between the optical normal mode $A_{k}$ and the mechanical normal mode $B_j=\sum_{j^{\prime}=1}^{2}(\textbf{U}_b)_{jj^{\prime}}b_{j^{\prime}}$.

\begin{figure}[tbp]
	\center
	\includegraphics[width=0.6 \textwidth]{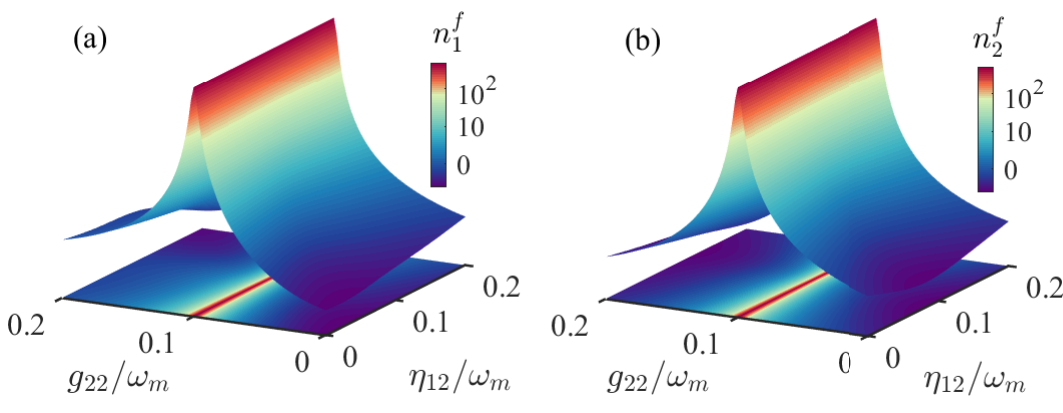}
	\caption{Final mean phonon numbers  (a) $n_{1}^{f}$  and (b) $n_{2}^{f}$ versus the scaled coupling strengths $\eta_{12}/\omega_{m}$ and  $g_{22}/\omega_{m}$.  Other parameters are $\delta_{1}/\omega_{m}=\delta_{2}/\omega_{m}=1$,  $\omega_{1}/\omega_{m}=\omega_{2}/\omega_{m}=1$, $\kappa_{1}/\omega_{m}=\kappa_{2}/\omega_{m}=0.1$, $\gamma_{1}/\omega_{m}=\gamma_{2}/\omega_{m}=10^{-5}$, $g_{11}/\omega_{m}=g_{12}/\omega_{m}=g_{21}/\omega_{m}=0.1$, $\xi_{12}/\omega_{m}=0.08$, and  $\bar{n}_{1}=\bar{n}_{2}=10^{3}$.}
	\label{FigS10}
\end{figure}

Below, we verify the dark-mode theorems by examining the cooling performance of the two mechanical modes. In Figs.~\ref{FigS10}(a) and~\ref{FigS10}(b), we display the final mean phonon numbers $n_{1}^{f}$  and $n_{2}^{f}$ versus $\eta_{12}/\omega_{m}$ and $g_{22}/\omega_{m}$.  We can find that the two mechanical modes cannot be cooled into their ground states when $g_{22}/\omega_{m}=0.1$, which corresponds to the case where a dark mode exists in the network. This result can be explained according to the dark-mode theorem. For the parameters used in Fig.~\ref{FigS10}, the coupling matrix $\textbf{C}_{AB}$ becomes
\begin{equation}
\mathbf{C}_{AB}=\frac{1}{2}\left(\begin{array}{cc}
	g_{22}-0.1\omega_{m} & g_{22}-0.1\omega_{m}\\
	g_{22}-0.1\omega_{m} & g_{22}+0.3\omega_{m}
\end{array}\right).~\label{CAB222}
\end{equation}
When $g_{22}/\omega_{m}=0.1$,
we find that all the elements in the first column of the coupling matrix (\ref{CAB222}) are zero. Based on Theorem 2(i), we know that the mechanical normal mode $B_{1}= \sqrt{2}(b_{2}-b_{1})/2$ is decoupled from the two optical normal modes $A_{1}$ and $A_{2}$. Consequently, the mode $B_{1}$ becomes a dark mode, and the simultaneous ground-state cooling of modes $b_{1}$ and $b_{2}$  cannot be realized at $g_{22}/\omega_{m}=0.1$.

\subsection{Two-optical-mode and four-mechanical-mode optomechanical networks \label{modeltwofour}}
In this section, we consider the case of  $M=2$ and $N=4$, i.e., the two-optical-mode and four-mechanical-mode optomechanical network [see Fig. \ref{FigS9}(b)]. For simplicity, below we consider the case  $\delta_1=\delta_2=\delta$, $\omega_1=\omega_2=\omega_3=\omega_m$, and $\eta_{ij}=\eta$.  In this case, the Hamiltonian (\ref{MNRWA}) is reduced to
\begin{equation}
	H^{[2,4]}_{\text{RWA}}=(a_{1}^{\dagger},a_{2}^{\dagger},b_{1}^{\dagger},b_{2}^{\dagger},b_{3}^{\dagger},b_{4}^{\dagger}) \mathbf{H}_{ab}^{[2,4]}( a_{1},a_{2},b_{1}, b_{2},b_{3},b_{4})^{T},
\end{equation}
where the coefficient matrix in the bare-mode representation reads
\begin{equation}
	\mathbf{H}^{[2,4]}_{ab}=\left(
	\begin{array}{cccccc}
		\delta  & \xi _{12} & g_{11} & g_{12} & g_{13}   & g_{14} \\
		\xi^{\ast} _{12} & \delta  & g_{21}  & g_{22} & g_{23}  & g_{24} \\
		g^{\ast}_{11} & g^{\ast}_{21}  & \omega _{m} &  \eta &  \eta &  \eta  \\
		g^{\ast}_{12} & g^{\ast}_{22}  &  \eta^{\ast}  & \omega _{m} & \eta &  \eta   \\
		g^{\ast}_{13} & g^{\ast}_{23} &   \eta^{\ast}  & \eta^{\ast}  & \omega _{m}& \eta _{34}\\
		g^{\ast}_{14} & g^{\ast}_{24} &   \eta ^{\ast} & \eta^{\ast}  & \eta^{\ast} & \omega _{m}
	\end{array}\right).
\end{equation}
In the normal-mode ($A_k,B_j$) representation, the Hamiltonian becomes
\begin{equation}
	H_{AB}^{[2,4]}=(A_{1}^{\dagger},A_{2}^{\dagger},B_{1}^{\dagger},B_{2}^{\dagger},B_{3}^{\dagger},B_{4}^{\dagger}) \mathbf{H}_{AB}^{[24]}( A_{1},A_{2},B_{1}, B_{2},B_{3},B_{4})^{T},
\end{equation}
where the coefficient matrix in the normal-mode representation becomes
\begin{eqnarray}
	\label{arrow24AB}
\mathbf{H}_{AB}^{[2,4]}=\left(\begin{array}{cc}
	\textbf{H}_{A} & \textbf{C}_{AB}\\
	\textbf{C}_{AB}^{\dagger} & \textbf{H}_{B}
\end{array}\right)=\left(\begin{array}{cccccc}
	\delta-\xi_{12} & 0 & G_{11} & G_{12} & G_{13} & G_{14}\\
	0 & \delta+\xi_{12} & G_{21} & G_{22} & G_{23} & G_{24}\\
	G_{11}^{\ast} & G_{21}^{\ast} & \omega_{m}-\eta & 0 & 0 & 0\\
	G_{12}^{\ast} & G_{22}^{\ast} & 0 & \omega_{m}-\eta & 0 & 0\\
	G_{13}^{\ast} & G_{23}^{\ast} & 0 & 0 & \omega_{m}-\eta & 0\\
	G_{14}^{\ast} & G_{24}^{\ast} & 0 & 0 & 0 & \omega_{m}+3\eta
\end{array}\right).
\end{eqnarray}
Here, $G_{kj}=\sum_{k^{\prime}=1}^{2}\sum_{j^{\prime}=1}^{4}(\textbf{U}_a)_{kk^{\prime}}g_{k^{\prime}j^{\prime}}(\textbf{U}_b^{\dagger})_{j^{\prime}j}$ is the coupling strength between the optical normal mode $A_{k}$ and the mechanical normal mode $B_j=\sum_{j^{\prime}=1}^{4}(\textbf{U}_b)_{jj^{\prime}}b_{j^{\prime}}$.

\begin{figure}[tbp]
	\center
	\includegraphics[width=0.56 \textwidth]{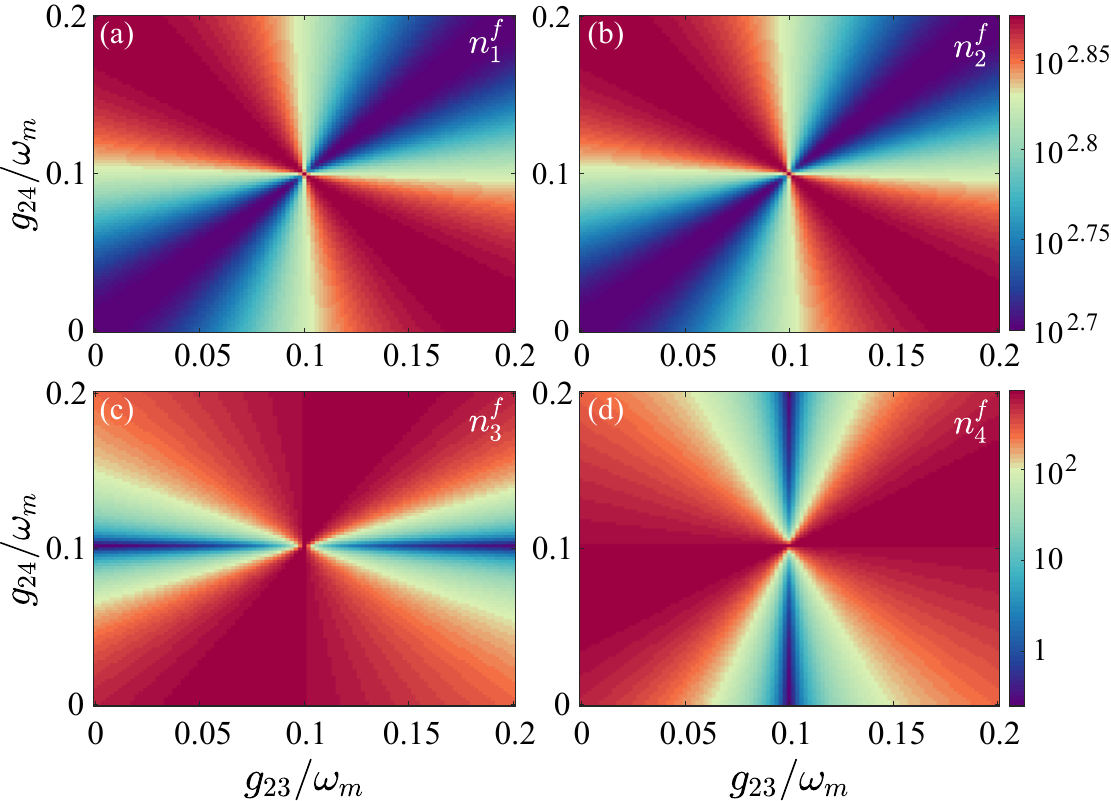}
	\caption{Final mean phonon numbers  (a) $n_{1}^{f}$,  (b) $n_{2}^{f}$,  (c) $n_{3}^{f}$, and (d) $n_{4}^{f}$  versus the scaled coupling strengths $g_{23}/\omega_{m}$ and  $g_{24}/\omega_{m}$.  Other parameters used are $\delta_{k=1\text{-}2}/\omega_{m}=1$,  $\omega_{j=1\text{-}4}/\omega_{m}=1$, $\kappa_{k=1\text{-}2}/\omega_{m}=0.1$, $\gamma_{j=1\text{-}4}/\omega_{m}=10^{-5}$, $g_{11}/\omega_{m}=g_{12}/\omega_{m}=g_{13}/\omega_{m}=g_{14}/\omega_{m}=g_{21}/\omega_{m}=g_{22}/\omega_{m}=0.1$, $\xi_{12}/\omega_{m}=0.08$, $\eta _{12}/\omega_{m}=\eta _{13}/\omega_{m}=\eta _{14}/\omega_{m}=\eta _{23}/\omega_{m}=\eta _{24}/\omega_{m}=\eta _{34}/\omega_{m}=0.09$, and  $\bar{n}_{j=1\text{-}4}=10^{3}$.}
	\label{FigS11}
\end{figure}

In Fig.~\ref{FigS11}, we plot the final mean phonon numbers $n_{1}^{f}$, $n_{2}^{f}$,  $n_{3}^{f}$, and  $n_{4}^{f}$ as functions of $g_{23}/\omega_{m}$ and $g_{24}/\omega_{m}$.  We see from Figs.~\ref{FigS11}(a) and \ref{FigS11}(b) that the simultaneous ground-state cooling of modes $b_{1}$ and $b_{2}$  are unaccessible no matter what values $g_{23}/\omega_{m}$ and $g_{24}/\omega_{m}$ take. Figure~\ref{FigS11}(c)  shows that the mode $b_{3}$ can be cooled into its ground state when $g_{24}/\omega_{m}=0.1$ and $g_{23}/\omega_{m}\neq0.1$. Similar to Fig.~\ref{FigS11}(c), Fig.~\ref{FigS11}(d)  shows that the ground-state cooling of mode $b_{4}$ can  be achieved when $g_{23}/\omega_{m}=0.1$ and $g_{24}/\omega_{m}\neq0.1$. These results can be explained based on the dark-mode theorem.
For the parameters used in Fig.~\ref{FigS11}, the coupling matrix $\textbf{C}_{AB}$ becomes
\begin{equation}
	\textbf{C}_{AB}=\left(\begin{array}{cccc}
		\frac{1}{2}(g_{24}-0.1\omega_{m}) & \frac{\sqrt{3}}{6}(2g_{23}-0.1\omega_{m}-g_{24}) & \frac{\sqrt{6}}{12}(0.2\omega_{m}-g_{23}-g_{24}) & \frac{\sqrt{2}}{4}(g_{23}+g_{24}-0.2\omega_{m})\\
		\frac{1}{2}(g_{24}-0.1\omega_{m}) & \frac{\sqrt{3}}{6}(2g_{23}-0.1\omega_{m}-g_{24}) & \frac{\sqrt{6}}{12}(0.2\omega_{m}-g_{23}-g_{24}) & \frac{\sqrt{2}}{4}(g_{23}+g_{24}+0.6\omega_{m})
	\end{array}\right).~\label{CAB244}
\end{equation}
The coupling matrix~(\ref{CAB244}) shows  that the elements in the first, second, and third columns are linearly dependent. Based on Theorem 2(ii), we know that two
modes superposed by the normal modes $B_{1}$,  $B_{2}$, and $B_{3}$ become dark modes, thus the simultaneous ground-state cooling of the modes $b_{1}$ and $b_{2}$ is strongly suppressed in all areas. However, when $g_{24}/\omega_{m}=0.1$ and $g_{23}/\omega_{m}\neq0.1$, the two dark modes can be expressed as $B_{2-}=( -b_{1}+b_{4}) /\sqrt{2}$ and $B_{3-}=( -b_{1}+2b_{2}-b_{4}) /\sqrt{6}$, which indicates that the mode $b_{3}$ is connected to the optical modes, so the ground state cooling of the mode $b_{3}$ can be realized in this case. Similarly, when $g_{23}/\omega_{m}=0.1$ and $g_{24}/\omega_{m}\neq0.1$, the two dark modes can be expressed as $B_{2-}=( -2b_{1}+3b_{2}-b_{3}) /\sqrt{14}$ and $B_{3-}=( -4b_{1}-b_{2}+5b_{3}) /\sqrt{42}$. In this case,  the  mode $b_{4}$ is connected to the optical modes, and it can be cooled into its ground state.

\subsection{Three-optical-mode and two-mechanical-mode optomechanical networks \label{modelthreetwo}}

\begin{figure}[tbp]
	\center
	\includegraphics[width=0.6 \textwidth]{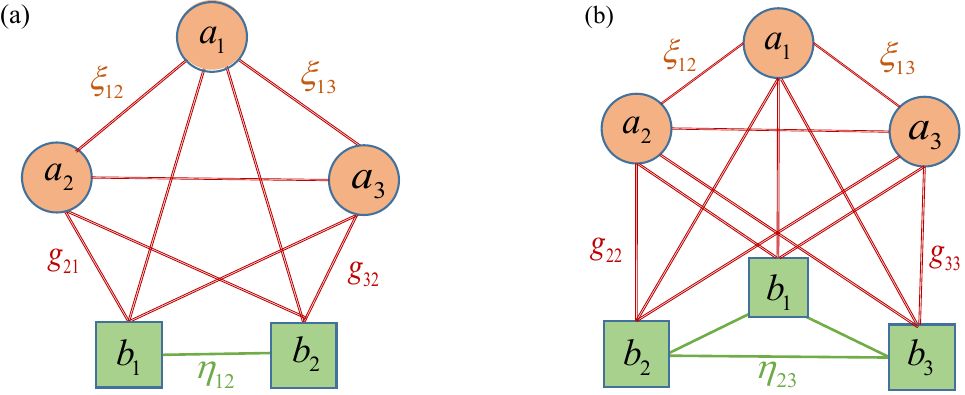}
	\caption{(a) Five-mode optomechanical network consisting of three optical modes optomechanically coupled to two mechanical modes, and all the two optical (mechanical) modes are coupled via the photon-hopping (phonon-hopping) interactions. (b) Six-mode optomechanical network consisting of three optical modes and three mechanical modes. Here the two-node couplings exist among three optical modes and three mechanical modes.}
	\label{FigS12}
\end{figure}

We next consider a five-mode optomechanical network consisting of three optical modes and two mechanical modes, as shown in Fig. \ref{FigS12}(a). Here the optical modes are coupled to the mechanical modes (optical modes) via the optomechanical interactions (photon-hopping interactions),  and the two mechanical modes are coupled to each other through the phonon-hopping interactions. Below, we will consider the case  $\delta_1=\delta_2=\delta_3=\delta$, $\xi_{12}=\xi_{13}=\xi_{23}=\xi$, and $\omega_{1}=\omega_{2}=\omega_{m}$. In the bare-mode ($a_{k},b_{j}$) representation, Hamiltonian (\ref{MNRWA}) is reduced to
\begin{equation}
	H_{\text{RWA}}^{[3,2]}=( a_{1}^{\dagger}, a_{2}^{\dagger}, a_{3}^{\dagger}, b_{1}^{\dagger}, b_{2}^{\dagger}) \mathbf{H}_{ab}^{[3,2]}( a_{1}, a_{2}, a_{3}, b_{1}, b_{2})^{T},
\end{equation}
where the coefficient matrix in the bare-mode representation is given by
\begin{equation}
	\mathbf{H}_{ab}^{[3,2]}=\left(
	\begin{array}{ccccc}
		\delta & \xi & \xi & g_{11} & g_{12}   \\
		\xi ^{\ast} & \delta  & \xi  & g_{21} & g_{22}  \\
		\xi^{\ast}  & \xi ^{\ast}  & \delta  & g_{31} & g_{32}  \\
		g^{\ast}_{11} & g^{\ast}_{21}  & g^{\ast}_{31} & \omega_m & \eta _{12} \\
		g^{\ast}_{12} & g^{\ast}_{22} &  g^{\ast}_{32} & \eta ^{\ast}_{12} & \omega_m
	\end{array}\right).
\end{equation}
In the normal-mode ($A_{k},B_{k}$) representation, the Hamiltonian becomes
\begin{equation}
	H_{\text{RWA}}^{[3,2]}=(A_{1}^{\dagger},A_{2}^{\dagger},A_{3}^{\dagger},B_{1}^{\dagger},B_{2}^{\dagger}) \mathbf{H}_{AB}^{[3,2]}( A_{1},A_{2},A_{3},B_{1}, B_{2})^{T},
\end{equation}
where the coefficient matrix in the normal-mode representation becomes
\small
\begin{equation}
	\label{arrow32AB}
\mathbf{H}_{AB}^{[3,2]}=\left(\begin{array}{cc}
	\textbf{H}_{A} & \textbf{C}_{AB}\\
	\textbf{C}_{AB}^{\dagger} & \textbf{H}_{B}
\end{array}\right)=\left(\begin{array}{ccccc}
	\delta-\xi & 0 & 0 & G_{11} & G_{12}\\
	0 & \delta-\xi & 0 & G_{21} & G_{22}\\
	0 & 0 & \delta+2\xi & G_{31} & G_{32}\\
	G_{11}^{\ast} & G_{21}^{\ast} & G_{31}^{\ast} & \omega_{m}-\eta_{12} & 0\\
	G_{12}^{\ast} & G_{22}^{\ast} & G_{32}^{\ast} & 0 & \omega_{m}+\eta_{12}
\end{array}\right).
\end{equation}
\normalsize
Here, $G_{kj}=\sum_{k^{\prime}=1}^{3}\sum_{j^{\prime}=1}^{2}(\textbf{U}_a)_{kk^{\prime}}g_{k^{\prime}j^{\prime}}(\textbf{U}_b^{\dagger})_{j^{\prime}j}$ is the coupling strength between the optical normal mode $A_{k}$ and the mechanical normal mode $B_j=\sum_{j^{\prime}=1}^{2}(\textbf{U}_b)_{jj^{\prime}}b_{j^{\prime}}$.

\begin{figure}[tbp]
	\center
	\includegraphics[width=0.6 \textwidth]{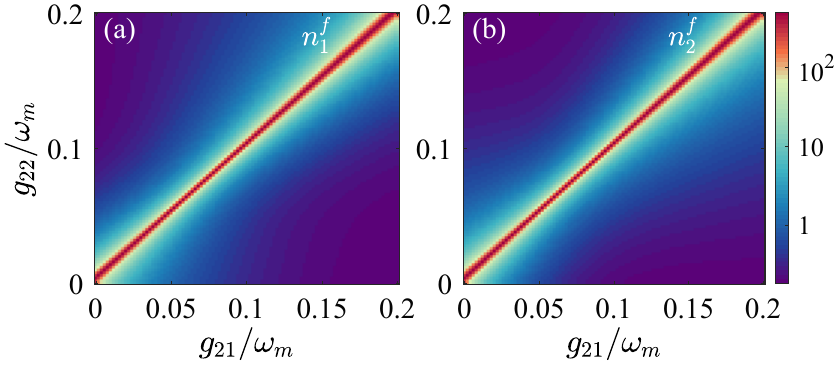}
	\caption{Final mean phonon numbers  (a) $n_{1}^{f}$  and (b) $n_{2}^{f}$ as functions of the coupling strengths $g_{21}/\omega_{m}$ and  $g_{22}/\omega_{m}$.  Other parameters used are $\delta_{k=1\text{-}3}/\omega_{m}=1$,  $\omega_{j=1\text{-}2}/\omega_{m}=1$, $\kappa_{k=1\text{-}3}/\omega_{m}=0.1$, $\gamma_{j=1\text{-}2}/\omega_{m}=10^{-5}$, $g_{11}/\omega_{m}=g_{12}/\omega_{m}=g_{31}/\omega_{m}=g_{32}/\omega_{m}=0.1$,  $\xi _{12}/\omega_{m}=\xi _{13}/\omega_{m}= \xi _{23}/\omega_{m}=0.08$, $\eta_{12}/\omega_{m}=0.09$, and  $\bar{n}_{1}=\bar{n}_{2}=10^{3}$.}
	\label{FigS13}
\end{figure}

To verify the dark-mode theorems, we show in Fig.~\ref{FigS13} the final mean phonon numbers $n_{1}^{f}$  and $n_{2}^{f}$ as functions of $g_{21}/\omega_{m}$ and $g_{22}/\omega_{m}$.  Here we can see that the two mechanical modes cannot be cooled into the ground states when $g_{21}/\omega_{m}$ and $g_{22}/\omega_{m}$ have equal or close values, as shown by the red area along the counter-diagonal line,  which corresponds to the appearance of dark modes in this network. This result can be explained according to the dark-mode theorem. For the  parameters used in Fig.~\ref{FigS13}, the coupling matrix $\textbf{C}_{AB}$ can be expressed as
\begin{equation}
\textbf{C}_{AB}=\left(\begin{array}{cc}
	0 & 0\\
	\frac{1}{\sqrt{3}}(g_{22}-g_{21}) & \frac{1}{\sqrt{3}}(g_{21}+g_{22}-0.2\omega_{m})\\
	\frac{1}{\sqrt{6}}(g_{22}-g_{21}) & \frac{1}{\sqrt{6}}(g_{21}+g_{22}+0.4\omega_{m})
\end{array}\right).~\label{CAB323}
\end{equation}
When $g_{21}=g_{22}$, all the elements in the first column of coupling matrix (\ref{CAB323}) are zero. Based on Theorem 2(i), we know that the normal mode $B_{1}=(b_{2}-b_{1})/\sqrt{2}$ becomes a dark mode, thus the simultaneous ground-state cooling of modes $b_{1}$ and $b_{2}$ cannot be achieved.

\subsection{Three-optical-mode and three-mechanical-mode optomechanical networks \label{modelthreethree}}

We finally consider a six-mode optomechanical network consisting of three optical modes and three mechanical modes, as shown in Fig.~\ref{FigS12}(b). In the following, we will consider the case $\delta_{1}=\delta_{2}=\delta_{2}=\delta$, $\xi_{12}=\xi_{13}=\xi_{23}=\xi$, $\eta_{12}=\eta_{13}=\eta_{23}=\eta$, and $\omega_{1}=\omega_{2}=\omega_{3}=\omega_{m}$. In this case, the Hamiltonian of the network in the bare-mode ($a_{k},b_{j}$) representation can be written as
\begin{equation}
	H_{\text{RWA}}^{[3,3]}=( a_{1}^{\dagger}, a_{2}^{\dagger}, a_{3}^{\dagger}, b_{1}^{\dagger}, b_{2}^{\dagger},b_{3}^{\dagger}) \mathbf{H}_{ab}^{[3,3]}( a_{1}, a_{2}, a_{3}, b_{1}, b_{2},b_{3})^{T},
\end{equation}
where the coefficient matrix in the bare-mode representation reads
\begin{equation}
	\mathbf{H}_{ab}^{[3,3]}=\left(
	\begin{array}{cccccc}
		\delta & \xi  & \xi & g_{11} & g_{12} & g_{13}  \\
		\xi ^{\ast} & \delta   & \xi  & g_{21} & g_{22} & g_{23} \\
		\xi ^{\ast}& \xi ^{\ast}  & \delta & g_{31} & g_{32}  & g_{33}\\
		g^{\ast}_{11} & g^{\ast}_{21}  & g^{\ast}_{31} & \omega _{m} & \eta _{12}& \eta _{13} \\
		g^{\ast}_{12} & g^{\ast}_{22} &  g^{\ast}_{32} & \eta^{\ast} & \omega _{m}& \eta _{23}\\
		g^{\ast}_{13} & g^{\ast}_{23} &  g^{\ast}_{33} & \eta ^{\ast} & \eta^{\ast}& \omega _{m}
	\end{array}\right).
\end{equation}
The Hamiltonian in the normal-mode ($A_{k},B_{j}$) representation can be expressed as
\begin{equation}
	H_{\text{RWA}}^{[3,3]}=(A_{1}^{\dagger},A_{2}^{\dagger},A_{3}^{\dagger},B_{1}^{\dagger},B_{2}^{\dagger},B_{3}^{\dagger}) \mathbf{H}_{AB}^{[33]}( A_{1},A_{2},A_{3},B_{1}, B_{2},B_{3})^{T},
\end{equation}
where the coefficient matrix in the normal-mode representation is given by
\begin{equation}
	\label{matrixm33}
\mathbf{H}_{AB}^{[3,3]}=\left(\begin{array}{cc}
	\textbf{H}_{A} & \textbf{C}_{AB}\\
	\textbf{C}_{AB}^{\dagger} & \textbf{H}_{B}
\end{array}\right)=\left(\begin{array}{cccccc}
	\delta-\xi & 0 & 0 & G_{11} & G_{12} & G_{13}\\
	0 & \delta-\xi & 0 & G_{21} & G_{22} & G_{23}\\
	0 & 0 & \delta+2\xi & G_{31} & G_{32} & G_{33}\\
	G_{11}^{\ast} & G_{21}^{\ast} & G_{31}^{\ast} & \omega_{m}-\eta & 0 & 0\\
	G_{12}^{\ast} & G_{22}^{\ast} & G_{32}^{\ast} & 0 & \omega_{m}-\eta & 0\\
	G_{13}^{\ast} & G_{23}^{\ast} & G_{33}^{\ast} & 0 & 0 & \omega_{m}+2\eta
\end{array}\right).
\end{equation}
Here, $G_{kj}=\sum_{k^{\prime}=1}^{3}\sum_{j^{\prime}=1}^{3}(\textbf{U}_a)_{kk^{\prime}}g_{k^{\prime}j^{\prime}}(\textbf{U}_b^{\dagger})_{j^{\prime}j}$ is the coupling strength between the optical normal mode $A_{k}$ and the mechanical normal mode $B_j=\sum_{j^{\prime}=1}^{3}(\textbf{U}_b)_{jj^{\prime}}b_{j^{\prime}}$.

\begin{figure}[tbp]
	\center
	\includegraphics[width=0.85 \textwidth]{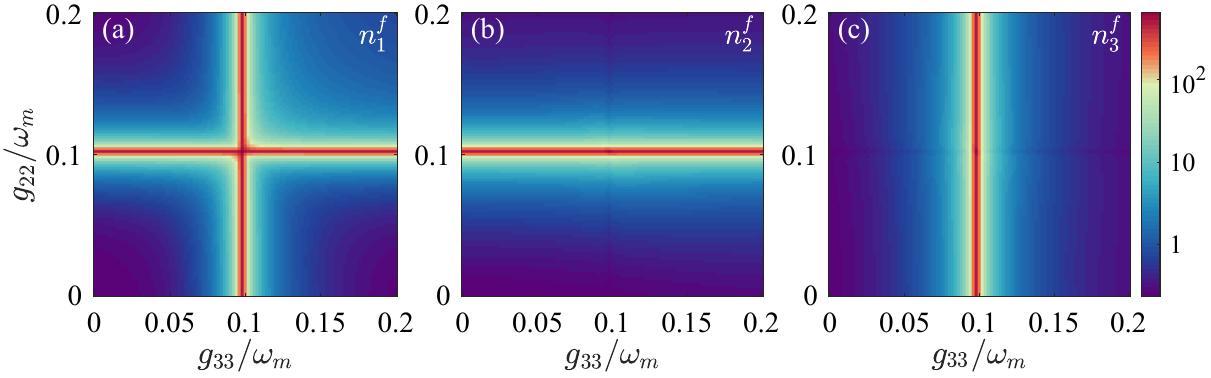}
	\caption{Final mean phonon numbers  (a) $n_{1}^{f}$, (b) $n_{2}^{f}$, and (c) $n_{3}^{f}$ as functions of the coupling strengths $g_{22}/\omega_{m}$ and  $g_{33}/\omega_{m}$.  Other parameters are $\delta_{k=1\text{-}3}/\omega_{m}=1$,  $\omega_{j=1\text{-}3}/\omega_{m}=1$, $\kappa_{k=1\text{-}3}/\omega_{m}=0.1$, $\gamma_{j=1\text{-}3}/\omega_{m}=10^{-5}$, $g_{11}/\omega_{m}=g_{12}/\omega_{m}=g_{13}/\omega_{m}=g_{21}/\omega_{m}=g_{23}/\omega_{m}=g_{31}/\omega_{m}=g_{32}/\omega_{m}=0.1$,  $\xi _{12}/\omega_{m}=\xi _{13}/\omega_{m}= \xi _{23}/\omega_{m}=0.08$, $\eta_{12}/\omega_{m}=\eta_{13}/\omega_{m}=\eta_{23}/\omega_{m}=0.09$, and  $\bar{n}_{j=1\text{-}3}=10^{3}$.}
	\label{FigS14}
\end{figure}

In Fig.~\ref{FigS14}, we plot the final mean phonon numbers $n_{1}^{f}$, $n_{2}^{f}$,  and  $n_{3}^{f}$ as functions of $g_{22}/\omega_{m}$ and $g_{33}/\omega_{m}$.  We see from Fig.~\ref{FigS14} that the simultaneous ground-state cooling of modes $b_{1}$ and $b_{3}$  cannot be realized when $g_{33}/\omega_{m}=0.1$. Similarly, the simultaneous ground-state cooling of modes $b_{1}$ and $b_{2}$ is unaccessible when $g_{22}/\omega_{m}=0.1$. These phenomena can be explained according to the dark-mode theorem.
Based on the parameters used in Fig.~\ref{FigS14}, the coupling matrix $\textbf{C}_{AB}$ becomes
\begin{equation}
	\label{matrixm33par}
	\textbf{C}_{AB}=\left(
	\begin{array}{ccc}
		\frac{1}{2}\left( g_{33}-0.1\omega_{m}\right)  & \frac{\sqrt{3}}{6}\left(
		0.1\omega_{m}-g_{33}\right)  & \frac{\sqrt{6}}{6}g_{33}-\frac{\sqrt{6}}{60}\omega_{m} \\
		-\frac{\sqrt{3}}{6}\left( g_{33}-0.1\omega_{m}\right)  & \frac{2}{3}g_{22}+\frac{1}{6}
		g_{33}-\frac{1}{12}\omega_{m} & \frac{\sqrt{2}}{3}g_{22}-\frac{\sqrt{2}}{6}g_{33}-
		\frac{\sqrt{2}}{60}\omega_{m} \\
		\frac{\sqrt{6}}{6}\left( g_{33}-0.1\omega_{m}\right)  & \frac{\sqrt{2}}{3}g_{22}-\frac{
			\sqrt{2}}{6}g_{33}-\frac{\sqrt{2}}{60}\omega_{m} & \frac{1}{30}\omega_{m}+\frac{1}{3}g_{22}+
		\frac{1}{3}g_{33}
	\end{array}
	\right).
\end{equation}
When $g_{33}/\omega_{m}=0.1$, we find that all the elements in the first column of the matrix~(\ref{matrixm33par}) are zero, which means that $B_{1}=(b_{3}-b_{1})/\sqrt{2}$  becomes a dark mode, so the  modes $b_{1}$ and $b_{3}$  cannot be simultaneously cooled into their ground states. When $g_{22}/\omega_{m}=0.1$, the elements in the first column  and the second column are linearly dependent,  based on Theorem 2(ii), the dark mode can be obtained as $B_{2-}=(b_{2}-b_{1})/\sqrt{2}$, so the simultaneous ground-state cooling of modes $b_{1}$ and $b_{2}$  cannot be realized.

\section{Analyzing the dark-state effect in driven atom systems based on the dark-mode theorems ~\label{thedarkstate}}
\begin{figure}[tbp]
	\center
	\includegraphics[width=0.6 \textwidth]{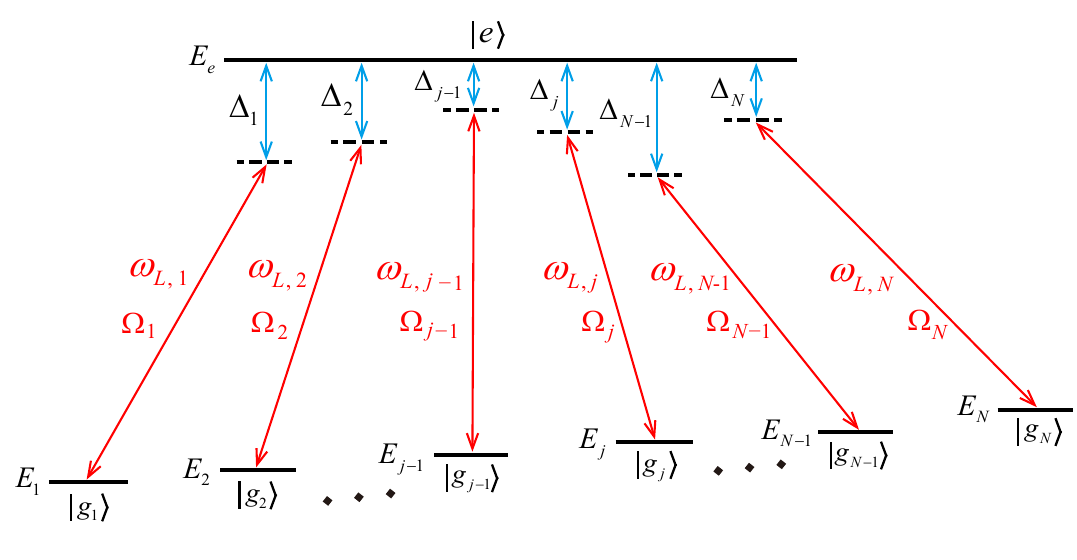}
	\caption{Schematic of a multi-level atomic system consisting
		of one excited state $\vert e\rangle$ and $N$ lower states $\vert g_{j}\rangle$ ($j=1,2,\dotsb,N$).
		The energy levels $\vert e\rangle$ and $\vert g_{j}\rangle$
		are excited by external driving fields with the driving frequency $\omega_{L,j}$
		and driving amplitude $\Omega_{j}$. The parameter $\Delta_j\equiv E_e-E_j-\omega_{L,j}$ is the detuning of the energy separation $E_e-E_j$ with respect to the driving frequency $\omega_{L,j}$.}
	\label{FigS15}
\end{figure}

In this section, we show that the dark-mode theorems can be used to evaluate the dark states in driven atom systems. As an example, we consider a multi-level atomic system consisting
of one excited state $\vert e\rangle$ and $N$ lower states $\vert g_{j}\rangle$ ($j=1,2,\dotsb,N$), and the transitions between the energy levels $\vert e\rangle$ and $\vert g_{j}\rangle$
are excited by external driving fields with the driving frequency $\omega_{L,j}$
and driving amplitude $\Omega_{j}$,  as shown in Fig.~\ref{FigS15}. The Hamiltonian of
the system reads
\begin{equation}
	H=\sum_{j=1}^{N}E_{j}\vert g_{j}\rangle\langle g_{j}\vert+E_{e}\vert e\rangle\langle e\vert+\sum_{j=1}^{N}(\Omega_{j}e^{-i\omega_{L,j}t}\vert e\rangle\langle g_{j}\vert+\Omega_{j}^{\ast}e^{i\omega_{L,j}t}\vert g_{j}\rangle\langle e\vert),\label{HEG}
\end{equation}
where $E_{j}$ ($E_{e}$) is the eigen energy of the energy level $\vert g_{j}\rangle$
($\vert e\rangle$). In a rotating frame defined by $\exp(-itE_{e}\vert e\rangle\langle e\vert-it\sum_{j=1}^{N}\varepsilon_{j}\vert g_{j}\rangle\langle g_{j}\vert)$,
the Hamiltonian (\ref{HEG}) becomes
\begin{eqnarray}
	H_{I} & = & \sum_{j=1}^{N}(E_{j}-\varepsilon_{j})\vert g_{j}\rangle\langle g_{j}\vert+\sum_{j=1}^{N}(\Omega_{j}\vert e\rangle\langle g_{j}\vert e^{it(E_{e}-\varepsilon_{j}-\omega_{L,j})}+\Omega_{j}^{\ast}\vert g_{j}\rangle\langle e\vert e^{-it(E_{e}-\varepsilon_{j}-\omega_{L,j})}).\label{HIATOM}
\end{eqnarray}
We consider the case $E_{e}-\varepsilon_{j}-\omega_{L,j}=0$ and define the
detuning $\Delta_{j}=E_{e}-E_{j}-\omega_{L,j}$, then the Hamiltonian
(\ref{HIATOM}) is reduced to
\begin{equation}
	H_{I}=-\sum_{j=1}^{N}\Delta_{j}\vert g_{j}\rangle\langle g_{j}\vert+\sum_{j=1}^{N}(\Omega_{j}\vert e\rangle\langle g_{j}\vert+\Omega_{j}^{\ast}\vert g_{j}\rangle\langle e\vert).
\end{equation}
By defining the bases
\begin{equation}
	\vert e\rangle=\left(\begin{array}{c}
		1\\
		0\\
		0\\
		0\\
		\vdots\\
		0
	\end{array}\right),\hspace{1em}\hspace{1em}\vert g_{1}\rangle=\left(\begin{array}{c}
		0\\
		1\\
		0\\
		0\\
		\vdots\\
		0
	\end{array}\right),\hspace{0.2em}\dotsb,\hspace{0.5em}\vert g_{j}\rangle=\left(\begin{array}{c}
		0\\
		0\\
		\vdots\\
		1_{j+1}\\
		\vdots\\
		0
	\end{array}\right),\hspace{0.2em}\dotsb,\hspace{0.5em}\vert g_{N}\rangle=\left(\begin{array}{c}
	0\\
	0\\
	\vdots\\
	0\\
	\vdots\\
	1
\end{array}\right),
\end{equation}
then the Hamiltonian (\ref{HIATOM}) can be expressed as an arrowhead matrix
\begin{equation}
	H_{I}=\left(\begin{array}{c|cccc}
		0 & \Omega_{1} & \Omega_{2} & \cdots & \Omega_{N}\\
		\hline \Omega_{1}^{\ast} & -\Delta_{1} & 0 & \cdots & 0\\
		\Omega_{2}^{\ast} & 0 & -\Delta_{2} & \cdots & 0\\
		\vdots & \vdots & \vdots & \ddots & \vdots\\
		\Omega_{N}^{\ast} & 0 & 0 & \cdots & -\Delta_{N}
	\end{array}\right).
\end{equation}
According to Theorem 1(ii), we know that, when $\Delta_{1}=\Delta_{2}=\dotsb=\Delta_{N}=\Delta$,
there exist one bright state
\begin{equation}
\vert g_{N+}\rangle=\frac{1}{\sqrt{\vert\Omega_{N}\vert^{2}+\vert\Omega_{(N-1)+}\vert^{2}}}(\Omega_{N}\vert g_{N}\rangle+\Omega_{(N-1)+}\vert g_{(N-1)+}\rangle)=\frac{1}{\sqrt{\sum_{j=1}^{N}\vert \Omega_{j}\vert^{2}}}\sum_{j=1}^{N}\Omega_{j}\vert g_j\rangle,
\end{equation}
and $N-1$ dark states: $\vert g_{2-}\rangle$, $\vert g_{3-}\rangle$,
..., and $\vert g_{N-}\rangle$, which are defined by
\begin{equation}
 \vert g_{j-}\rangle=\frac{1}{\sqrt{\vert\Omega_{j}\vert^{2}+\vert\Omega_{(j-1)+}\vert^{2}}}(\Omega_{j}^{\ast}\vert g_{(j-1)+}\rangle-\Omega_{(j-1)+}\vert g_{j}\rangle),
\end{equation}
with
\begin{equation}
\vert g_{(j-1)+}\rangle=\frac{1}{\sqrt{\vert\Omega_{(j-2)+}\vert^{2}+\vert\Omega_{(j-1)}\vert^{2}}}(\Omega_{(j-2)+}\vert g_{(j-2)+}\rangle+\Omega_{(j-1)}\vert g_{(j-1)}\rangle),
\end{equation}
and $\Omega_{N+}=\sqrt{\vert\Omega_{1N}\vert^{2}+\vert\Omega_{(N-1)+}\vert^{2}}=\sqrt{\vert\Omega_{11}\vert^{2}+\vert\Omega_{12}\vert^{2}+\dotsb+\vert\Omega_{1N}\vert^{2}}>0$.
Note that the forms of these dark states are not unique. Their forms depend on the grouping order of these lower levels $\vert g_{j=1-N}\rangle$.

\section{Analyzing the decoherence-free-subspace based on the dark-mode theorems~\label{DFS}}
In this section, we show that the decoherence-free subspace can be understood
as an atomic dark state with respect to their common environment
coupled to the two atoms. Concretely, we consider two two-level atoms coupled
to a common vacuum bath, which is formed by many bosonic modes (i.e., electromagnetic fields). The
Hamiltonian of the system reads
\begin{equation}
	H_{\text{DFS}}=\omega_{01}\sigma_{+}^{(1)}\sigma_{-}^{(1)}+\omega_{02}\sigma_{+}^{(2)}\sigma_{-}^{(2)}+\sum\limits _{k=1}^{M}\omega_{k}a_{k}^{\dagger}a_{k}+\sum\limits _{k=1}^{M}(J_{1k}a_{k}^{\dagger}\sigma_{-}^{(1)}+J_{1k}^{\ast}\sigma_{+}^{(1)}a_{k})+\sum\limits _{k=1}^{M}(J_{2k}a_{k}^{\dagger}\sigma_{-}^{(2)}+J^{*}_{2k}\sigma_{+}^{(2)}a_{k}),
\end{equation}
where $\sigma_{+}^{(j=1,2)}$ and $\sigma_{-}^{(j=1,2)}$ are the raising and lowering operators of the $j$th atom, with the energy separation $\omega_{0j}$, $a_{k}$ ($a_{k}^{\dagger}$) is the annihilation (creation) operator of the $k$th bath mode, and $J_{jk}$  is the coupling strength between the $j$th atom and the $k$th bath mode.

In this system, the total excitation number operator $\hat{N}=\sum_{j=1}^{2}\sigma_{+}^{(j)}\sigma_{-}^{(j)}+\sum_{k}a_{k}^{\dagger}a_{k}$ is a conserved quantity because of $[\hat{N},H_{\text{DFS}}]=0$,
thus the single-excitation subspace is spanned by the following bases
\{$\left\vert e\right\rangle _{1}\left\vert g\right\rangle _{2}\left\vert \emptyset\right\rangle ,\left\vert g\right\rangle _{1}\left\vert e\right\rangle _{2}\left\vert \emptyset\right\rangle ,\left\vert g\right\rangle _{1}\left\vert g\right\rangle _{2}\left\vert 1_{k}\right\rangle $\}
with $\left\vert 1_{k}\right\rangle =a_{k}^{\dagger}\left\vert \emptyset\right\rangle $.
By defining the following matrix vectors for these bases,
\begin{equation}
	\left\vert g\right\rangle _{1}\left\vert g\right\rangle _{2}\left\vert 1_{1}\right\rangle  =  \left(\begin{array}{c}
		1\\
		0\\
		\cdots\\
		0\\
		\cdots\\
		0\\
		0
	\end{array}\right),\dotsb,\left\vert g\right\rangle _{1}\left\vert g\right\rangle _{2}\left\vert 1_{k}\right\rangle =\left(\begin{array}{c}
		0\\
		0\\
		0\\
		\cdots\\
		1_{\dotsb k}\\
		\cdots\\
		0\\
		0
	\end{array}\right),\dotsb,
	\left\vert e\right\rangle _{1}\left\vert g\right\rangle _{2}\left\vert \emptyset\right\rangle   =  \left(\begin{array}{c}
		0\\
		0\\
		\cdots\\
		0\\
		\cdots\\
		1\\
		0
	\end{array}\right),\hspace{0.5em}\left\vert g\right\rangle _{1}\left\vert e\right\rangle _{2}\left\vert \emptyset\right\rangle =\left(\begin{array}{c}
		0\\
		0\\
		\cdots\\
		0\\
		\cdots\\
		0\\
		1
	\end{array}\right),
\end{equation}
the Hamiltonian $H_{\text{DFS}}$ in the single-excitation space can be expressed as
\begin{equation}
	H=\left(\begin{array}{ccccc|cc}
		\omega_{1} & 0 & \cdots & 0 & \cdots & J_{11} & J_{21}\\
		0 & \omega_{2} & \cdots & 0 & \cdots & J_{12} & J_{22}\\
		\cdots & \cdots & \cdots & \cdots & \cdots & \cdots & \cdots\\
		0 & 0 & \cdots & \omega_{k} & \cdots & J_{1k} & J_{2k}\\
		\cdots & \cdots & \cdots & \cdots & \cdots & \cdots & \cdots\\\hline
		J^{\ast}_{11} & J^{\ast}_{12} & \cdots & J^{\ast}_{1k} & \cdots & \omega_{01} & 0\\
		J^{\ast}_{21} & J^{\ast}_{22} & \cdots & J^{\ast}_{2k} & \cdots & 0 & \omega_{02}
	\end{array}\right).
\end{equation}
We know that when the two atoms are degenerate ($\omega_{01}=\omega_{02}=\omega_{0}$), and the two column vectors in the coupling matrix are linearly dependent ($J_{2k}=\lambda J_{1k}$
for $k=1,2,\dotsb$), then $(1+\vert\lambda\vert^{2})^{-1/2}\left(\left\vert e\right\rangle _{1}\left\vert g\right\rangle _{2}+\lambda\left\vert g\right\rangle _{1}\left\vert e\right\rangle _{2}\right)\left\vert \emptyset\right\rangle $ becomes a bright state,
and  $(1+\vert\lambda\vert^{2})^{-1/2}\left(\lambda^{\ast}\left\vert e\right\rangle _{1}\left\vert g\right\rangle _{2}-\left\vert g\right\rangle _{1}\left\vert e\right\rangle _{2}\right)\left\vert \emptyset\right\rangle $ becomes a dark state with respect to all the bath modes. The state $(1+\vert\lambda\vert^{2})^{-1/2}\left(\lambda^{\ast}\left\vert e\right\rangle _{1}\left\vert g\right\rangle _{2}-\left\vert g\right\rangle _{1}\left\vert e\right\rangle _{2}\right)\left\vert \emptyset\right\rangle $ is the basis of the decoherence-free subspace.

\end{document}